\documentclass[a4paper,fleqn,usenatbib]{mnras}

\usepackage{times}

\usepackage[T1]{fontenc}
\usepackage{ae,aecompl}

\usepackage{graphicx}	
\usepackage{amsmath}	
\usepackage{amssymb}	
\usepackage{multirow}
\usepackage{pdflscape}
\usepackage{nicefrac}
\usepackage{epstopdf}
\usepackage{url}
\usepackage{longtable}

\title[New WR stars in the Scutum-Crux Arm]
{A deep near-infrared spectroscopic survey of the Scutum-Crux arm for Wolf-Rayet 
stars\thanks{Based on  observations with ESO telescopes at the La Silla Paranal 
Observatory under programme 094.D-0839(A)}}

\author[C. K. Rosslowe and P. A. Crowther]{C. K. 
Rosslowe and Paul A. Crowther\thanks{Email: paul.crowther@sheffield.ac.uk}
\\
Dept of Physics and Astronomy, University of Sheffield, Hicks Building,
 Hounsfield Road, Sheffield, S3 7RH, United Kingdom}

\date{\today}

\pubyear{2017}
\begin{document}\label{firstpage}
\pagerange{\pageref{firstpage}--\pageref{lastpage}} 
\maketitle


\begin{abstract} \noindent We present an NTT/SOFI spectroscopic survey of infrared 
selected Wolf-Rayet  candidates in the Scutum-Crux spiral arm (298$^{\circ}\leq 
l\leq340^{\circ}$, $|b|\leq 0.5^{\circ}$). We obtained near-IR spectra of 
127  candidates, revealing 17 Wolf-Rayet stars -- a $\sim 13\%$ success rate --  
of which 16 are newly identified here. The majority of the new Wolf-Rayet stars
are classified as narrow-lined WN5--7 stars, with 2 broad-lined WN4--6 stars and 3 WC6--8 stars. 
The new stars, with distances estimated from previous absolute magnitude calibrations, have
  no obvious association with the Scutum-Crux arm. Refined near-infrared (YHJK) 
classification  criteria based on over a hundred Galactic and Magellanic Cloud WR stars,
providing diagnostics for hydrogen in WN stars, plus the identification of WO stars and intermediate WN/C stars.
Finally, we find that only a quarter of WR stars in the survey region are 
associated with star clusters and/or H\,{\sc ii} regions,
with similar statistics found for Luminous Blue Variables in the Milky Way.
  The relative isolation of evolved massive stars is discussed, together with the significance of the co-location of LBVs and WR stars
  in young star clusters.
\end{abstract}

\begin{keywords}
stars: emission-line -- stars: Wolf-Rayet -- stars: evolution -- galaxy: stellar content -- infrared: stars.
\end{keywords}

\section{Introduction}


Wolf-Rayet (WR) stars -- the progeny of massive O-type stars -- are excellent tracers of young stellar 
populations in galaxies owing to their unique spectroscopic signatures of strong, broad emission lines 
\citep{crowther07}. However, whilst WR surveys of nearby galaxies are nearing completeness \citep{massey14}, 
the Wolf-Rayet content of the Milky Way remains woefully incomplete \citep[e.g.][]{shara09} due to high
dust obscuration at visual wavelengths. Our detailed knowledge of the evolution of massive stars remains 
unclear, with inaccuracies in earlier evolutionary phases magnified in the WR phase. 

In addition, it is becoming clear that the conventional view of $\geq$20--25 $M_{\odot}$ O stars advancing 
through the Luminous Blue Variable (LBV) stage en route to the nitrogen- (WN) and carbon- (WC) sequence 
Wolf-Rayet  phase and ultimately a  stripped  envelope core-collapse supernova (ccSN) is incomplete if not 
incorrect. First, a high fraction of massive stars are now known to be multiple \citep{sana12}, so the major 
effects of close  binary evolution needs to be considered. Second, it has been proposed that LBVs are lower
mass binary products, from an inspection of their spatial location in the Milky Way and Large Magellanic Cloud
(LMC) with respect to Wolf-Rayet and O stars \citep{smith15}. Third, \citet{sander12} argue 
from a spectroscopic analysis of Milky Way  WC stars
that the most massive stars do not pass through  this phase. Finally, it is 
not clear whether the most massive stars will undergo a bright SN explosion after core-collapse, since they may 
collapse  directly to a black hole or produce a faint SN and fallback to a black hole \citep{langer12}.

Still, our Galaxy contains the largest \emph{spatially resolved} population of WR stars, predicted to number 
${\sim}1200$ \citep[][hereafter RC15]{rosslowe15a,rosslowe15b}. The confirmed population has doubled over the previous 
decade, and currently stands at ${\sim}640$\footnote{\url{http://pacrowther.staff.shef.ac.uk/WRcat/}}. The 
Galactic disk therefore presents a  rich hunting ground for further discoveries. Due to the large 
foreground interstellar dust extinction towards stars in the galactic disk, near and mid- infrared surveys are 
required.

The dense, ionized stellar wind of WR stars facilitate two approaches to infrared surveys. First, their strong, 
broad emission lines are amenable to near-IR narrow-band imaging \citep{shara09, shara12, kanarek15}. Second, 
their dense winds exhibit a free-free excess leading to unusual infrared colours, which have been exploited
in the near-IR \citep{homeier03, hadfield07} and mid-IR \citep{mauerhan11, messineo12, faherty14}. To 
date, the majority of spectroscopic follow-up has been carried out to an approximate depth of $K_{\rm S} \lesssim$11\,mag.
However, the (coarse) model of the Galactic Wolf-Rayet distribution developed by RC15 suggests
follow-up spectroscopy is needed to $K_{\rm S} \sim$13\,mag in order to sample the majority of Wolf-Rayet stars.

Here we exploit prior photometric approaches to spectroscopically survey a region of the Galactic 
disk to fainter limits than to date ($K_{\rm S}\sim$13\,mag). This has two interrelated 
goals:  1) the refinement and development of techniques that can be used to classify Wolf-Rayet stars using only infrared 
spectroscopy; 2) comprehensive searches for Wolf-Rayet stars in the Milky Way to allow more robust comparisons between their 
spatial locations and other massive stars; longer term goals involve the second data release (DR2) of {\it Gaia} which
will provide parallaxes for hundreds of Wolf-Rayet stars, permitting their use as tracers of Galactic structure, 
and will 
be combined with upcoming large fibre-fed spectroscopic surveys, including WHT/WEAVE \citep{weave} and VISTA/4MOST 
\citep{4most}.

This paper is structured as follows. In Section~\ref{sec:cand}, we describe our photometric selection criteria
and survey region, namely the Scutum-Crux spiral arm, the tangent to which lies at approximately 
$l\,{\simeq}\,310^\circ$ \citep{georgelin76}. Spectroscopic observations of Wolf-Rayet candidates, plus some 
previously known Galactic WR templates, are presented in Section~\ref{sec:data}, including a brief
  description of non-WR stars. Refined near-IR classification 
criteria for Wolf-Rayet stars are presented in Section~\ref{sec:class}. Results for newly identified WR stars are 
presented in  Section~\ref{sec:new}, including distance estimates. We consider the 
spatial location of Wolf-Rayet and other massive stars in Section~\ref{sec:disc}, including discussion of prior 
inferences about the nature of Luminous Blue Variables. Finally, in  Section~\ref{sec:conc} we reflect on the low
success rate of the methodology employed, and share some motivating points for future IR surveys targeting WR stars.



\begin{figure*}
\begin{center}
\includegraphics[width=2\columnwidth]{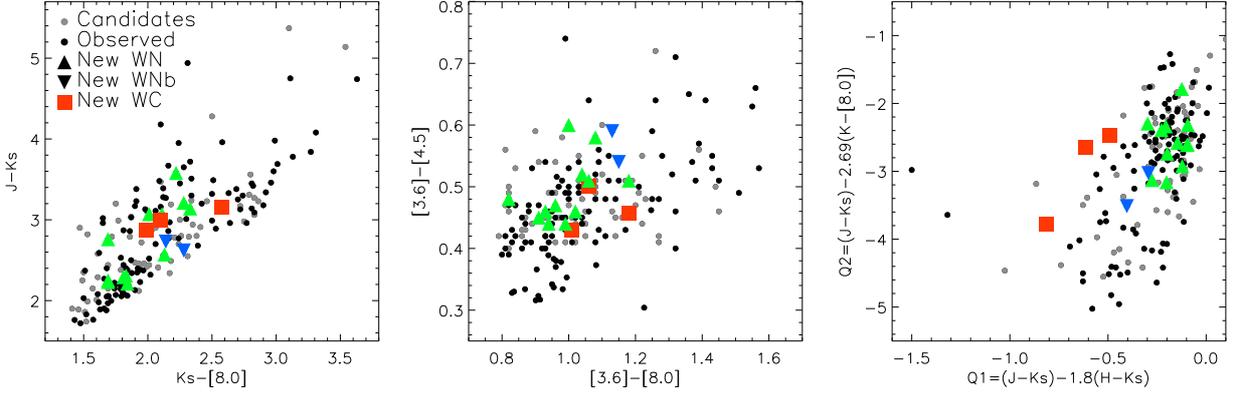}
\caption{
Colour-colour diagrams from 2MASS and GLIMPSE-I, showing 191 candidates in our survey area (circles)
we obtained NTT/SOFI HK spectra for 127 stars (black), leaving 64 unobserved (grey). Newly discovered WR stars 
are indicated: triangles for WN stars and squares for WC stars. The left panel shows K$_{\rm s}$--[8.0] versus J--K$_{\rm 
s}$, central panel indicates [3.6]--[8.0] versus [3.6]--[4.5] while the right panel presents the reddening free
parameters Q1 versus Q2 from \citet{messineo12}. 
}
\label{fig:c-c}
\end{center}
\end{figure*}

\section{Candidate Selection}\label{sec:cand}

Here we discuss our selection of sight lines towards the Scutum-Crux arm, plus our photometric criteria for the 
selection of candidate Wolf-Rayet stars.  Specifically, we focus on $l\,{=}\,\mbox{298--340}^\circ$, which
\citet{Russeil05} have previously highlighted in determining Galactic structure, since this intersects three proposed 
spiral arm features -- Sagittarius-Carina, Scutum-Crux and Norma-Cygnus \citep[][their Fig.~5]{russeil03}. The 
majority of known Wolf-Rayet stars in this region lie at distances ${<}6$\,kpc, consistent with  the nearby 
Sagittarius-Carina arm.  However, assuming typical $M_{K_S}\,{=}\,{-}\,5$  and  $A_{K_S}\,{=}\,2$ for Galactic WR 
stars, it is possible to  probe  heliocentric distances $\mbox{6--15}\,$kpc by identifying WR  stars in the 
magnitude range $K_S\,{=}\,\mbox{11--13}$\,mag. Therefore
WR stars may provide a comparable and complimentary tracer of Galactic structure to  commonly used non-stellar 
objects, i.e., H\,{\sc ii} regions and atomic H gas. We confined our search to  latitudes $|b|{<}0.5^\circ$, 
ensuring that at the furthest expected distances, candidate WR stars remain  within a few scale heights of the 
Galactic plane (FWHM${=}80\,$pc for WRs, RC15).



\begin{table*}
\begin{center}
\caption{Catalogue of newly identified Wolf-Rayet stars, including WR75-30 (1083--1765) from \citet{kanarek15}}
\label{tab:obs}
\begin{tabular}{
    l@{\hspace{2mm}}l@{\hspace{2mm}}c@{\hspace{2mm}}c@{\hspace{2mm}}c@{\hspace{2mm}}
    c@{\hspace{2mm}}r@{\hspace{2mm}}c@{\hspace{2mm}}c@{\hspace{2mm}}c@{\hspace{2mm}}
  c@{\hspace{2mm}}c@{\hspace{2mm}}r@{\hspace{2mm}}l}
\hline
ID & WR              & RA       & Dec       & $l$ & $b$ & K$_S$ & J--K$_S$ & H--K$_S$ & K$_S$--[3.6] & [3.6]--[4.5] &
K$_S$--[8.0] & SOFI   & Spectral\\
   & Number          & \multicolumn{2}{c}{-- J2000 --}   
&     &     & mag & mag & mag   &  mag  &  mag  &  mag   & Grisms & Type \\ 
\hline
E\#3   & WR46-18 & 12:08:52.49 &--62:50:54.9 & 298.0981 &--0.3769 & 10.47 & 2.87 & 1.20 & 0.80 & 0.46 & 1.98 & GB,GR &WC6-7\\ 
B\#13  & WR47-5  & 12:50:48.98 &--62:24:39.8 & 302.8599 & +0.4606 & 11.09 & 2.23 & 0.83 & 0.70 & 0.44 & 1.69 & GB,GR & WN6(h)\\
B\#37  & WR56-1  & 13:41:50.01 &--62:20:25.8 & 308.7434 &--0.0387 & 12.18 & 2.76 & 1.03 & 0.87 & 0.48 & 1.69 & GB,GR & WN5o\\ 
B\#51  & WR60-7  & 14:02:33.44 &--61:20:27.2 & 311.3533 & +0.3626 & 10.28 & 3.00 & 1.29 & 1.09 & 0.43 & 2.10 & GB,GR & WC7-8\\ 
B\#56  & WR60-8  & 14:12:15.19 &--61:42:45.2 & 312.3517 &--0.3277 & 11.76 & 3.14 & 1.22 & 1.15 & 0.51 & 2.33 & GB,GR & WN6o \\
B\#85  & WR64-2  & 15:01:14.05 &--58:49:07.4 & 319.0607 &--0.0724 & 11.89 & 2.57 & 0.99 & 1.07 & 0.51 & 2.13 & GB,GR & WN6o \\ 
B\#87  & WR64-3  & 15:02:46.14 &--58:27:06.5 & 319.4120 & +0.1535 & 10.18 & 2.21 & 0.86 & 0.88 & 0.47 & 1.84 & GB,GR & WN6o \\ 
B\#88  & WR64-4  & 15:04:11.15 &--58:27:21.5 & 319.5721 & +0.0601 &  9.10 & 2.25 & 0.91 & 0.76 & 0.46 & 1.69 & GB,GR & WN6o+OB\\
B\#91  & WR64-5  & 15:07:31.84 &--58:15:09.6 & 320.0540 & +0.0209 & 10.80 & 2.31 & 0.86 & 0.89 & 0.44 & 1.83 & GB,GR & WN6o\\
B\#93  & WR64-6  & 15:10:57.65 &--57:57:28.5 & 320.5939 & +0.0474 & 11.11 & 2.73 & 1.08 & 0.99 & 0.54 & 2.14 & GB,GR & WN6b \\
B\#105 & WR70-13 & 15:37:46.51 &--56:08:45.2 & 324.6325 &--0.4487 &  9.96 & 3.16 & 1.42 & 1.52 & 0.50 & 2.58 & GB,GR & WC8d\\ 
B\#107 & WR70-14 & 15:39:17.02 &--55:49:18.9 & 324.9945 &--0.3129 & 11.50 & 2.62 & 1.08 & 1.15 & 0.59 & 2.28 & GB,GR & WN4b \\ 
B\#123 & WR70-15 & 15:58:57.97 &--52:46:05.4 & 329.1414 & +0.2865 & 12.40 & 3.21 & 1.19 & 1.20 & 0.58 & 2.28 & GB,GR & WN5o \\
B\#132 & WR72-5  & 16:07:01.45 &--51:58:18.3 & 330.5909 & +0.0725 & 10.27 & 2.29 & 0.87 & 0.77 & 0.52 & 1.81 & GB,GR & WN6o \\
A\#11  & WR75-31 & 16:25:13.60 &--48:58:22.3 & 334.7557 & +0.2255 & 12.21 & 3.07 & 1.13 & 1.09 & 0.46 & 2.11 & GB,GR & WN7o \\
A\#13  & WR75-30 & 16:32:25.70 &--47:50:45.8 & 336.3959 & +0.1395 & 11.57 & 3.58 & 1.36 & 1.22 & 0.60 & 2.22 & GR & WN7o \\
B\#154 & WR76-11 & 16:40:12.92 &--46:08:54.0 & 338.5451 & +0.2996 & 11.98 & 3.07 & 1.17 & 1.10 & 0.45 & 2.01 & GR & WN7o \\ 
\hline
\end{tabular}
\end{center}
\end{table*}







We selected candidate Wolf-Rayet stars for which $298^{\circ} \leq l \leq 340^{\circ}$ and $|b| \leq 0.5^{\circ}$ from their near-IR 
\citep[2MASS][]{skrutskie06} and  mid-IR \citep[GLIMPSE-I][]{benjamin03} photometry. We limited our survey to GLIMPSE-I point 
sources with a corresponing 2MASS detection\footnote{via the IPAC/NASA Infrared Science Archive: 
\url{http://irsa.ipac.caltech.edu}}, requiring a minimum 2MASS quality flag of `C' in the $K_{\rm S}$ filter, and 
rejected sources with one or more Source Quality Flags ${>}\,10^5$ in the GLIMPSE-I catalogue. 
We then used the TOPCAT\footnote{Available at: \url{http://www.starlink.ac.uk/topcat/}} tool to apply various 
cuts in colour and magnitude. 

\citet{mauerhan11} identified several regions of colour space (their grey shaded region in Fig.~1) favoured by Wolf-Rayet
stars, which we have adapted as follows:
\begin{gather*}
1.25(\mathrm{K}_\mathrm{S}-[8.0]) \leq (J-K_S) + 0.5 <2.5(\mathrm{K}_\mathrm{S}-[8.0]), \\
\mathrm{K}_\mathrm{S}-[8.0] \geq 1.3, \\
0.8 \leq ([3.6]-[8.0]) \leq 1.6, \\
0.3 \leq ([3.6]-[4.5]) \leq 0.75.
\end{gather*}
In addition, \citet{messineo12} introduced additional reddening-free parameters $Q1 = 
(J-H) - 1.8 (H-K_{\rm S})$ and $Q2 = (J-K_{\rm S}) - 2.69 (K - [8.0])$ which we also utilise:
\begin{gather*}
(11.25\mathrm{Q}1-2.38)<\mathrm{Q}2<-1.0. 
\end{gather*}

It is necessary to emphasise that not all WR stars 
occupy this parameter space, although there is no bias towards either WN or WC subtypes.
Still, some dusty WC stars are offset from the majority of WR stars in the J-K$_S$ vs $\mathrm{K}_\mathrm{S}-[8.0]$ colour colour
  diagram \citep[][their Fig.~1]{mauerhan11}, so our survey criteria are potentially biased against such stars.
Approximately 250 sources satisfied these criteria. We subsequently  cross-checked the co-ordinates of these 
with the SIMBAD\footnote{\url{http://simbad.u-strasbg.fr/simbad/}} database, to  find any with previous 
identifications. Encouragingly, 14\% of these were known WR stars (23\% of known WRs in the survey area), 4\% 
had  non-WR classifications (mostly Be stars or young stellar objects),  leaving $\sim$200 candidates with no 
previous spectral classification. Colour-colour diagrams for 191 candidates involving $\mathrm{K}_\mathrm{S}-[8.0]$ vs 
$(\mathrm{J}-\mathrm{K}_\mathrm{S}$) and $[3.6]-[8.0]$ vs $[3.6]-[4.5]$ are presented in 
Figure~\ref{fig:c-c} together with reddening-free parameters Q1 and Q2 from \citet{messineo12}.

\begin{table}
\begin{center}
\caption{Spectroscopic datasets exploited for the updated near-IR classification scheme}
\label{tab:log}
\begin{tabular}{
    l@{\hspace{-2mm}}c@{\hspace{0mm}}c@{\hspace{0mm}}c@{\hspace{0mm}}c@{\hspace{0mm}}
    l@{\hspace{1mm}}l@{\hspace{2mm}}l@{\hspace{2mm}}l}
\hline
Tel/Inst & R & \multicolumn{4}{l}{Spect}    & Epoch      & Ref & Dataset/\\
         &     & \multicolumn{4}{l}{Range} &            &     & Programme\\
\hline
CTIO 4m/IRS  & 3000     & Y &   &   & K & Mar 1996   & a & C1 \\
ESO 1.5m/B\&C&          & Y &   &   &   & Feb 1982   & b & E1 \\
INT/IDS      & 3000     & Y &   &   &   & Aug 1990   & c & I1 \\
Lick 3m/UCLA & 525      &   &   &   & K & 1994-1995  & d & L1 \\
MSO/CGS      &600--850  & Y & J & H & K &            & e & M1 \\
NTT/SOFI     & 1000     &  Y& J & H & K & Sep 1999   & f & N1/63.H-0683     \\ 
NTT/SOFI     & 600--1300&  Y& J &   & K & May 2002   & g & N2/69.B-0030     \\ 
NTT/SOFI     &600--1000 & Y& J &H   &  K & Nov 2003  &   & N3/71.D-0272     \\ 
NTT/SOFI     &  600     & Y& J &    &    & Nov 2004  &   & N4/74.D-0696     \\ 
NTT/SOFI     &  600     & Y& J & H  & K    & Jun 2005 &  h   & N5/75.D-0469     \\ 
NTT/SOFI     & 600      & Y& J & H  & K    & Mar 2015  &  i   & N6/94.D-0839     \\ 
OHP/CARELEC  &          & Y&   &    &      & Sep 1989   & j  & O1 \\
UKIRT/CGS2   &400--600  &Y & J & H &  K    &            & k  & U1 \\
UKIRT/CGS4   &1000--1400&  &  &  H& K     & Jan 1994    & l   & U2 \\ 
UKIRT/CGS4   &600--850 & Y& J & H & K     & Aug 1994   & m & U3   \\ 
UKIRT/CGS4   & 1500    & Y& J &   & K     & Sep 1995   &  & U4  \\ 
UKIRT/CGS4   & 3000    &  &   &   & K      & May 1996  &  a & U5 \\ 
UKIRT/CGS4   & 1500    & Y& J &   & K     & Apr 1997   &  & U6 \\ 
VLT/XSHOOTER & 5000    & Y&J& H   & K     & 2013-2014  & n & V1 \\
\hline
\multicolumn{9}{l}{
  \begin{minipage}{\columnwidth} 
a: \citep{bohannan99}, b: \citep{vreux83}; c: \citep{howarth92}, d: \citep{figer97}, 
e: \citep{hillier83}, f: \citep{crowther00}, g: \citep{homeier03}; h: \citep{crowther06}, 
i: This study; j: \citep{vreux90}, k: \citep{eenens91}, l: \cite{crowther95c}, m: \citep{crowther96}: n: \citep{tramper15}
  \end{minipage}
}
\end{tabular}
\end{center}
\end{table}


\section{NTT/SOFI spectroscopy of Wolf-Rayet candidates}\label{sec:data}

We obtained near-IR spectroscopy of 127 WR candidates between 29--31 March 2015 
(program ID 094.D-0839) using the Son-of-Isaac (SOFI) spectrograph at the New Technology Telescope (NTT). 
These represent 66\% of the IR selected candidates presented in Fig.~\ref{fig:c-c}. Candidates were
observed with the red grism (GR) covering the 1.53--2.52$\mu$m spectral region, a dispersion of 10.2\AA/pix, and 
a slit width of 1 arcsec, 
providing a spectral resolution of $R\sim$600. All sources were observed using a standard ABBA sequence, 
including a small random offset in the A and B positions between exposures. 

Before extracting 1D spectra, we subtracted a median dark frame from each individual frame, then subtracted 
adjacent AB pairs from one another. The result of this was 4 dark frame-corrected spectra for each source, free 
from sky lines. We extracted these 4 spectra for each object using IRAF. Wavelength calibration was performed 
using strong and isolated sky lines at known wavelengths \citep{rousselot00} present in each raw frame, after 
which all spectra for each object were co-added. 

Throughout each night, we periodically observed bright Vega-type telluric standard stars, at similar airmasses to 
the WR candidates. The removal of telluric spectral features was achieved using \texttt{telluric} in IRAF. We 
also used these telluric standards, together with Kurucz models of the same spectral types, to perform 
relative flux calibration, which are subsequently adjusted to match 2MASS photometry.

Of the 127 candidates, 17 stars were identified as WR stars. Of these, one candidate was subsequently matched to the 
recently discovered WN6 star 1093-1765 (= WR75-30) from  \citet{kanarek15}, such that 16 stars are newly identified as 
WR stars in this study. Previous surveys of Wolf-Rayet stars from IR photometric criteria have achieved similar 
efficiencies \citep{mauerhan11, faherty14}. We briefly discuss the nature of the non-WR stars in Sect.~\ref{sec:reject}
and discuss the newly identified WR stars in Sect.~\ref{sec:class}. New WR stars are indicated in Fig.~\ref{fig:c-c}, with
a subtype dependence apparent in the reddening-free Q1 vs Q2 diagram. Table~\ref{tab:obs} provides basic observational 
properties for the new Wolf-Rayet stars, for which we obtained NTT/SOFI spectroscopy, while Table~B1 (available online) 
provides a list of all candidates for which we obtained spectroscopy, together with a brief note 
describing the nature of each source. 

In addition to the candidate WR stars, we have also obtained NTT/SOFI 
spectroscopy of 14 Wolf-Rayet stars for  which optical classifications have been undertaken, in order to refine 
near-IR based classification criteria (see Sect.~\ref{sec:class}). Finally, we
also obtained blue grism (GB) spectroscopy with SOFI for the majority of newly 
identified  WR stars. The GB observations cover the spectral region 0.95--1.64$\mu$m, a dispersion of 
7.0\AA/pix, and  an identical slit width of 1 arcsec, again providing $R\sim$600. Data reduction was undertaken 
in an identical manner to the GR datasets.

\subsection{Non Wolf-Rayet stars}\label{sec:reject}

A significant subset of the 110 candidates that were not confirmed to be WR stars exhibited a hydrogen emission line spectrum, with 
Br$\gamma$ observed in 60 (55\%) cases, plus often higher Brackett series (Br10, 11), and He\,{\sc i} 2.058$\mu$m emission present 
in a quarter of instances. These sources are likely to be massive young stellar objects (mYSOs) or Herbig AeBe stars
\citep[see e.g.][]{porter98, cooper13}. 
Br$\gamma$ emission equivalent widths are typically 10--30\AA, with He\,{\sc i}/Br$\gamma$ ratios of 0.3--1. The majority of Br$\gamma$ emission lines are unresolved (FWHM$\sim$30\AA) although several stars (e.g. B\#127, 147, 149) possess broad emission (FWHM$\sim$50-60\AA). Unusually, B\#66 exhibits strong He\,{\sc i} 2.058$\mu$m emission, without significant Br$\gamma$ emission, warranting follow up observations.

Mid-IR 
imaging has revealed circumstellar ring nebulae around many evolved stars (e.g., \citealt{wachter10,toala15}). Such nebulae appear 
prominently in Spitzer 8.0\micron\ images, owing to thermal emission from dust swept up by stellar winds. Mindful of this, we 
inspected $5^\prime\times5^\prime$ Spitzer 8.0\micron\ images centred on all candidates. One of the Br$\gamma$ emission line 
sources, A\#9 (SSTGLMC G330.7900-00.4539), is the central star of a striking oval mid-IR ring nebula, S65, which was identified by 
\citet{churchwell06} and studied by \citet{simpson12}.

Strong absorption lines in the Brackett series are observed in one candidate, B\#3, indicating an A- or late-B type star, with Br$\gamma$ 
absorption observed in another object, B\#54, albeit without other prominent features suggesting an early type star in this instance. Four candidates 
-- B\#21, B\#100, B\#122 and B\#153 -- exhibit prominent CO 2.3$\mu$m bandhead absorption features, although none of these involve Br$\gamma$ 
emission line sources, indicating a late-type star origin. The remaining 44 candidates (40\% of the non WR stars) either have no prominent 
absorption or emission  features, or the S/N achieved was insufficient to identify their nature. 


\section{Near-IR classification of Wolf-Rayet stars}\label{sec:class}

The switch from optical to near-IR spectroscopy for the overwhelming majority of new Galactic Wolf-Rayet stars
requires a reassessment of spectral classification criteria. \citet{vreux90} provided an classification 
scheme
based upon Y-band observations of northern WR stars, while \citet{eenens91} devised a near-IR scheme for WC stars
from 1--5$\mu$m spectroscopy. More recently \citet[][hereafter C06]{crowther06} provide 
near-IR classification diagnostics for WN and WC stars, based on equivalent width ratios. Qualitatively, early-type WC4--6
stars possess broader emission lines than later WC7--9 subtypes, although exceptions do exist \citep{eenens94}.

An updated a quantitative near-IR classification of Wolf-Rayet stars is made feasible by access to a greatly
  expanded sample of Galactic and Magellanic Cloud WR stars for which 
optical classifications have been made, primarily \citet{smith96} for WN stars and \citet{smith90} for WC stars.
The datasets utilised were drawn from various sources, primarily NTT/SOFI and UKIRT/CGS4, as summarised in Table~\ref{tab:log}.
We have also inspected high resolution, intermediate resolution IRTF/SpeX spectroscopy of northern Galactic WR stars,
provided by W.D.~Vacca (priv. comm.), although we focus our criteria on moderate resolution (R = 600 -- 1000), modest signal-to-noise
spectroscopy in the 0.95 -- 2.5$\mu$m near infrared. 

To date, no criteria for the identification of WN/C or WO stars from near-IR spectroscopy have been considered,
nor has an attempt to distinguish between H-rich and H-deficient WN stars, 
although C06 did separate broad-lined WN stars
  (FWHM He\,{\sc ii} 1.01\micron\ $\geq$65\AA) from narrow-lined counterparts. In our revised
near-IR classification scheme we attempt to utilise pairs of lines from adjacent ionization stages of helium for WN stars, and 
adjacent ionization stages of carbon for WC stars. In some instances nitrogen lines are required for WN stars, in common  with C06,
plus ratios of carbon to helium lines are utilised for WN/C and WC stars, which will also depend upon their relative abundances.
We omit from our discussion the near-IR classification of transition Of/WN stars, which has been considered by \citet{bohannan99} 
and \citet{crowther11}. WN stars with intrinsic absorption features (WNh+abs, WNha) also offer specific challenges which will
need to be considered separately. 
 

A detailed description of the updated classification scheme is provided in Appendix A (available online), while
we present a summary of Y, J, H and K-band classification diagnostics for WN, WN/C, WC and WO subtypes
in Table~\ref{tab:class}. In Figs~A1--A2 (available online) we present YJ-band and HK-band
spectroscopy of template  (optically classified) WN, WN/C, WC and WO stars, with line measurements provided in Tables~A1--A3, 
also online. Overall, the use of solely IR diagnostics
provide satisfactory classifications, although confidence in resulting spectral types requires
multi-band spectral coverage, a minimum spectral resolution of $R\sim$500 and moderate signal-to-noise. In many 
instances, observations at a single band prevent a refined classification. For example, WC4--7 stars may not be distinguished using
solely K-band spectroscopy, while it is not possible to differentiate between broad-lined WN4--7 stars on the basis of low S/N 
spectroscopy in the Y-band. Still, reliable, Wolf-Rayet subtypes can be obtained from complete 1--2.5$\mu$m spectroscopy, with the 
exception of  broad-lined WN4--6 stars and WC5--6 stars. In addition, WO stars have a very distinctive near-IR spectrum, and WN/C 
stars possess characteristics in each of Y, J, H and K-bands which distinguish them from normal WN stars. In addition, the presence 
of hydrogen in WN stars can be identified in most subtypes, although very late subtypes are challenging since a low He\,{\sc ii} 
1.163/P$\beta$ or He\,{\sc ii} 2.189/Br$\gamma$ ratio may indicate either a high hydrogen content or a low ionization atmosphere.

\subsection{Robustness of near-IR classification}

In order to assess the robustness of the new scheme, we reclassify several WN and WC stars which have been discovered and classified from red
optical spectroscopy. We utilise NTT/SOFI spectroscopy of four WN stars, WR62a, WR68a,
WR93a from  \citet{homeier03} (dataset N2 in Table~\ref{tab:log}), 
plus WR75-30 from our own observations (dataset N5 in Table~\ref{tab:log}), together with three WC stars, 
WR107a from   \citet{homeier03} plus WR75aa and WR75c from our own observations.
Near-IR spectra are presented in Figs~\ref{fig:new_wr_blue}--\ref{fig:new_wr_red}.
Individual line measurements are provided in Table~\ref{newbies_wn} and \ref{newbies_wc} for WN and WC stars, respectively, while
line ratios are presented in Table~\ref{tab:new_wrstars}. Measurements have employed Gaussian fits, using the  \texttt{elf} suite of commands
within the Starlink \texttt{DIPSO}  package\footnote{Available at: \url{http://starlink.eao.hawaii.edu/starlink}}. 

\subsubsection{WN stars}

WR62a was classified as WN5o by \citet[][their source \#11]{shara99}, and we support its classification as a narrow-lined WN 
star. Consequently, the primary diagnostics are the He\,{\sc i} 1.08/He\,{\sc ii} 1.01$\mu$m ratio and K-band morphology. The 
former indicates a WN6 subtype, while He\,{\sc i} + N\,{\sc iii} 2.11 $>$ N\,{\sc v} 2.10, favours a WN5--6 subtype. The 
P$\beta$/He\,{\sc ii} 1.16$\mu$m ratio indicates WR62a is hydrogen-free, while the Br$\gamma$/He\,{\sc ii} 2.19$\mu$m 
suggests a borderline o/(h) classification, so overall we favour WN6o for WR62a. The same arguments and ratios apply to WR68a 
for which \citet[][their source \#13]{shara99} assigned WN6o. We support this classification owing to its morphological 
similarity to WR62a.

WR93a (Th~3--28), was originally classified as WN2.5--3 by \citet{acker90} and revised to WN6 by \citet{miszalski13} from 
optical spectroscopy. This is also a narrow-lined WN star so again we focus on its He\,{\sc i} 1.08/He\,{\sc ii} 1.01$\mu$m 
ratio and K-band morphology. Both favour a WN6 subtype, with a significant hydrogen content from our multiple diagnostics 
(darker shaded regions in Fig.~A4, available online), so we adopt WN6h for WR93a.

\citet[][their source 1083-1765]{kanarek15} originally classified WR75-30 as a WN6 star from near-IR spectroscopy. The 
He\,{\sc i} 1.70/He\,{\sc ii} 1.69$\mu$m ratio favours a WN7 subtype, as does (He\,{\sc i} + N\,{\sc iii} 2.11)/He\,{\sc i} 
2.19$\mu$m $\sim$ 1, while the Br$\gamma$/He\,{\sc ii} 2.19$\mu$m ratio lies in the hydrogen-free region of 
Fig.~A4 so we favour WN7o for this star.

\subsubsection{WC stars}

WR75aa and WR75c were identified as WC9 stars by \citet{hopewell05} from red optical spectroscopy. All our primary near-IR 
diagnostics support this assessment, as do the secondary criteria involving helium for WR75c. WR75aa has a  borderline WC8--9
classification from the   He\,{\sc i-ii} 1.7/C\,{\sc iv} 1.74$\mu$m ratio, but overall both stars are unambiguous WC9 stars. 

Finally, WR107a \cite[\#18 
from][]{shara99} was originally classified as a WC6 star from red optical spectroscopy. Our primary criteria indicate the 
follow for WR107a: WC6$\pm$1 from both C\,{\sc iii} 1.20/C\,{\sc iv} 1.19 and C\,{\sc iii} 2.11/C\,{\sc iv} 2.07, WC5--8 from 
C\,{\sc ii} 0.99/C\,{\sc iii} 0.97. Our secondary criteria indicate WC5 from He\,{\sc i} 1.08/He\,{\sc ii} 1.01, and WC7 from 
C\,{\sc iii} 0.97/He\,{\sc ii} 1.01$\mu$m (H-band spectra are unavailable), so although WC6 is plausible we provide a more 
cautious WC5--7 classification. Indeed, the primary optical diagnostic ratio (C\,{\sc iii} 5696/C\,{\sc iv} 5808) also 
favoured WC6--7 according to \citet{shara99}.

In general, the K-band is preferred to shorter wavelengths for classification of highly reddened WR stars, but K-band
spectral features of dusty WC stars are often masked by host dust emission. Extremely high S/N is required to identify K-band spectral features of 
the Quintuplet Q stars. By way of example, \citet{liermann09} assign a WC8/9d+OB subtype to Q3 (WR102ha) from K-band spectroscopy,  whereas WC 
features are relatively prominent in deep H- and J-band spectroscopy. We confirm a WC9d subtype for Q3 on the basis of high S/N Gemini spectroscopy 
presented by \citet{najarro15}, owing to C\,{\sc iii}  1.20/C\,{\sc iv} 1.19$\gg$1, C\,{\sc ii} 1.78/C\,{\sc iv} 1.74$>$1 and C\,{\sc iii} 
2.11/C\,{\sc iv} 2.07$\gg$1.




\begin{landscape}

  \begin{table}
    \begin{center}
\caption{Classification of Wolf-Rayet stars based on emission equivalent
width ratios of diagnostics in the Y,J,H,K spectral regions, updated from C06, primary diagnostics in bold font.
The majority of diagnostics are blended with other lines in
broad-lined Wolf-Rayet stars, including O\,{\sc vi} 1.075 with C\,{\sc iv} 1.054 and He\,{\sc ii} 1.093, 
He\,{\sc i} 1.083 with P$\gamma$, C\,{\sc iv} 1.191 with C\,{\sc iii} 1.198, 
He\,{\sc ii} 1.692 with He\,{\sc i} 1.700, C\,{\sc ii} 1.785 with C\,{\sc iv} 1.801,  C\,{\sc iv} 2.070--2.080 with 
C\,{\sc iii} 2.115, and He\,{\sc ii} 2.189 with Br$\gamma$. For near-IR classification of transition Of/WN stars, 
see \citet{bohannan99} and/or \citet{crowther11}.}
\label{tab:class} 
\begin{tabular}{r@{\hspace{2mm}}c@{\hspace{2mm}}c@{\hspace{2mm}}c@{\hspace{2mm}}c@{\hspace{2mm}}c@{\hspace{2mm}}c@{\hspace{2mm}}l@{\hspace{2mm}}l}
\hline   
Subtype & FWHM         & Diagnostic  & Diagnostic & Diagnostic & Diagnostic & \multicolumn{2}{l}{Notes} & Templates \\
        & km\,s$^{-1}$ &  Y-band      & J-band    & H-band      & K-band     &            &      & \\ 
\hline
\multicolumn{9}{c}{\bf Narrow-lined WN stars (FWHM(He\,{\sc ii} 1.01$\mu$m) $\lesssim$ 1900 km\,s$^{-1}$ and $\log W_{\lambda}$(He\,{\sc ii} 1.01$\mu$m/\AA) $\lesssim$  2.5)} \\
WN      & He\,{\sc ii}    & log $W_{\lambda}$(He\,{\sc i} 1.08/  & log $W_{\lambda}$(P$\beta$/ & $\log W_{\lambda}$(He\,{\sc i} 1.70/ & $\log W_{\lambda}$(Br$\gamma$/ & \multicolumn{2}{l}{Notes} & Templates\\
Subtype & 1.012$\mu$m &   He\,{\sc ii} 1.01)   & He\,{\sc ii} 1.16) & He\,{\sc ii} 1.69)   & He\,{\sc ii} 2.19)   \\
9      & 300             &  {\bf $\geq$1.4}  & $\sim$1.5 (h) & {\bf $\geq$1.4}   & {\bf $\sim$1.5 (h)} 
& \multicolumn{2}{l}{\bf He\,{\sc i}+N\,{\sc iii} 2.11 $\gg$ He\,{\sc ii} 2.19   } & WR105, BAT76\\
8      & 700             & {\bf 1$\pm$0.4} & $\leq$0.3 (o); $\geq$0.1 (h) & {\bf 0.9$\pm$0.5} &  $\leq$0.5 (o); $\geq$0.4 (h) 
& \multicolumn{2}{l}{\bf He\,{\sc i}+N\,{\sc iii} 2.11 $\gg$ He\,{\sc ii} 2.19}  & WR40, WR123 \\
7      & 800             & {\bf 0.4$\pm$0.2}   & $\leq$0 (o); $\geq$0 (h) & {\bf 0.2$\pm$0.2}  & $\leq$0.1 (o); $\geq$0.1 (h) 
& \multicolumn{2}{l}{\bf He\,{\sc i}+N\,{\sc iii} 2.11 $\sim$ He\,{\sc ii} 2.19}   & WR78, WR120 \\
6      & 1200            & {\bf 0.0$\pm$0.2}   & $\leq$--0.1 (o); $\geq$--0.1 (h) & $\sim$ --0.1    & $\leq$--0.1 (o); $\geq$--0.1 (h) 
& \multicolumn{2}{l}{\bf He\,{\sc ii} 2.19 $>$ He\,{\sc i}+N\,{\sc iii} 2.11 $\gg$ N\,{\sc v} 2.10} & WR115 \\
5      & 1400            & {\bf --0.3$\pm$0.1}    & $\leq$--0.2 (o); $\geq$--0.2 (h) & $\sim$ --0.3 & $\leq$--0.2 (o); $\geq$--0.2 (h)  
& \multicolumn{2}{l}{\bf He\,{\sc ii} 2.19 $\gg$ He\,{\sc i}+N\,{\sc iii} 2.11 $\geq$ N\,{\sc v} 2.10} & BAT122 \\
4      & 1500            & {\bf --0.7$\pm$0.3}    &  $\leq$--0.3 (o); $\geq$--0.3 (h)  & $\sim$ --0.6 & $\leq$--0.3 (o); $\geq$--0.3 (h) 
& \multicolumn{2}{l}{\bf He\,{\sc ii} 2.19 $\gg$ N\,{\sc v} 2.10 $>$ He\,{\sc i}+N\,{\sc iii} 2.11)} & WR128, BAT75 \\
3      & 1600            & {\bf $\leq$ --1.0}     & $\leq$--0.4 (o); $\geq$--0.4 (h)      &   $\leq$--0.8 &  $\leq$--0.4 (o); $\geq$--0.4 (h)  
& \multicolumn{2}{l}{\bf He\,{\sc ii} 2.19 $\gg$ N\,{\sc v} 2.10 $\gg$ He\,{\sc i}+N\,{\sc iii} 2.11} & WR46, WR152 \\
\hline
\multicolumn{9}{c}{\bf Broad-lined WNb stars  (FWHM(He\,{\sc ii} 1.01$\mu$m) $\gtrsim$ 1900 km\,s$^{-1}$ and $\log W_{\lambda}$(He\,{\sc ii} 1.01$\mu$m/\AA)  $\gtrsim$ 2.5)} \\
WN      & He\,{\sc ii}    & log $W_{\lambda}$(He\,{\sc i} 1.08)/  & log $W_{\lambda}$(P$\beta$/ & $\log W_{\lambda}$(He\,{\sc i} 1.70/ & $\log W_{\lambda}$(Br$\gamma$/ & \multicolumn{2}{l}{Notes} & Templates\\
Subtype & 1.012$\mu$m &   He\,{\sc ii} 1.01)   & He\,{\sc ii} 1.16) & He\,{\sc ii} 1.69)   & He\,{\sc ii} 2.19)   \\
7    & 3300   & {\bf $\geq$0.2} & --0.3 (o) & +0.4: &  --0.2 (o) 
& \multicolumn{2}{l}{\bf He\,{\sc ii} 2.19 $>$ He\,{\sc i}+N\,{\sc iii} 2.11} &  WR77sc \\
6    & 2600  &  {\bf +0$\pm$0.2} & $\leq$--0.3 (o); $\geq$--0.3 (h) & --0.3: & $\leq$--0.2 (o); $\geq$--0.2 (h)  
& \multicolumn{2}{l}{\bf He\,{\sc ii} 2.19 $\gg$ He\,{\sc i}+N\,{\sc iii} 2.11 $>$ N\,{\sc v} 2.10} &  WR75, WR134 \\
4     & 2400            & {\bf --0.5$\pm$0.5}  & $\leq$--0.3 (o) & --0.5: & $\leq$--0.2 (o); $\geq$--0.2 (h) 
& \multicolumn{2}{l}{\bf He\,{\sc ii} 2.19 $\gg$ He\,{\sc i}+N\,{\sc iii} 2.11 $\sim$ N\,{\sc v} 2.10} & WR6, WR18 \\
 2--3   & 2550            & {\bf $\leq$--1}     & $\leq$--0.5 (o); $\geq$--0.5 (h) & $\leq$--1.0       &  $\leq$--0.4 (o); 
$\geq$--0.4 (h)       
& \multicolumn{2}{l}{\bf He\,{\sc ii} 2.19 $\gg$ N\,{\sc v} 2.10 $\gg$ He\,{\sc i}+N\,{\sc iii} 2.11} & WR2, BAT51 \\
\hline
\multicolumn{9}{c}{\bf WN/C stars} \\
WN/C  & & $\log W_{\lambda}$(C\,{\sc iii} 0.97/   & $\log W_{\lambda}$(C\,{\sc iv} 1.19/ & $\log W_{\lambda}$(C\,{\sc iv} 1.74/   & $\log W_{\lambda}$(C\,{\sc iv} 2.07/  & $\log W_{\lambda}$(C\,{\sc iv} 2.43/ & Notes & Templates\\
Subtype&               & He\,{\sc ii} 1.01)    & He\,{\sc ii} 1.16) & He\,{\sc i-ii} 1.7)  & He\,{\sc i}+C\,{\sc iii} 2.11) & He\,{\sc ii} 2.34) \\
All   &                & --0.5              & {\bf $\geq$--0.7} & {\bf $\geq$--0.7}            & {\bf $\geq$--0.3}            & $\geq$0 & & WR8, WR26\\
\hline
\multicolumn{9}{c}{\bf WC stars} \\
WC      & He\,{\sc ii} & $\log W_{\lambda}$(He\,{\sc i} 1.08/   & $\log W_{\lambda}$(C\,{\sc iii} 1.20/  & $\log W_{\lambda}$(He\,{\sc i-ii} 1.7/ & $\log W_{\lambda}$(C\,{\sc iii} 2.11/  & $\log W_{\lambda}$(C\,{\sc ii} 0.99/ & Notes & Templates \\
Subtype & 1.190$\mu$m &He\,{\sc ii} 1.01)& C\,{\sc iv} 1.19)   & C\,{\sc iv} 1.74)  & C\,{\sc iv} 2.07)  &C\,{\sc iii} 0.97) & \\
9      & 850  & +1.1 &      {\bf +0.6$^{+0.2}_{-0.5}$}  & 0.3$^{+0.1}_{-0.5}$ & {\bf  +0.1$\pm$0.3}  &    {\bf --1.1$\pm$0.2} & {\bf C\,{\sc ii} 1.78 $>$ C\,{\sc iv} 1.74} &  WR92, WR103\\
8      & 1800  & +0.4$\pm$0.2 & {\bf --0.2$^{+0.3}_{-0.2}$}  &  --0.4$\pm$0.2  & {\bf --0.35$\pm$0.1}   & {\bf --1.3$\pm$0.1}  & & WR135\\
7      & 1900  & +0.2$\pm$0.2   & {\bf --0.5$\pm$0.1}  &  --0.7$\pm$0.1  &  {\bf  --0.6$^{+0.2}_{-0.15}$ }    & {\bf --1.35$\pm$0.15} & & WR90\\
6      & 2900  & +0.2    &    {\bf  --0.7$\pm$0.2 } & --0.8$\pm$0.1&   {\bf --0.7$\pm$0.1 }  & {\bf --1.6$\pm$0.1} &    & WR15, WR23\\
5      & 2300  & +0.0    & {\bf --0.6$\pm$0.1}      & --1.0$\pm$0.2          &  {\bf --0.7$\pm$0.1}& {\bf --1.5$\pm$0.1} &                           & WR111\\
4   & 3300 &  --0.1    & {\bf --0.6$\pm$0.1}  &  --1.2$\pm$0.2        &  {\bf $<$ --0.7}             & {\bf $<$--1.5 } & {\bf C\,{\sc ii} 0.99 absent} & WR143,BAT11 \\
\hline
\multicolumn{9}{c}{\bf WO stars} \\
WO     & C\,{\sc iv} & $\log W_{\lambda}$(O\,{\sc vi} 1.07/   & $\log W_{\lambda}$(He\,{\sc ii} 1.16/&$\log W_{\lambda}$(O\,{\sc vi} 1.46+He\,{\sc ii} 1.47)/  
& $\log W_{\lambda}$(C\,{\sc iv-iii} 2.07-2.11/  & $\log W_{\lambda}$(C\,{\sc iii} 0.97/ & Notes & Templates\\
Subtype    & 1.74$\mu$m & C\,{\sc iv} 1.19) & C\,{\sc iv} 1.19)& C\,{\sc iv} 1.74) & (C\,{\sc iv} 2.43 + O\,{\sc vi} 2.46) & He\,{\sc ii} 1.01)   & \\
4    & 3600 & --0.8 &  --0.7     & --0.3 &   0.2  & --0.7 & {\bf O\,{\sc vi} 1.075, 1.46, 2.46} & LH41-1042 \\
3    & 4200 & --0.8 &  --0.7     & --0.5 &$\leq$0.0 & C\,{\sc iii} weak & {\bf O\,{\sc vi} 1.075, 1.46, 2.46} & WR93b \\
2    & 6300 & --0.8 &  $\leq$--1 & --0.5 & --1.0 & {\bf C\,{\sc iii} absent} & {\bf O\,{\sc vi} 1.075, 1.46, 2.46} & WR102 \\
\hline
\end{tabular}
\end{center}
\end{table}

\end{landscape}

\begin{landscape}

\begin{table}
  \begin{center}
    \begin{footnotesize}
      \caption{Near-IR equivalent width and FWHM measurements for newly identified Galactic WN stars plus previously discovered WN stars lacking
      optical spectroscopy. Equivalent widths (in \AA) are generally robust to $\pm$0.05 dex, except for weak lines $\pm$0.1 dex, while measured FWHM (in km\,s$^{-1}$) are generally
reliable to $\pm$50 km\,s$^{-1}$ (approximate values are indicated with colons). The key to the spectroscopic datasets utilised is provided in Table~\ref{tab:log}.}\label{newbies_wn}
\begin{tabular}{l@{\hspace{1mm}}l@{\hspace{1mm}}c@{\hspace{1mm}}
    c@{\hspace{1mm}}c@{\hspace{1mm}}c@{\hspace{1mm}}c@{\hspace{1mm}}c@{\hspace{1mm}}
    c@{\hspace{1mm}}c@{\hspace{1mm}}c@{\hspace{1mm}}c@{\hspace{1mm}}c@{\hspace{1mm}}c@{\hspace{1mm}}
    c@{\hspace{1mm}}c@{\hspace{1mm}}c@{\hspace{1mm}}c@{\hspace{1mm}}c@{\hspace{1mm}}l}
\hline
WR & WN     & \multicolumn{2}{c}{He\,{\sc ii} 1.01} & He\,{\sc i} 1.08 & 
P$\gamma$ & N\,{\sc v} 1.11   & He\,{\sc ii} 1.16 & P$\beta$          & He\,{\sc ii} 1.48 & N\,{\sc 
v} 1.55 & He\,{\sc ii} 1.69 & He\,{\sc i} 1.70 & He\,{\sc i} 2.06 & N\,{\sc iii-v} 
2.11 & Br$\gamma$ & \multicolumn{2}{c}{He\,{\sc ii} 2.19} & Note & Data \\
& SpType & FWHM & $\log W_{\lambda}$ & $\log W_{\lambda}$& $\log W_{\lambda}$  & $\log W_{\lambda}$ & $\log W_{\lambda}$
& $\log W_{\lambda}$ & $\log W_{\lambda}$  & $\log W_{\lambda}$ & $\log W_{\lambda}$ & $\log W_{\lambda}$ & $\log W_{\lambda}$&  
$\log W_{\lambda}$ & $\log W_{\lambda}$ & FWHM & $\log W_{\lambda}$   \\
\hline
WR47-5&WN6(h)&  1000 & 1.82 &1.90  &1.1 &     & 1.76 & 1.59 & 1.54 &      & 0.95 &$<$0.4&     & 1.3: &1.58 &1630 & 1.59& 2.11$>$2.10    & N6   \\ 
WR56-1&WN5o&   1220 & 2.09 & 1.78 &    &     & 1.73 & 1.32 & 1.50 &      & 1.18 & 0.9: &     & 1.45 &1.30 & 1200& 1.56& 2.11$>$2.10  & N6   \\ 
WR60-8&WN6o&     1510 &2.28  & 2.27 &1.31&0.7: & 2.10 & 1.72 & 1.82 &      & 1.53 & 1.26 &     & 1.46 & 1.63& 1670& 1.83& 2.11$>$2.10  & N6   \\ 
WR62a&WN6o &     1670 & 1.82 & 1.73 &0.7:&     & 1.55 & 1.09 & 1.30 &      &      &      &     & 1.18 & 1.28& 1650& 1.48& 2.11$>$2.10    & N2   \\ 
WR64-2&WN6o&     1480 & 2.29 & 2.26 &1.33&     & 2.08 & 1.67 & 1.83 &      & 1.50 & 1.18 &     & 1.54 & 1.64&1640 & 1.97&2.11$>$2.10     & N6   \\ 
WR64-3&WN6o  &   1340 & 2.18 & 2.20 &1.40&     & 2.02 & 1.80 & 1.75 &      & 1.40 & 1.08 &     & 1.45 & 1.60&1710 &1.79 &2.11$>$2.10   & N6   \\ 
WR64-4&WN6o+&    1930 & 1.87 & 2.00 &1.04&     & 1.71 & 1.43 & 1.48 & 0.5: & 1.20 & 0.94 &     & 1.08 &1.20 & 2220&1.56 &2.11$\gg$2.10   & N6   \\ 
WR64-5&WN6o&     1680 & 2.09 & 2.25 &1.18&0.8: & 1.91 & 1.62 & 1.69 & 0.6: & 1.30 & 1.20 &     & 1.48 & 1.54&1670 & 1.79&2.11$\gg$2.10   & N6   \\ 
WR64-6&WN6b  &   2130 & 2.39 & 2.54 &1.27&     & 2.17 & 1.84 & 1.91 & 0.7  & 1.56 &1.40  &     & 1.55 & 1.61&2300 & 1.99&2.11$\gg$2.10   & N6   \\ 
WR68a&WN6o&      1680 & 1.92 & 1.81 &0.8 &0.5: & 1.68 & 1.31 & 1.46 & 0.6  &      &      &     & 1.30 & 1.41&1570 & 1.54&2.11$>$2.10     & N2   \\ 
WR70-14&WN4b&    2760 & 2.49 & 2.33 &    &     & 2.38 & 1.95 & 2.11 &      & 1.65 &1.3:  &     & 1.69 & 1.89&2800 & 2.23&2.11$\sim$2.10  & N6   \\ 
WR70-15&WN5o&    1450:& 2.3: & 1.85 &1.2:&     & 2.17 & 1.67 & 1.90 & 0.6  & 1.58 &$<$0.5&     & 1.38 & 1.43&1740 &1.98 &2.11$>$2.10     & N6   \\ 
WR72-5&WN6o&     1370 & 2.23 & 2.25 &1.2 &     & 2.07 & 1.69 & 1.82 & 0.5  & 1.52 & 1.26 & 0.6:& 1.63 &1.66 &1500 &1.92 &2.11$\gg$2.10   & N6   \\ 
WR75-31&WN7o  &  1250:& 1.7: & 2.3: &    &     & 1.95 & 1.78 & 1.67  &      & 1.23 &1.62  &1.1: & 1.72 &1.71 &1370 & 1.72&2.11$\sim$2.19  & N6   \\ 
WR75-30&WN7o  &       &      &      &    &     &      &      &      &      &  1.34& 1.59 & 0.8:& 1.84 & 1.79& 1880& 1.82&2.11$\sim$2.19  &  N6  \\ 
WR76-11&WN7o&         &      &      &    &     &      &      &      &      & 1.15 & 1.57 & 1.1 &  1.64& 1.67& 1260& 1.59& 2.11$\sim$2.19 &  N6  \\ 
WR93a&WN6h&     1800 & 1.91  &1.85 &1.71 &     &1.71  & 1.98 &1.38  &      &      &      &     & 1.38 &1.88 &1770:&1.59 &2.11$>$2.10     &  N2  \\
\hline
\multicolumn{20}{l}{
\begin{minipage}{\columnwidth}~\\
Note: 2.10 = N\,{\sc v} 2.100; 2.11 = He\,{\sc i} 2.112 + N\,{\sc iii} 2.116; 2.19 = He\,{\sc ii} 2.189 \\
   \end{minipage}
 }
\end{tabular}
\end{footnotesize}
\end{center}
\end{table}

\begin{table}
  \begin{center}
    \begin{footnotesize}
      \caption{Near-IR equivalent width and FWHM measurements for newly identified Galactic WC stars plus previously discovered WC stars lacking
      optical spectroscopy. Equivalent widths (in \AA) are generally robust to $\pm$0.05 dex, except for weak lines $\pm$0.1 dex, while measured FWHM (in km\,s$^{-1}$) are generally
reliable to $\pm$50 km\,s$^{-1}$. The key to the spectroscopic datasets utilised is provided in Table~\ref{tab:log}.}\label{newbies_wc}
\begin{tabular}{l@{\hspace{1mm}}l@{\hspace{1mm}}c@{\hspace{1mm}}
    c@{\hspace{1mm}}c@{\hspace{1mm}}c@{\hspace{1mm}}c@{\hspace{1mm}}c@{\hspace{1mm}}
    c@{\hspace{1mm}}c@{\hspace{1mm}}c@{\hspace{1mm}}c@{\hspace{1mm}}c@{\hspace{1mm}}c@{\hspace{1mm}}
    c@{\hspace{1mm}}c@{\hspace{1mm}}c@{\hspace{1mm}}c@{\hspace{1mm}}c@{\hspace{1mm}}l}
\hline
WR & WC & C\,{\sc iii} 0.97 & C\,{\sc ii} 0.99 & \multicolumn{2}{c}{He\,{\sc ii} 1.01} & He\,{\sc i} 1.08 & He\,{\sc ii} 1.16 &
C\,{\sc iv} 1.19 & C\,{\sc iii} 1.20 & C\,{\sc iv} 1.43 & He\,{\sc i-ii} 1.70 & \multicolumn{2}{c}{C\,{\sc iv} 1.74} & C\,{\sc ii} 1.78 & He\,{\sc i} 2.06 & C\,{\sc iv} 2.07 & 
C\,{\sc iii} 2.11 & Data \\ 
& SpType & $\log W_{\lambda}$ & $\log W_{\lambda}$& FWHM & $\log W_{\lambda}$ & $\log W_{\lambda}$ & $\log W_{\lambda}$ & $\log W_{\lambda}$ & $\log W_{\lambda}$ & $\log W_{\lambda}$ & $\log W_{\lambda}$ & 
FWHM & $\log W_{\lambda}$ & $\log W_{\lambda}$ & $\log W_{\lambda}$& $\log W_{\lambda}$ & $\log W_{\lambda}$ \\
\hline
WR46-18  & WC6--7  & 2.72 &$<$1.1& 3290 & 2.24 & 2.38 & 2.09 & 2.26 & 1.81 & 1.88 & 1.62 & 3040 & 2.33 &      &      & 3.02 & 2.36       
& N6\\ 
WR60-7   & WC7--8  & 2.99 & 1.63 & 1640 & 2.08 & 2.36 & 2.05 & 2.25 & 1.89 & 2.25 & 1.56 & 1770 & 2.26 &      &      & 2.89 & 2.34       & N6\\ 
WR70-13  & WC8d & 2.89 & 1.68 & 1350 & 1.72 & 2.34 & 1.85 & 1.94 & 2.01 & 2.01 & 1.54 & 1300 & 1.90 & 1.84 &      & 2.31 & 2.03       & N6\\ 
WR75aa   & WC9d & 2.72 & 1.67 & 1060 & 1.41 & 2.26 & 1.76 & 1.75 & 1.98 & 1.75 & 1.11 & 1330 & 1.30 & 1.54 &      & 1.64 & 1.57       & N5\\ 
WR75c    & WC9  & 2.58 & 1.70 &  990 & 1.30 & 2.63 & 1.69 & 1.60 & 2.12 & 1.82 & 1.81 & 1190 & 1.58 & 2.14 & 2.43 & 2.00 & 2.16       & N5\\ 
WR107a   & WC5--7& 3.11 & 1.72 & 2170 & 2.24 & 2.25 & 2.12 & 2.47 & 1.83 & 2.34 &      &      &      &      &      & 2.94 & 2.25       & N2\\ 
\hline 
\end{tabular}
\end{footnotesize}
\end{center}
\end{table}

\end{landscape}

\begin{figure*}
\begin{center}
\includegraphics[width=0.85\textwidth]{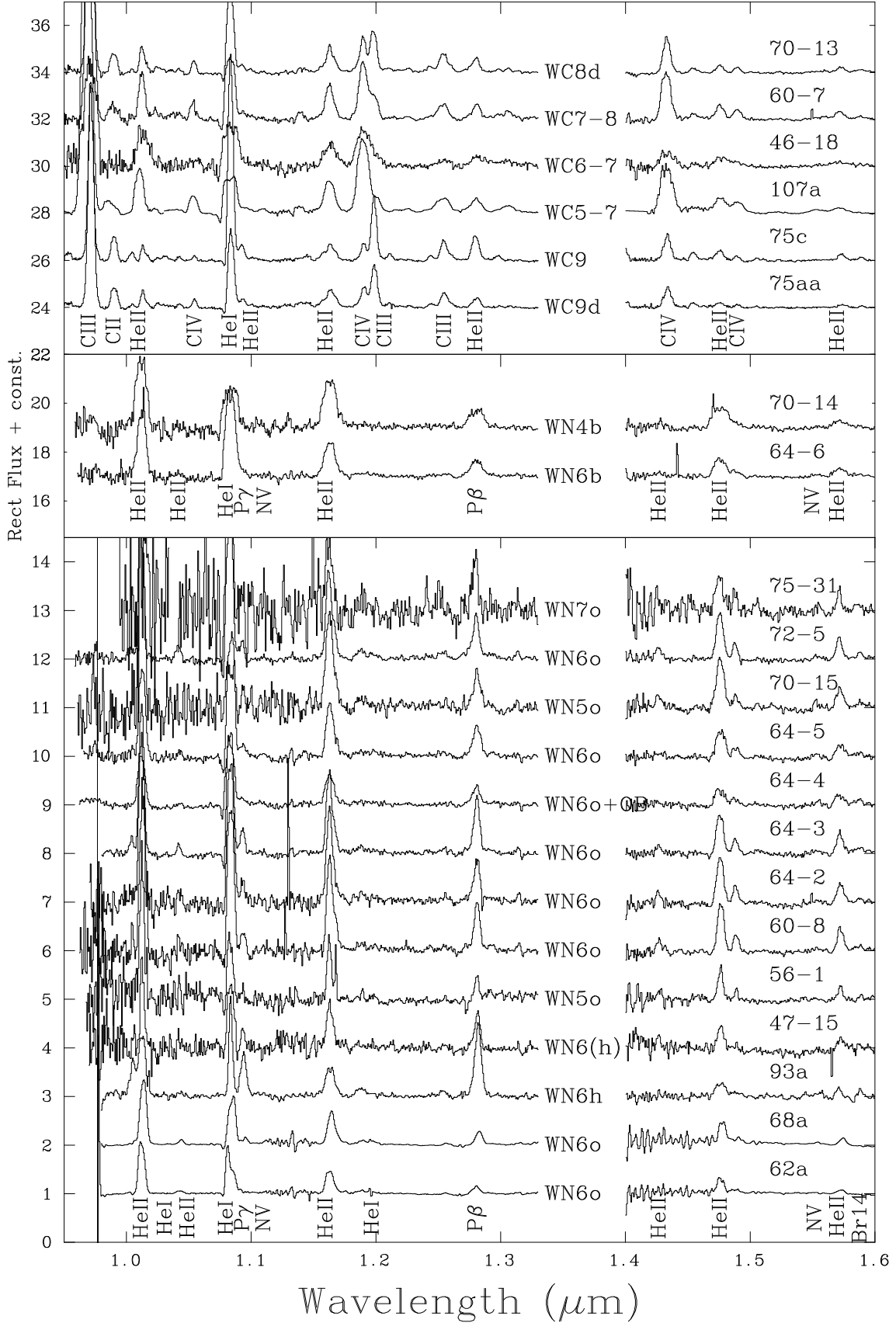}
\caption{YJ-band spectra of new WR stars, plus unpublished NTT/SOFI spectroscopy of previously identified WR stars (WR62a, 
WR68a, WR75aa, WR75c, WR93a and WR107a).}
\label{fig:new_wr_blue}
\end{center}
\end{figure*}

\begin{figure*}
\begin{center}
\includegraphics[width=0.85\textwidth]{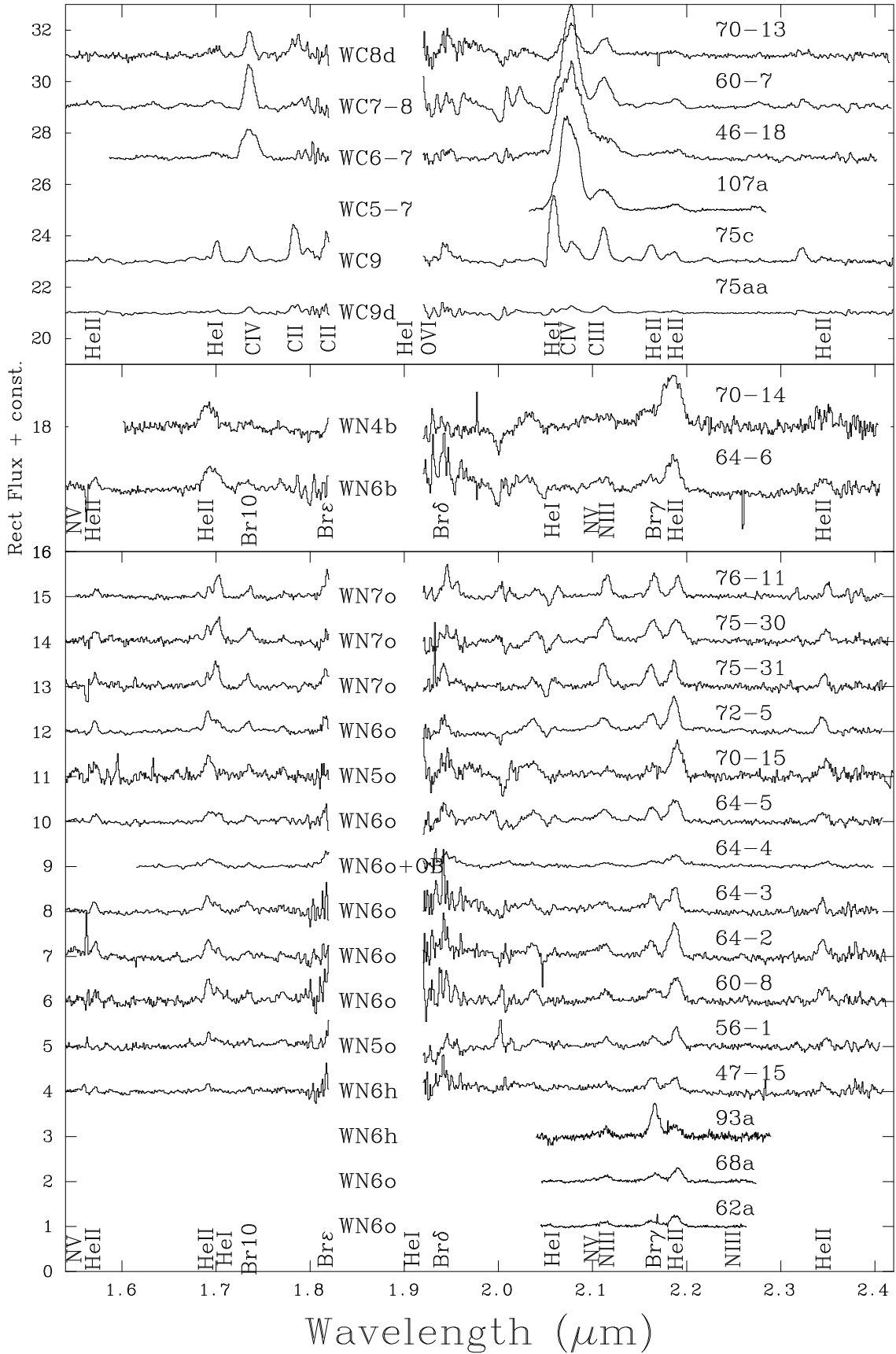}
\caption{HK-band spectra of new WR stars, plus unpublished NTT/SOFI spectroscopy of previously identified WR stars (WR62a, 
WR68a, WR75aa, WR75c, WR93a and WR107a) plus our NTT/SOFI spectroscopy of WR75--30.}
\label{fig:new_wr_red}
\end{center}
\end{figure*}

\begin{table*}
\begin{center}
\caption{Ratios of near-IR diagnostic lines for newly identified Wolf-Rayet stars from NTT/SOFI spectroscopy plus
  revised near-IR classifications for previously known stars. Line strengths/widths are provided for He\,{\sc ii} 1.012$\mu$m
(2.189$\mu$m in parenthesis) or C\,{\sc iv} 1.736$\mu$m (1.190$\mu$m in parenthesis) for WN and WC stars, respectively.}
\label{tab:new_wrstars}
\begin{tabular}{
l@{\hspace{2mm}}l@{\hspace{2mm}}c@{\hspace{2mm}}
r@{\hspace{2mm}}l@{\hspace{1mm}}
r@{\hspace{2mm}}l@{\hspace{1mm}}
r@{\hspace{2mm}}l@{\hspace{1mm}}
r@{\hspace{2mm}}l@{\hspace{1mm}}
r@{\hspace{2mm}}l@{\hspace{1mm}}
l@{\hspace{2mm}}l@{\hspace{1mm}}l}
\hline
ID & WR  & 1.01$\mu$m & 
\multicolumn{2}{c}{$\log W_{\lambda}$} & 
\multicolumn{2}{c}{$\log W_{\lambda}$} &  
\multicolumn{2}{c}{$\log W_{\lambda}$} &  
\multicolumn{2}{c}{$\log W_{\lambda}$} & 
\multicolumn{2}{c}{$\log W_{\lambda}$} & 
Old & Ref & New \\   
    &   & FWHM & 
\multicolumn{2}{c}{(He\,{\sc i} 1.08/ } & 
\multicolumn{2}{c}{(P$\beta$/ } & 
\multicolumn{2}{c}{(He\,{\sc i} 1.70/ } & 
\multicolumn{2}{c}{(Br$\gamma$/ }  & 
\multicolumn{2}{c}{(He\,{\sc i}+N\,{\sc iii} 2.11/ } & 
SpT &  & SpT\\
&    & km\,s$^{-1}$ & 
\multicolumn{2}{c}{He\,{\sc ii} 1.01)} &
\multicolumn{2}{c}{He\,{\sc ii} 1.16)} & 
\multicolumn{2}{c}{He\,{\sc ii} 1.69)} & 
\multicolumn{2}{c}{He\,{\sc ii} 2.19)} & 
\multicolumn{2}{c}{N\,{\sc v} 2.10)} &  
&  &  \\
\hline
B\#13 & WR47-5 &1000     & +0.08 & WN6      & --0.17 & (h)   &  --0.55  & WN4 &  --0.01 & (h) & $>$0 & WN5--6 & -- & --& WN6(h) \\
B\#37 & WR56-1  &1220    &--0.31 & WN5      & --0.41 & o     &  --0.28  & WN5 &  --0.26 & o   & $>$0  & WN5--6 & -- & --& WN5o \\
B\#56 & WR60-8  &1510 &   --0.01 & WN6      & --0.38 & o     &  --0.27  & WN5 &  --0.20 & o   & $>$0 & WN5--6 & -- & --& WN6o \\
      & WR62a   &1670&    --0.09 & WN6      & --0.46 & o  & \multicolumn{2}{c}{---} & --0.20 & o/(h) & $>$0 & WN5--6 &WN5o&S99& WN6o\\
B\#85 & WR64-2  &1480&  --0.03 & WN6        & --0.41 & o     &  --0.32 & WN5  & --0.33 & o    & $>$0 & WN5--6 & -- & --& WN6o \\
B\#87 & WR64-3  &1340&   +0.02 & WN6        & --0.22 & o/(h) &  --0.32 & WN5  & --0.19 & o    & $>$0 & WN5--6 & -- & --& WN6o\\
B\#88 & WR64-4  &1930&   +0.13 & WN6        & --0.28 & o     &  --0.26 & WN5  &  --0.36 & o   & $\gg$0 & WN6 & -- & --& 
WN6o+OB\\
B\#91 & WR64-5  &1680&   +0.16 & WN6    &   --0.29 & o     &  --0.10 & WN6  &  --0.25 & o   & $\gg$0 & WN6 & -- & --& WN6o\\
B\#93 & WR64-6  &2130&   +0.15 & WN6b   &  --0.33 & o     &  --0.16 & WN6b &  --0.38 & o   & $>$0 & WN6b & -- & --& WN6b\\
      & WR68a   &1680&   --0.11& WN6   &  --0.37 & o     & \multicolumn{2}{c}{--- } & --0.13 & o/(h) & $>$0 & WN5--6 &WN6o&S99& WN6o\\
B\#107& WR70-14 &2760&   --0.16& WN4--6b & --0.43 & o  &  --0.35 & WN6b &  --0.34 & o    & $\simeq$0 & WN4b& -- & --& 
WN4b\\
B\#123& WR70-15 &1450&  --0.45 & WN4--5  &  --0.50 & o     &  --1.08 & WN2--4 & --0.55 & o  & $>$0 & WN5--6 & -- & 
--& WN5o\\
B\#132& WR72-5  &1370 &  +0.02 & WN6          & --0.38 & o &  --0.26 & WN5:  &  --0.26 & o   & $\gg$0 & WN6   & -- & --& 
WN6o\\
  A\#11 & WR75-31 &1250&  +0.60: & WN7--8    &  --0.17 & o &  +0.39 & WN7--8 &  --0.01 & o   & 0.00$\ddag$ & WN7  & -- & 
--& WN7o\\
    A\#13 & WR75-30 &(1880)& \multicolumn{2}{c}{---} & \multicolumn{2}{c}{---} &  +0.25 &  WN7 &  --0.03 & o   & 
0.02$\ddag$ & WN7 & WN6 &K15& WN7o\\
B\#154& WR76-11 &(1260)  & \multicolumn{2}{c}{---} & \multicolumn{2}{c}{---} & +0.42 & WN7--8 & --0.08 & o & 0.05$\ddag$ & WN7  & 
-- & --& WN7o\\
& WR93a         &1800 &   --0.06 & WN6   &  +0.27 & h  & \multicolumn{2}{c}{---} & +0.29 & h &  $\gg$0 & WN6  &WN6 &M13& WN6h\\
\hline
ID   & WR   & 1.74$\mu$m & 
\multicolumn{2}{c}{$\log W_{\lambda}$} & 
\multicolumn{2}{c}{$\log W_{\lambda}$} & 
\multicolumn{2}{c}{$\log W_{\lambda}$} & 
\multicolumn{2}{c}{$\log W_{\lambda}$} & 
\multicolumn{2}{c}{$\log W_{\lambda}$} & 
Old & Ref & New\\
  &    & FWHM  & 
\multicolumn{2}{c}{(He\,{\sc i} 1.08/ }  & 
\multicolumn{2}{c}{(C\,{\sc iii} 1.20/ }  & 
\multicolumn{2}{c}{(He\,{\sc i-ii} 1.70/ } & 
\multicolumn{2}{c}{(C\,{\sc iii} 2.11/ } & 
\multicolumn{2}{c}{(C\,{\sc ii} 0.99/ } & 
SpT &  & SpT\\
&    & km\,s$^{-1}$ &
\multicolumn{2}{c}{He\,{\sc ii} 1.01)}& 
\multicolumn{2}{c}{C\,{\sc iv} 1.19)}  & 
\multicolumn{2}{c}{C\,{\sc iv} 1.74)} & 
\multicolumn{2}{c}{C\,{\sc iv} 2.07)}  &
\multicolumn{2}{c}{C\,{\sc iii} 0.97)} &  
& &  \\
\hline
E\#3 & WR46-18  & 3100 & +0.14 & WC6--7 & --0.45 & WC7   & --0.71 & WC6--7 & --0.66 & WC4--7 &  --1.62 & WC4--6 &--&--& WC6--7\\
B\#51& WR60-7   & 1800 & +0.28 & WC7--8 & --0.36 & WC8   & --0.70 & WC6--7 & --0.55 & WC7--8 &  --1.36 & WC7--8 &--&--& WC7--8\\
B\#105& WR70-13 & 1300 & +0.62 & WC8--9 & +0.07  & WC8--9& --0.36 & WC8    & --0.28 & WC8    & --1.21 & WC8--9 & -- & --& WC8d\\
      & WR75aa & 1400 &  +0.85 & WC9    & +0.23  & WC9   & --0.19 & WC8--9 & --0.07 & WC9    & --1.05 & WC9    &WC9d&H05& WC9\\
      & WR75c  & 1350 & +1.33 & WC9     & +0.52  & WC9   & +0.23 & WC9     & +0.19 & WC9     & --0.88 & WC9    &WC9 &H05 & WC9\\
      & WR107a &(2400)& +0.01 & WC5     & --0.64 & WC5--7 & \multicolumn{2}{c}{---} 
                                                                           & --0.69 & WC5--7 & --1.39 & WC5--8 &WC6&S99& WC5--7\\
\hline
\multicolumn{15}{l}{
\begin{minipage}{2\columnwidth}~\\
 S99 \citep{shara99}, H05 \citep{hopewell05}, M13 
 \citep{miszalski13}, K15 \citep{kanarek15} \\
$ \ddag: \log W_{\lambda}$ (He\,{\sc i} 2.112 + N\,{\sc iii} 2.116/He\,{\sc ii} 2.189)\\ 
   \end{minipage}
 }
\end{tabular}
\end{center}
\end{table*}

\section{New Galactic Wolf-Rayet stars}\label{sec:new}

We have identified 16 new Wolf-Rayet stars, which we have assigned Galactic WR numbers, in accordance with the 
current IAU convention (see Appendix of RC15). Here we discuss their spectral types, spatial location and 
their potential association with Scutum-Crux or other spiral arms. Near-IR spectra of the new WR stars are 
presented in Figures~\ref{fig:new_wr_blue} (IJ) and \ref{fig:new_wr_red} (HK), together with our NTT/SOFI observations of WR75aa, WR75c, the 
recently discovered WN star WR75-30  \citep{kanarek15}, plus previously unpublished NTT/SOFI spectroscopy of WR62a, WR68a, WR93a, WR107a, as discussed above.
line measurements are provided in Table~\ref{newbies_wn} and \ref{newbies_wc} for WN and WC stars, respectively, with diagnostic line ratios
presented in Table~\ref{tab:new_wrstars}.

\begin{figure*}
\begin{center}
  \includegraphics[width=2\columnwidth]{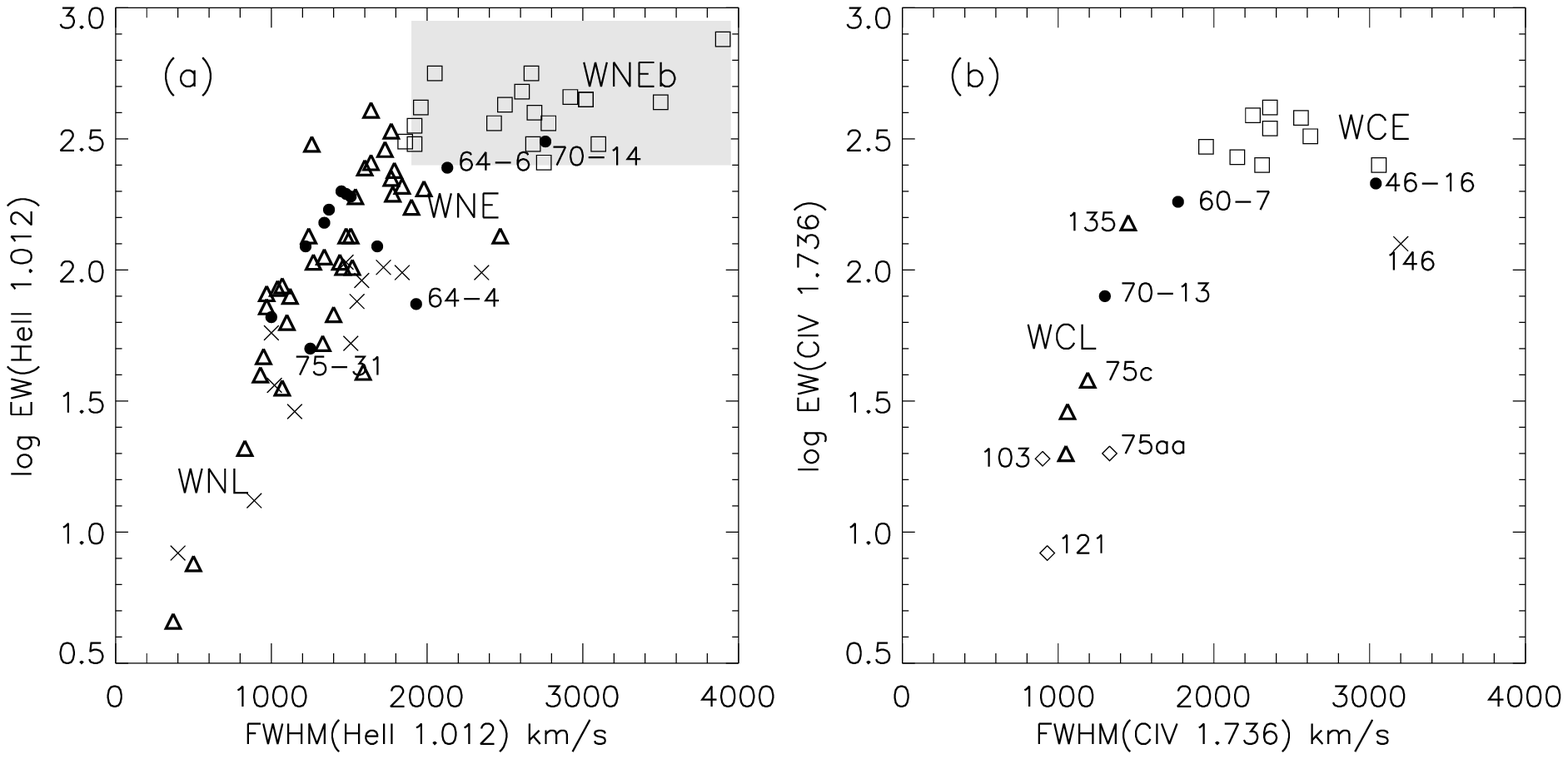}
\caption{{\bf (a):} FWHM (in km\,s$^{-1}$) vs equivalent width (in \AA) for He\,{\sc ii}\,1.012\micron\ in
  apparently single Galactic broad-lined WN stars (open squares), weak-lined WN stars (open triangles) and WNha/+abs stars (crosses)
  together with the newly identified WN stars (filled circles). The grey region indicates the parameter space covered by broad-lined WN
  stars. {\bf (b)}: FWHM (in km\,s$^{-1}$) vs equivalent width (in \AA) for C\,{\sc iv}\,1.74\micron\ in
  apparently single Galactic WC4--7 stars (open squares), WC8--9 stars (open triangles) plus WCd/WC+O systems (crosses)
  together with the newly identified WC stars (filled circles).} 
\label{fig:wn-lines}
\end{center}
\end{figure*}

\subsection{Classification of the new WR stars}

\subsubsection{Broad-lined WN stars}

Only two of the new WN stars, WR64--6 and WR70--14, are identified as broad-lined WN stars, owing to their He\,{\sc ii}
1.01$\mu$m line widths (FWHM $>$ 1900 km\,s$^{-1}$) and strengths ($\log W_{\lambda}$/\AA\ $\geq$ 2.4), albeit WR64-6 only
narrowly complies with the second criterion. A WN6b subtype is favoured for WR64--6 from its He\,{\sc i} 1.08/He\,{\sc ii} 1.01$\mu$m
ratio which is supported by (N\,{\sc iii} + He\,{\sc i} 2.11) $>$ N\,{\sc v} 2.10, while both hydrogen criteria (involving
P$\beta$ and Br$\gamma$) indicate no hydrogen, so we adopt WN6b for this star. For WR70--14, the He\,{\sc i} 1.08/He\,{\sc ii} 
1.01$\mu$m ratio is somewhat ambiguous, consistent with WN4--6b, but (N\,{\sc iii} + He\,{\sc i} 2.11) $\sim$ N\,{\sc v} 2.10
favours WN4b. This is supported by weak N\,{\sc v} 1.11$\mu$m in the J-band (Fig.~\ref{fig:new_wr_blue}). Again, there is no 
evidence for atmospheric hydrogen from  our criteria (Fig.~A4, available online), so WN4b is assigned to WR70--14.

\subsubsection{Narrow-lined WN stars}

The remaining 11 WN stars are a relatively homogeneous group, almost all classified as either WN5o, WN6o or WN7o stars, with 
only WR47--5 showing evidence of hydrogen so we consider these according to their subtype.

The two highest ionization narrow-lined stars are WR56-1 and WR70-15, according to their He\,{\sc i} 
1.08/He\,{\sc ii} 1.01 ratios (Fig.~A3, available online), which indicate WN5 for both stars. This is supported by the He\,{\sc 
i}  1.70/He\,{\sc ii} 1.69 ratio for WR56-1, although an earlier WN3--4 subtype is favoured by He\,{\sc i} 1.70/He\,{\sc ii} 1.69 
for WR70-15. We also consider  their K-band morphologies, for which (N\,{\sc iii} + He\,{\sc i} 2.11) $>$ N\,{\sc v} 2.10 in 
both cases, indicating WN5--6. Neither star shows any evidence for atmospheric hydrogen from Fig.~A4 so we 
adopt WN5o for both stars.

The majority of our narrow-lined WN stars are WN6 stars according to their He\,{\sc i}
1.08/He\,{\sc ii} 1.01 ratios (Fig.~A3), with He\,{\sc i} 1.70/He\,{\sc ii} 1.69 suggesting WN4, 5 or 6. As with
the WN5o stars considered above, we also consider the K-band morphology, for which either (N\,{\sc iii} + He\,{\sc i} 2.11) 
$>$ N\,{\sc v} 2.10, implying WN5--6 or (N\,{\sc iii} + He\,{\sc i} 2.11) $\gg$ N\,{\sc v} 2.10, implying WN6. Only WR47--5
indicates the presence of (modest) hydrogen from Fig.~A4, such that we classify it as WN6(h), but favour
WN6o for WR60-8, WR64-2, -3, -4, -5 and WR72-5.

Of the remaining stars only WR75-31 was observed in the IJ-band with SOFI, for which He\,{\sc i} 1.08/He\,{\sc ii} 1.01 
indicates a WN7--8 subtype, with He\,{\sc i} 1.70/He\,{\sc ii} 1.69 also providing an ambiguous WN7--8 classification. Its 
K-band morphology strongly favours WN7 since (N\,{\sc iii} + He\,{\sc i} 2.11) $\sim$ He\,{\sc ii} 2.19, while there is no 
evidence for atmospheric hydrogen in WR75-31 from Fig.~A4, such that we assign WN7o to this star. WR76-11 
was observed solely in the H and K bands, but closely resembles WR75-31 such that we classify it as WN7o, as with WR75-30. 
None of the new WN stars qualify as WN/C stars, since C\,{\sc iii} 0.971$\mu$m, C\,{\sc iv} 1.19$\mu$m, 1.74$\mu$m, 
2.07$\mu$m are weak/absent.



\subsubsection{WC stars}

Three of the new WR stars are carbon sequence WC stars. Considering the primary diagnostics for WR46--16, WC7 is favoured 
from the C\,{\sc iv} 1.19/C\,{\sc iii} 1.20 ratio, 
WC4--6 from the C\,{\sc iii} 0.97/C\,{\sc ii} 0.99 ratio, and WC5--7 from the C\,{\sc iv} 2.07/C\,{\sc iii} 2.11 ratio.
Secondary 
indicators suggest WC5--7 from He\,{\sc i} 1.08/He\,{\sc ii} 1.01, WC5--6 from C\,{\sc iii} 0.97/He\,{\sc ii} 
1.01 and WC6--7 from He\,{\sc i-ii} 1.7/C\,{\sc iv} 1.74. Overall, we adopt WC6--7 reflecting the tension in primary
indicators for WR46--16 (Fig.~A7, available online). 

WR60--7 is classified as WC8, WC7, WC7--8 from primary diagnostics C\,{\sc iv} 1.19/C\,{\sc iii} 1.20, C\,{\sc iv} 
2.07/C\,{\sc iii} 2.11 and C\,{\sc iii} 0.97/C\,{\sc ii} 0.99, respectively. Secondary criteria 
C\,{\sc iii} 0.97/He\,{\sc ii} 1.01 and  He\,{\sc i-ii} 1.7/C\,{\sc iv} 1.74 indicate WC6--7, while
He\,{\sc i} 1.08/He\,{\sc ii} 1.01 favours WC7--8. Overall, WC7--8 is selected for WR60--7, reflecting the lack of a 
consensus amongst primary criteria (Fig.~A7). 

Finally, primary diagnostics C\,{\sc iv} 1.19/C\,{\sc iii} 1.20, C\,{\sc iv} 
2.07/C\,{\sc iii} 2.11 and C\,{\sc iii} 0.97/C\,{\sc ii} 0.99, imply WC8--9, WC8 and WC8--9 for WR70--13, while C\,{\sc iv} 
1.74 $\geq$ C\,{\sc ii} 1.78 indicates WC8. Consequently we adopt WC8 for WR70--13, which is supported by our secondary 
indicator He\,{\sc i-ii} 1.7/C\,{\sc iv} 1.74, with He\,{\sc i} 1.08/He\,{\sc ii} 1.01 and C\,{\sc iii} 0.97/He\,{\sc 
ii} 1.01 consistent with either WC8 or WC9 (Fig.~A7).


\subsection{Binarity}\label{binary}


Approximately 40\% of the Galactic WR population are observed in multiple
systems \citep{vdh01}. This is a lower limit on the true binary fraction, since no
systematic survey has been carried out. It is therefore highly likely that some
of the newly discovered WR stars are in fact multiple systems. Direct detection
of companion stars, usually main-sequence OB stars, is not
possible with the current dataset since their absorption lines are generally weak
with respect to the strong WR emission lines.

It is, however, possible to infer the presence
of a companion star by considering the equivalent width of near-IR emission lines which
will be diluted by the continuum of a companion star, and/or dust for the
case of some WC+OB systems since dust formation is an indicator of binarity in WC stars.

Since a companion star and/or thermal dust emission will not reduce line widths, a weak line
compared to single stars at a specific FWHM is suggestive of binarity.
In Figure~\ref{fig:wn-lines} we compare the FWHM (km\,s$^{-1}$) and equivalent widths (in \AA)
of strong, isolated lines in apparently single Galactic WN stars
(He\,{\sc ii}\,1.012\micron) and WC stars (C\,{\sc iv} 1.736\micron) 
with newly discovered WR stars. We also include  weak-lined WN stars with intrinsic absorption
lines (WR24, WR87, WR108) which could be mistaken for WN+OB stars,
plus dusty WC stars (WR121, WR75aa), whose near-IR emission lines are diluted by hot dust.

Of the newly identified stars, the majority of WR stars possess emission line strengths
which are charactistic of single stars. From Fig.~\ref{fig:wn-lines}(a) two
exceptions are WR75-31 (WN7(h)) and WR64-4 (WN6o) which
possess weak emission for their He\,{\sc ii} 1.012$\mu$m FWHM. Both are potential binaries, although WR64-4
is the strongest candidate, such that we revise its spectral type to WN6o+OB. In contrast, WR75-31
has an overall relatively strong emission line spectrum, albeit with an anomalously weak (and low S/N)
He\,{\sc ii} 1.0124$\mu$m line.

Of the WC stars, none possess unusually weak emission lines based on their
C\,{\sc iv} 1.736$\mu$m FWHM (Fig.~\ref{fig:wn-lines}(b)). However, the increased dilution of WC emission lines from 1$\mu$m 
to 2.5$\mu$m arising from hot dust in WCd systems also severely modifies equivalent width ratios of C\,{\sc iii-iv} lines.
For example, $W_{\lambda}$(C\,{\sc iii} 2.11)/$W_{\lambda}$(C\,{\sc iii} 0.97) = 0.3 for WR88 (WC9) but hot
dust in WR121 (WC9d) reduces this ratio to 0.05. A similar reduction in line strength is observed for prominent
He\,{\sc ii} lines, with $W_{\lambda}$(He\,{\sc ii} 2.19)/$W_{\lambda}$(He\,{\sc ii} 1.28) = 0.5 for WR88 and 
0.17 for WR121. WR135 is a prototypical non-dusty WC8 star with
$W_{\lambda}$(C\,{\sc iii} 2.11)/$W_{\lambda}$(C\,{\sc iii} 0.97) = 0.2, with a ratio of 0.2 for WR60-7
but only 0.1 for WR70-13, suggestive of dust dilution in the latter. Indeed, WR60-7 (WC7--8) and WR70-13 (WC8) possess
similar J-H colours, yet the latter has 0.5 mag higher K$_S$--[8] colour (Table~\ref{tab:obs}), so we amend its spectral type 
to WC8d. Indeed, WR70-13 is offset from the other WC stars in the reddening free Q1 vs Q2 comparison in Fig.~\ref{fig:c-c}.
Turning to WR46-18, $W_{\lambda}$(C\,{\sc iii} 2.11)/$W_{\lambda}$(C\,{\sc iii} 0.97) $\sim$0.3 for non-dusty WC6--7 stars, with
a ratio of 0.4 for WR46-18, arguing against dust emission in this instance.


Discovery of a hard X-ray source associated with any of the WR stars would be 
highly indicative of stellar wind collision in a massive binary. Indeed, several WR 
stars towards the Galactic Centre, coincide with hard X-ray sources 
(e.g., \citealt{mauerhan10}, \citealt{nebot15}). From a search of the XMM Newton 
science archive and the Chandra source catalogue (1.0), fields including
WR60-7 (WC7--8), WR60-8 (WN6o), and WR64-4 (WN6o) had been imaged by Chandra ACIS, although none
revealed a source at the location of the WR star.

\subsection{Spatial location of the new WR stars}

\begin{figure}
\begin{center}
\includegraphics[width=0.9\columnwidth]{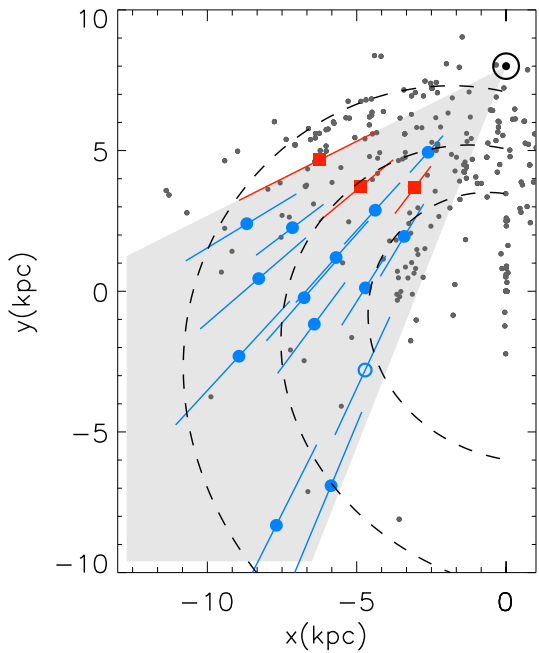}
\caption{
A top-down view of the Galactic disk showing the 4th quadrant. The x-axis is parallel to the 
line-of-sight in the direction $l\,{=}\,270^\circ$. Shaded grey is the swathe of Galactic longitude 
constituting the survey area ($298^\circ\,{<}\,l\,{<}\,340^\circ$). New WN stars are
shown as filled blue circles (WR75-30 is an open symbol), new WC stars are shown as filled red 
squares, while 355 previously identified WR stars are small grey circles \citep{rosslowe15a}. Dashed lines 
show the measured locations of prominent spiral arms in this Galactic quadrant, represented as 
logarithmic spirals. }
\label{fig:spiral}
\end{center}
\end{figure}


We have estimated distances to the new WR stars by adopting an absolute K$_S$-band magnitude based on the 
assigned spectral subtype. To do this, we followed the approach of RC15, which we briefly summarise here. 
To calculate the foreground dust extinction to each new WR star, we used subtype-specific intrinsic J$-$H 
and H$-$K$_S$ colours to measure a colour excess. Using the near-IR extinction law of \citet{stead09}, we 
thereby obtained two measures of extinction in the K$_S$-band, of which an average was taken. Distances 
were then calculated using the standard relation between absolute and apparent magnitude. We used the 
uncertainties on calibrated absolute magnitudes, given by RC15, to calculate upper and lower bounds on the 
distances calculated.  In Table~\ref{tab:dist} we provide interstellar extinctions, distance 
moduli/distances, with uncertainties, for every new WR star. Typical extinctions are $A_{\rm K} = 
1.2\pm$0.2 mag, with characteristic distances of 9.7$\pm$3.8 kpc.

This method inherently assumes the WR star is the sole (or dominant) contributor of near-IR flux to each 
source. Recalling Sect.~\ref{binary}, this assumption is justified for new WN stars with the exception of 
WR64-4, the emission line strengths of which suggest a significant contribution from a companion source, 
implying a larger distance. In addition, we provide a second distance estimate to WR70-13 in 
Table~\ref{tab:dist} since there is evidence for a contribution by circumstellar dust to the K-band. 
Adopting the absolute magnitude of a WC8 star which is dominated by hot dust would significantly increase 
the distance to WR70-13 from 5.3 to 10.7 kpc, though in reality the dust contribution is likely to be modest 
such that an intermediate distance is more realistic.



\begin{table}
  \begin{center}
\caption{Estimated interstellar extinctions and distances to the newly discovered WR stars, following
the methodology of C15. Two entries are provided for the WR70-13, the second appropriate for a 
dusty WC8 star. We also provide a distance estimate to the recently identified star WR75-30 
\citep{kanarek15} resulting from our new spectroscopic classification.}
\begin{footnotesize}
\begin{tabular}{l@{\hspace{1mm}}l@{\hspace{0.5mm}}c@{\hspace{1mm}}c@{\hspace{2mm}}c@{\hspace{2mm}}r}
\hline
WR              & Spec. & M$_{\mathrm{K}_\mathrm{S}}$ & A$_{\mathrm{K}_\mathrm{S}}$ & DM & \multicolumn{1}{c}{d} \\
                & Type & mag & mag & mag & \multicolumn{1}{c}{kpc} \\
\hline
WR46-18 & WC6--7 & $-4.75 \pm 0.77$ & 0.97$\pm$0.10 & 14.25$\pm$0.78 & $7.1^{+3.0}_{-2.1}$ \\
WR47-5 & WN6(h)  & $-4.94 \pm 0.46$ & 0.96$\pm$0.02 & 15.07$\pm$0.46 & $10.3^{+2.4}_{-2.0}$ \\
WR56-1 & WN5o    & $-3.86 \pm 0.34$ & 1.23$\pm$0.00 & 14.81$\pm$0.34 & $9.2^{+1.6}_{-1.3}$ \\
WR60-7 & WC7--8  & $-4.94 \pm 0.55$ & 1.16$\pm$0.02 & 14.06$\pm$0.55 & $6.5^{+1.9}_{-1.5}$ \\
WR60-8 & WN6o    & $-4.94 \pm 0.46$ & 1.45$\pm$0.05 & 15.25$\pm$0.46 & $11.2^{+2.6}_{-2.1}$ \\
WR64-2 & WN6o    & $-4.94 \pm 0.46$ & 1.15$\pm$0.02 & 15.68$\pm$0.46 & $13.7^{+3.2}_{-2.6}$ \\
WR64-3 & WN6o    & $-4.94 \pm 0.46$ & 0.98$\pm$0.01 & 14.14$\pm$0.46 & $6.7^{+1.6}_{-1.3}$ \\
WR64-4 & WN6o+ & $-4.94 \pm 0.46$ & 1.02$\pm$0.04 & 13.02$\pm$0.46 & $4.0^{+0.9}_{-0.8}$ \\
WR64-5 & WN6o    & $-4.94 \pm 0.46$ & 1.00$\pm$0.01 & 14.74$\pm$0.46 & $8.9^{+2.1}_{-1.7}$ \\
WR64-6 & WN6b    & $-5.16 \pm 0.37$ & 1.13$\pm$0.01 & 15.14$\pm$0.37 & $10.7^{+2.0}_{-1.7}$ \\
WR70-13 & WC8d   & $-5.04 \pm 0.41$ & 1.38$\pm$0.08 & 13.62$\pm$0.42 & $5.3^{+1.1}_{-0.9}$ \\
        &        & $-6.57 \pm 0.41$ & 1.38$\pm$0.08 & 15.15$\pm$0.42 & $10.7^{+2.3}_{-1.9}$ \\
WR70-14 & WN4b   & $-4.85 \pm 0.38$ & 1.11$\pm$0.04 & 15.24$\pm$0.38 & $11.2^{+2.1}_{-1.8}$ \\
WR70-15 & WN5o   & $-3.86 \pm 0.34$ & 1.45$\pm$0.01 & 14.81$\pm$0.34 & $9.2^{+1.6}_{-1.3}$ \\
WR72-5  & WN6o   & $-4.94 \pm 0.46$ & 1.00$\pm$0.00 & 14.21$\pm$0.46 & $6.9^{+1.6}_{-1.3}$ \\
WR75-31 & WN7o   & $-5.49 \pm 0.42$ & 1.42$\pm$0.02 & 16.28$\pm$0.42 & $18.0^{+3.9}_{-3.2}$ \\
WR75-30 & WN7o   & $-5.49 \pm 0.42$ & 1.70$\pm$0.06 & 15.36$\pm$0.42 & $11.8^{+2.5}_{-2.1}$ \\
WR76-11 & WN7o   & $-5.49 \pm 0.42$ & 1.45$\pm$0.05 & 16.02$\pm$0.42 & $16.0^{+3.4}_{-2.8}$ \\
\hline
\end{tabular}
\end{footnotesize}
\label{tab:dist}
\end{center}
\end{table}


\subsection{Association of WR stars with the Scutum-Crux and other spiral arms}

In Figure~\ref{fig:spiral} we present the locations of the new WR stars on a top-down
  view of the Galactic disk, together with WR stars mapped by RC15
  (including approximately half of the currently known population). Over-plotted are the locations of the three 
main spiral features detected in the 4th Galactic quadrant 
($270\,{<}\,l\,{<}\,360$), assuming $R_G\,{=}\,8.0$\,kpc. We assume each arm is
a logarithmic spiral, parameterised as: 
\begin{equation}
x\,{=}\,r\cos(\theta),~y\,{=}\,r\sin(\theta),~r\,{=}\,r_t\exp[(\theta-\theta_0)\tan(p)].
\end{equation}
in which $r_t$ is the Galactocentric radius of the observed tangent,
$\theta$ is the angle measured 
anti-clockwise about the origin from the positive x-axis, and $\theta_t$ is the 
angle at which the observed tangent is located. The parameter $p$ is the pitch 
angle of the arm. The calculation of $r_t$ requires a measurement of the 
longitude of the tangent to each arm ($l_t$), and the Galactocentric radius of 
the Sun ($R_G$). 
\begin{equation}
r_t=R_G\sin(360^\circ-l_t)/\sin(90^\circ-p).
\end{equation}  
\citet{vallee14} catalogued the observed tangents to spiral arms in the Galaxy, and calculate 
averages of measurements using different tracers. Subsequently, \citet{vallee15} 
use these observations to measure pitch angles for 
individual spiral arms. From these studies we adopt 
$l_t\,{=}\,284^\circ\,\&\,p\,{=}\,14^\circ$ for the Sagittarius-Carina arm, 
$310^\circ\,\&\,13.3^\circ$ for the Scutum-Crux arm, and $328^\circ\,\&\,9.9^\circ$ 
for the Norma (3kpc) arm. 

The new WR stars appear to be evenly distributed throughout the section of Galactic disk observed; neither 
they, nor the previously mapped WR stars, show any obvious association to the spiral features in the 
region. Indeed, no pattern resembling a spiral arm can be seen in the distribution of WR stars, as would 
be expected if they were tightly confined to spiral arms, albeit affected by a systematic offset in distance 
measurements. The upcoming second data release (DR2) from {\it Gaia} should address this general question, 
although the majority of the newly discovered WR stars are too faint for reliable parallaxes with {\it Gaia}. Indeed, only
  7 of the 17 stars are included in the {\it Gaia} first data release \citep[DR1][]{GAIA-DR1}, with G = 17.4 -- 20.1 mag.

\begin{table*}
\begin{center}
\caption{Galactic WR stars hosted by a star cluster in the range 298$^{\circ} \leq l \leq 340^{\circ}$, updated from \citet{lundstrom84}.
We consider WR stars to be associated with a cluster if $r \leq 4 R$, representing 27\% of the known WR content of this region of the Milky Way.}
\label{tab:clusters}
\begin{tabular}{
l@{\hspace{2mm}}l@{\hspace{2mm}} 
c@{\hspace{2mm}}c@{\hspace{2mm}} 
r@{\hspace{2mm}}r@{\hspace{2mm}} 
l@{\hspace{2mm}} 
l@{\hspace{2mm}}l@{\hspace{2mm}} 
l}
\hline
Cluster     & Alias        & $l$     & $b$     & $d$  & $R$   &Ref& WN ($r$, arcmin)           & WC      ($r$, arcmin)         & Ref \\
            &              & deg     & deg     & kpc  & arcmin&   &                            &                               &  \\
\hline
            & VVV CL 011   & 298.506 & --0.170 & 5.0: & 0.1  & d, l               & WR46-17 (0.0)               &                           & d \\
            & Mercer 30    & 298.756 & --0.408 & 7.2  & 0.3  & a, j            & WR46-3 (0.2), WR46-4 (0.1)  &                           & a \\ 
            &              &         &         &      &      &                 & WR46-5 (0.1), WR46-6 (0.2)  &                           & a \\
C 1240-628  & Hogg 15      & 302.047 & --0.242 & 3.0  & 3.5  & k               & WR47 (1.6)                  &                           & b \\ 
C 1309-624  & Danks 1       & 305.339 &  +0.080 & 4.2 & 0.75 & c, k            & WR48-8 (0.6), WR48-9 (0.6) & WR48a (1.3), WR48-3 (1.9) & c \\ 
            &               &         &         &     &      &                 & WR48-10 (0.6), WR48-6 (2.7)& WR48-4 (2.4)              & c \\ 
            &               &         &         &     &      &                 & WR48-7 (2.5)               &                           & c \\
C 1310-624  & Danks 2       & 305.393 &  +0.088 & 4.2 & 0.75 & c, k            &                            & WR48-2 (0.6)              & c \\
             & VVV CL 036   & 312.124 &  +0.212 & 2.0:& 0.8 & d, l                & WR60-6 (0.1)                &                           & d \\
             & VVV CL 041   & 317.109&  +0.281 &  4.2 & 0.5 & e, l                & WR62-2 (0.2)                &                           & e \\
             &              &          &        &     &     &                  &                      &                           &   \\
             & Pismis 20    & 320.516 &--1.200  & 3.6 & 2.0 & k                & WR67 (2.1)                  &                           & i \\ 
             & Mercer 70    & 329.697 & +0.584  & 7.0  & 0.4 & f, j               &                             & WR70-12 (0.4)             & f\\
             & VVV CL 073   & 335.894  & +0.133 & 4.0:& 0.3 & d, l                & WR75-25 (0.1), WR75-26 (0.1)&                           & d \\
             & VVV CL 074   & 336.373  & +0.194 & 6.0:& 0.55 & d, l               & WR75-28 (0.1), WR75-29 (0.0)&  WR75-27 (0.3)          & d \\
             & Mercer 81    & 338.384 & +0.111  & 11.0 & 0.6 & g, j                & WR76-2 (0.6), WR76-3 (0.7) &                           & g \\
             &              &         &         &     &     &               & WR76-4 (0.9), WR76-5 (0.8) &                           & g \\
             &              &        &          &     &      &              & WR76-6 (0.9), WR76-7 (0.9) &                           & g \\
             &              &        &          &     &      &              & WR76-8 (0.9), WR76-9 (1.0) &                           & g \\
C 1644-457   & Westerlund 1 & 339.555 & --0.399 & 3.8 & 1.2 & m, n, k            & WR77a (1.7), WR77c (0.8) &  WR77aa (4.1), WR77b (4.8)  & h \\ 
             &              &         &         &     &     &               & WR77d (1.1), WR77e (0.4) & WR77g (0.1), WR77i (0.9)    & h  \\
             &              &        &        &     &       &               & WR77f (0.4), WR77h (0.1) & WR77l (0.6), WR77m (0.4)    & h \\
             &              &        &        &     &       &               & WR77j (0.7), WR77k (0.4) & WR77n (1.7), WR77p (0.8)    & h \\
             &              &        &        &     &       &               & WR77o (0.4), WR77q (0.5) &                            & h \\
             &              &        &        &     &       &               & WR77r (0.8), WR77s (0.5)  &                            & h \\
             &              &        &        &     &       &               & WR77sa (0.4), WR77sb (2.0) &                           & h \\
             &              &        &        &     &       &               & WR77sc (0.8), WR77sd  (2.8)&                           & h \\
\hline
\multicolumn{9}{l}{
\begin{minipage}{2\columnwidth}~\\
a: \citet{kurtev07}; 
b: \citet{sagar01};  
c: \citet{davies12a};  
d: \citet{chene13};  
e: \citet{chene15};  
f: \citet{delaFuente15}; 
g: \citet{davies12b}; 
h: \citet{crowther06} 
i: \citet{lundstrom84}
j: \citet{mercer05}
k: \citet{dias02}
l: \citet{borissova11}
m: \citet{kothes07}
n: \citet{koumpia12}
  \end{minipage}
}
\end{tabular}
\end{center}
\end{table*}

%

\begin{table*}
\begin{center}
\caption{Galactic WR stars coincident with star-forming regions in the range 298$^{\circ} \leq l \leq 340^{\circ}$. We consider a WR star to be 
associated with a star-forming region \citep[WISE v1.4]{anderson14} if $r \leq 1.5 R_{\rm HII}$, although some WR stars will doubtless be foreground sources, such that
53/206 (= 26\%) WR stars associated with star-forming regions represents a strict upper limit.}
\label{tab:SFRegions}
\begin{tabular}{
l@{\hspace{2mm}} 
r@{\hspace{1mm}}r@{\hspace{2mm}} 
l@{\hspace{2mm}} 
l@{\hspace{1mm}}                 
l@{\hspace{1mm}}l@{\hspace{2mm}} 
l}
\hline
WISE        & $d$  & $R_{\rm HII}$   &Ref& Cluster? & WN ($r$, arcmin)           & WC      ($r$, arcmin)         & Ref \\
H\,{\sc ii} & kpc  & arcmin&   &          &                            &                               &  \\
\hline
G298.224--0.334 & 11.1  & 5.0   & a &                   &                                             & WR46-7 (5), WR46-18 (8) & b, 
p  \\ 
G298.529--0.251 & 10.5  & 16.2  & a & VVV CL 011        &  WR46-8 (15.6), WR46-9 (3.5), WR46-17 (4.9) & WR46-10 (7.6)       & b, c, d, e  \\ 
                &       &       &   &                   &  WR46-15 (16.9), WR46-2 (22.4)              &                     & g \\
G298.862--0.432 & 10.7  & 4.8   & a & Mercer 30         & WR46-3 (6.6), WR46-4 (6.6), WR46-5 (6.6),   &                     & f \\ 
                &       &       &   &                   & WR46-6 (6.6)                                &                     & f \\
G302.503--0.762 & 12.1  & 4.2   & a &                   & WR47b (4.0)                                 &                     & h \\
G302.631+0.030  &  4.6  & 14.3  & a &                   &  WR47-1 (18.9)                              &                     & e \\ 
G303.445-0.745  & 12.5  & 7.2   & a &                   &                                             & WR47-2 (2.6)        & d \\ 
G305.233+0.110  & 4.9   & 11.3  & a & Danks 1/2         & WR48-6 (5.5), WR48-10 (6.5), WR48-7 (7.7)   & WR48-1 (8.6), WR48-3 (6.0) & e, g, h, i \\ 
                &       &       &   &                   & WR48-9 (6.8), WR48-8 (6.8)                  & WR48-4 (8.2), WR48a (8.4)  & h, i \\
                &       &       &   &                   &                                             & WR48-2 (10.3)       & b \\
G305.322--0.255 & 4.9   & 13.4  & a &                   & WR48-5 (4.7)                                &                     & h \\ 
%
%
G311.991+0.219 &       & 6.0   & a & VVV CL 036       & WR60-6 (7.8)                                  &                     & c \\ 
G317.030+0.028$^{\ast}$ &       & 24.4 & a & VVV CL 041       & WR62-2 (15.7), WR62-1 (19.4), WR63 (34)               &                     & h, j \\ 
G321.015--0.527& 4.1  & 4.0    & a &                  & WR67-3 (6.1)                                  &                     & k \\ 
G321.115--0.546& 3.8  & 4.2    & a &                  & WR67-1 (1.8)                                  &                     & l \\ 
G326.474--0.292&      & 20.9   & a &                  &                                               & WR70-5 (17)         & m \\ 
G327.824+0.117 & 7.2  & 3.9    & a &                  & WR70-6 (5.2)                                  &                     & g \\ 
G331.020--0.143&      & 3.8    & a &                  &                                               & WR72-4 (3.5)        & n \\ 
G335.794+0.153 &      & 13.7   & a & VVV CL 073       & WR75-6 (19.7), WR75-25 (6.1), WR75-26 (6.1)   & WR75b (16.1), WR75-15 (12.4)  & h \\ 
               &      &        &   &                  &                                               & WR75-16 (4.2)        & n \\ 
G336.446--0.198&      & 10.4   & a &                  &                                               & WR75-9 (11.7), WR75-21 (10.3)       & n \\ 
               &      &        &   &                  &                                               & WR75-5 (14.3)        & d \\
G338.350+0.221 &      & 6.9    & a &                  &                                               & WR76-10 (4.7)        & n \\ 
G338.430+0.048 &      & 5.8    & a & Mercer 81        & WR76-2 (3.5), WR76-3 (3.5), WR76-4 (3.5)      &                      & o \\ 
               &      &        &   &                  & WR76-5 (3.5), WR76-6 (3.5), WR76-7 (3.5)      &                      & o \\
               &      &        &   &                  & WR76-8 (3.5), WR76-9 (3.5)                    &                      & o \\
G338.911+0.615 & 4.4  & 3.8    & a &                  &                                               & WR76 (1.9)           & h \\ 
\hline
\multicolumn{8}{l}{
\begin{minipage}{2\columnwidth}~\\
a: \citet{anderson14}
b: \citet{mauerhan09}
c: \citet{chene13}
d: \citet{shara09}
e: \citet{hadfield07}
f: \citet{kurtev07}
g: \citet{mauerhan11}
h: \citet{vdh01}
i: \citet{davies12a};  
j: \citet{chene15}
k: \citet{marston13}
l: \citet{roman11}
m: \citet{wachter10}
n: \citet{shara12}
o: \citet{davies12b};  
p: This work \\
$\ast$: Alternatively WR62-2 and WR62-1 may be associated with WISE H\,{\sc ii} region G317.236+0.516 
  \end{minipage}
}
\end{tabular}
\end{center}
\end{table*}

\section{Spatial distribution of Wolf-Rayet stars and other massive stars in the Milky Way}\label{sec:disc}

\subsection{Association of WR stars with star clusters?}

If the majority of WR progenitors are born in relatively massive star-forming regions, one might expect them to be in close proximity
to their natal star cluster. We have compared the spatial location of the new WR stars with young star clusters from \citet{dutra03},
\citet{mercer05} and \citet{borissova11}. None of the new WR stars are located within known clusters. At face value, this
might be considered to be surprising, but \citet{lundstrom84} in the most extensive study of the environment of Galactic
WR stars to date, found that only 10--30\% of WR stars are located in star clusters. 

Since then, the known WR population has quadrupled, so we provide revised statistics for the  298$^{\circ} \leq l \leq 340^{\circ}$
survey region as a whole, involving 190 + 16 = 206  Galactic Wolf-Rayet stars,
comprising 119 WN, 1 WN/C and 86 WC stars. \citet{lundstrom84} considered a WR star to be associated with a star cluster if its projected
distance was within two cluster radii. Here, we soften this requirement, extending the potential association to 4 cluster radii,
utilising published cluster centres and radii from \citet{dias02}, \citet{dutra03}, \citet{mercer05}, \citet{borissova11}. Overall, 55 WR stars (27\%) 
are associated with a total of 12 star clusters, as shown in Table~\ref{tab:clusters}, although it is 
notable that 44\% of all the WR cluster members in our survey range are located within a single open cluster, Westerlund~1 (C06).
Consequently 73\% of WR stars in the survey region are {\it not} associated with a star cluster.

If the majority of WR progenitors are born in relatively massive star forming regions, why are so few currently associated with clusters? 
The lower WR mass limit for single rotating stars at Solar metallicity is $\sim$22 $M_{\odot}$ star \citep{meynet03}. Empirically, such stars
  are observed in clusters with $\geq 500 M_{\odot}$  \citep{larson03, weidner10}.
Stochasticity in the sampling of the initial mass function will result in some massive stars originating in low mass 
(10$^{2} M_{\odot}$) 
clusters/star-forming regions, as demonstrated theoretically by \citet{parker07}. Therefore, some WR stars could be
associated with low mass clusters, with other members of the
star forming region inconspicuous. Indeed, $\gamma$ Vel (WC8+O), the closest Galactic WR star, is located in a very  low mass star-forming region \citep{jeffries14}. Other members of such star-forming region would be very difficult to identify at the typical distance of Galactic WR stars.

There is evidence that some dense, young massive clusters rapidly achieve virial equilibrium 
\citep[e.g.][]{henault12}, such that they would retain the bulk of their stellar content over the representative WR ages of 5--10 Myr \citep{meynet03}. Indeed, the bulk of the 
WR stars associated with Westerlund 1 lie within the 1.2 arcmin radius -- corresponding to 1.4 pc at a distance of 3.8 kpc -- such that our 4 cluster radius limit is equivalent 
to $\sim$5.5 pc. Of course not all open clusters are dense or bound. For example, Hogg 15 is a sparsely populated cluster with a much larger radius of 3 pc, at its distance of 3 
kpc, so will be in the process of dispersing. Consequently the general lack of an association with star clusters is not wholly unsurprising. Indeed, inspection of the  Galactic 
O Star  Catalogue \citep[GOSC v3.0][]{gosc13} reveals 50 (optically visible) O stars in our survey  region. Of these, only 18 (36\%) are
associated with a star cluster, not significantly larger than WR stars since the bulk of these are late-type O stars with comparable ages to many WR stars.

Of course, not all massive stars originate from star clusters. \citet{bressert10} demonstrated that only a small fraction of star formation in the Solar Neighbourhood
originates from dense environments. It is probable that the majority of OB stars originate from intermediate density regions, i.e. OB asociations and/or extended star-forming
regions \citep[see][]{parker17}. Indeed, \citet{wright16} have unambiguously demonstrated that the distributed massive stars in Cyg OB2 did not originate from a dense star cluster.

\citet{lundstrom84} found that $\geq$50\% of optically visible WR stars were located in OB associations. It is not possible to calculate the fraction of IR-selected WR
stars that lie within OB associations since the latter are limited to nearby optical surveys. Instead, it is necessary to compare the location of Wolf-Rayet stars with
infrared catalogues of star-forming regions. We have compared the locations of all 206 WR stars in our survey region with confirmed H\,{\sc ii} regions from the 
all-sky WISE catalogue of Galactic H\,{\sc ii} regions from \citet{anderson14}. In  total, 53 stars are located within $\approx$1.5 $R_{\rm HII}$ of the H\,{\sc ii} region, 
representing 26\% of the  total WR population, as shown in Table~\ref{tab:SFRegions}.  Of course,  a subset of these WR stars will be foreground sources, so the quoted 
statistics represent strict upper limits.  

The majority of these WR stars are associated with complexes at G298 (Dragonfish nebula), G305 and G338. By way of example, as many as twelve WR stars are associated with 
G298 `Dragonfish'  complex \citep{rahman11}. The VVV CL011 and Mercer 30 clusters  are coincident with this region, although radio distances of $\sim$10.5 kpc significantly 
exceed spectrophotometric cluster distances. Similar issues affect the association of Mercer 81 with G338.430+0.048, and VVV CL 041 with G317.030+0.028. 
Finally, stellar winds and/or supernovae associated with Westerlund~1  have sculpted an IR cavity, such that  there 
is no longer an associated H\,{\sc ii} region. 

Overall, it is apparent that WR stars are rarely located within known H\,{\sc ii} regions, albeit with some notable exceptions 
which includes
the Carina Nebula in the Milky Way. Typical radii of star-forming regions are 9 arcmin, so for representative radio-derived distances of 7.5 kpc to star forming 
regions, we include stars within a projected distance of 30 parsec from the centre of the H\,{\sc  ii} region.   Again, it is 
possible that stars have migrated away from  where they formed. Typical velocity dispersions of stars in such regions are 
several km/s, corresponding  to several pc/Myr, such that the WR progenitor could have travelled several 
ten's of parsec. It is important to stress that the parent star-forming region of a WR star would not necessarily display significant radio free-free emission after 5--10 Myr 
\citep{meynet03} since most O stars will have died if there had been a single burst of star formation.  Indeed, only 23 (46\%) of the 50 GOSC optically  visible O stars in our 
survey region lie in an OB association (Cen OB1, Nor OB1, Ara OB1).

Regardless of whether WR progenitors form in clusters or lower density environments, there are other explanations for their relative isolation.
To have travelled more than a few 10s of parsec from their birth site, the WR progenitor may have been ejected via dynamical effects or following a 
supernova kick  from a close companion. The former, involving 3-body dynamical interactions, is favoured in dense stellar environments, in which
the fraction of massive stars ejected is inversely proportional to the stellar mass of the cluster \citep{fujii11}. 
Therefore, the population of WR stars dynamically ejected 
in this way will be dominated by those from low to  intermediate mass clusters, explaining why the 10$^{5} M_{\odot}$ cluster Westerlund~1 has successfully retained the 
majority of its WR population. 
Historically, a supernova origin for runaway WR stars has been considered to be a major factor affecting  their spatial 
distribution \citep{eldridge11}. This
requires a close binary companion, and assumes that all massive stars whose initial masses exceed $\sim20 M_{\odot}$ undergo a core-collapse SN. Of course, such
stars  may collapse directly to a black hole, or form a black hole via fallback, so their companion would not necessarily receive a significant 
supernova kick \cite[e.g.][]{oConnor11}. Therefore, only a small fraction of isolated WR stars likely originate in this way.

Overall, the most promising scenario for the low observed frequency of WR stars associated with star clusters is via dynamical ejection, but this alone 
doesn't explain the very high fraction of WR stars in the field. Instead, it is likely that a majority of Galactic WR progenitors do not form 
within dense, high mass star clusters. The observed high fraction of WR stars located in the Galactic field can therefore be attributed to a combination of 
dynamical ejection from star clusters, and their origins in modest star-forming regions which subsequent dissolve into the field. The latter will not necessarily be 
recognisable as an OB association during the WR  stage since the majority of O stars  will have ended their lives after 5--10 Myr, with similar arguments applying to isolated 
H\,{\sc ii} regions. Some distant WR stars will not be recognised as being associated with a star forming region if other stars in the region possess low masses, as is the case for $\gamma$ Vel \citep{jeffries14}.

\subsection{Relative isolation of WR stars and Luminous Blue Variables}

The general lack of association between Wolf-Rayet stars and O stars -- except for the 
minority located in young, dense star clusters (e.g. NGC~3603, Westerlund 1) -- is relevant to the ongoing 
debate about the nature of Luminous Blue Variables \citep{humphreys16, smith16}. Historically LBVs were
considered to be massive stars in transition towards the WR stage. In part, this association arose from
the spectroscopic evolution of LBVs to the WN phase \citep[e.g. AG Car][]{smith94} and LBV outbursts from WN stars
\citep[e.g. R127, HDE\,269582,][]{walborn17}. \citet{smith15} have challenged this view, finding that LBVs in the Milky Way and LMC are more isolated (from O stars) 
than Wolf-Rayet stars. They have argued that they possess lower masses, with their high luminosities arising from being the mass gainers 
(former secondaries) within close binary systems. 

Their conclusions were largely based upon the spatial  distribution of O stars, WR stars and LBVs in the LMC, owing to visual studies of Galactic massive stars 
being severely restricted by dust extinction. Still, the reliance on SIMBAD for catalogues of O stars hinders  their conclusions owing to severe incompleteness for both 
galaxies, while \citet{humphreys16} argued for a  mass separation between high luminosity (classical) and low luminosity LBVs \citep[though see][]{smith16}.

\citet{kniazev15, kniazev16} provide the current census of 17 confirmed LBVs in the Milky Way, which is
restricted to confirmed spectroscopic variables. Their criteria therefore exclude candidate LBVs possessing ejecta nebulae, which 
\citet{bohannan97} had previously argued should be an additional discriminator. Only three spectroscopically variable LBVs
lie within our Galactic survey region --  WS1 \citep{kniazev15}, Wray 16-137 \citep{gvaramadze15} and Wd1-W243 \citep{clark05} --
so we need to consider the global Milky Way population of LBVs and WR stars. 

Of the known WR content in  the Milky Way, 27\% are 
members of star clusters \citep[][]{crowther15}. Meanwhile,
4 of the 17 spectroscopically variable LBVs are located in star clusters -- W243 in Westerlund 1,  $\eta$ Car in
Trumpler 16, GCIRS 34W in the Galactic Centre cluster and qF 362 in the Quintuplet cluster \citep{geballe00}) -- comprising 24\% of 
the total, so their global statistics are comparable. Indeed,
a number of widely accepted LBVs known to be associated with star clusters are omitted from the compilation of 
\citet{kniazev15}. 
These include the Pistol star (qF 134), another member of the Quintuplet cluster \citep{figer98}, and the LBV in the 1806-20 
cluster 
\citep{eikenberry04, figer04}.


5 of the 17 spectroscopically variable LBVs are potentially associated with star-forming regions \citep{anderson14}
from a similar exercise to that discussed above for WR stars, namely $\eta$ Car, Wray 16-137, HD~168607, AFGL 2298 and G24.73+0.69, 
namely 29\% of the total. Consequently, the overwhelming majority of WR stars and LBVs are located in the Galactic field, away from 
star clusters or star-forming regions. Overall, there is no significant difference in the spatial distribution of WR and LBVs in our Galaxy, with a quarter of both populations associated with  young massive star clusters.

\citet{smith15} proposed that LBVs generally arise from significantly lower mass progenitors than WR stars, challenging the hitherto 
scenario that LBVs are transition objects between the O and WR phases. However, those star clusters which host LBVs are relatively 
young \citep[4--6 Myr][]{clark05, bibby08, liermann12, schneider14}, putting aside $\eta$ Car\footnote{$\eta$ Car is unusual since 
it is associated with an even younger star cluster, Trumpler~16, which also hosts the massive main-sequence O2.5\,If/WN6 star WR25 
\citep{massey93, crowther11}}. Not only are the statistics of WR and LBV association with star clusters comparable, but crucially 4 
young Milky Way clusters hosting LBVs -- Westerlund~1, CL 1806-20, the Galactic Centre and the Quintuplet -- also contain 
(classical) WR stars.

 \citet{smith15} argued that LBVs arise from mass gainers in close
binary systems \citep{langer12, demink14}. Mass gainers will be rejuvenated, yielding an apparently younger star than the
rest of the population \citep{schneider14}. The presence of  spectroscopically variable LBVs (Wd1-W243, GCIRS 34W, qF 362) plus 
leading LBV candidates (qF 134, LBV 1806-20) in young clusters with progenitor masses in the range 30--40  $M_{\odot}$ and coexisting 
with O stars and WR stars, should permit their scenario to be tested. Of course, LBVs in such systems might be the mass-gaining
  secondaries whose primaries have already undergone core-collapse. However, LBV 1806-20 is an SB2 system, whose companion is not
  a WR star since its exhibits He\,{\sc i}
  absorption lines. This appears to rule out the \citet{smith15} scenario in this instance.

  If LBVs are rejuvenated secondaries in close binaries spanning a wide range of masses, they should
  also be present in older, massive star clusters
  such as those hosting large red supergiant (RSG) populations. However, \citet{smith15} report that LBVs and RSGs do not share a common parent population, and LBVs are not known within RSG-rich clusters at the base of the Scutum-Crux arm \citep[RSGC 1--3,][]{figer06, davies07, clark09}.
  The presence of $\sim$50 RSG in these clusters and absence
  of LBVs argues either argues against the \citet{smith15} scenario, or requires a short ($\sim 2 \times 10^{4}$ yr) lifetime for such LBV,
  in view of the $\sim$1 Myr RSG lifetime. The only potential LBV within these RSG rich clusters identified to date is
  IRAS 18367---0556 in RSGC 2, which possesses a
  mid-IR ejecta nebula \citep{davies07}. Overall, it is apparent that LBVs span a wide range of physical properties \citep{smith04}, though
the same is true for WR stars, some of which are located in significantly older star clusters (recall Table~9 from C06), 
albeit solely Westerlund  1 hosts RSG and WR stars.

%
%

\section{Conclusions}\label{sec:conc}

We have undertaken a near-IR spectroscopic search for Wolf-Rayet stars 
along the Scutum-Crux spiral arm, based upon 2MASS and GLIMPSE photometric
criteria \citep{mauerhan11, faherty14}. Observations of 127 candidate stars
($K_{\rm S} \sim 10-13$ mag) resulted in the confirmation of 17 WR stars 
(14 WN, 3 WC), representing a success rate of $\sim$13\%, comparable to previous IR-selected studies based on similar criteria
\citep{hadfield07, mauerhan11}.
More sophisticated techniques are clearly required for optimising future spectroscopic searches.
 As we have found, large numbers of other stellar types (young stellar objects, Be stars) lie in the same location as WR
  stars within individual colour-colour diagrams, but may not do so in a multi-dimensional parameter space.
  Therefore, future progress might entail a Bayesian approach to optimising searches for candidate WR stars.

We have extended our earlier near-IR classification scheme (C06) to cover all YJHK bands
and all subtypes, including WN/C and WO subtypes. This has been tested on several recently discovered WR 
stars for which optical classifications have previously been obtained. In general, the near-IR criteria
are successful in obtaining reliable subtypes, including the presence of atmospheric
hydrogen in WN stars, and identifying transition WN/C stars. However, inferences are weaker if limited 
wavebands are available, the spectral resolution is modest, and/or the signal-to-noise obtained is low.

The majority of newly discovered WR stars are weak-lined WN5--7 stars, with 
two broad-lined WN4--6 stars and three WC6--8 stars. Therefore, despite the low success rate, our goal of 
extending the  spectroscopic confirmation of WR stars to the far side of the Milky Way has 
been successful. Based on the absolute magnitude calibration of C15, inferred distances ($\sim$10 
kpc) extend beyond previous spectroscopic surveys, with $A_{\rm K_{s}}{\sim}1.2$ mag. 
Spectroscopic searches beyond the  Galactic Centre ($A_{\rm K_{s}}{\sim}3$ mag) will be 
significantly more challenging. 

Only a quarter of WR stars in the selected Galactic longitude range are associated with star clusters and/or
H\,{\sc ii} regions. We suggest that this arises from a combination of dynamical ejection from (modest
stellar mass) clusters and formation within lower density environments (OB associations/extended H\,{\sc ii} regions). 
We also revisit  the recent controversy about the association between LBVs and WR stars, or lack thereof.
Considering the whole Milky Way, 27\% of WR stars are hosted by clusters,  comparable to that of spectroscopically variable LBVs (4 
from 17 stars). More significantly, several young clusters with main sequence turn-off masses close to 30--40\,$M_{\odot}$ host 
classical WR stars {\it  and} LBVs, permitting \citet{smith15}'s suggestion that some LBVs are rejuvenated
mass gainers of close binaries to be tested. Specifically, the non-WR companion to LBV 1806-20 and absence of LBVs in
  RSG-rich star clusters argue against this scenario.

Returning to the main focus of this study, until more robust distances to WR stars - and in turn absolute 
magnitudes - can be established by {\it Gaia} there are legitimate concerns  about the completeness of 
surveys for different subtypes, especially the  challenge of identifying faint, weak emission line WN 
stars with respect to WC stars \citep{Massey15b}. Near-IR narrow-band photometric searches suffer 
from dilution of emission lines by companion stars and hot dust emission 
from WC binaries, while our broad-band near to mid-IR photometric approach 
is limited by the low spatial resolution of Spitzer. \citet{Massey15a} have undertaken a deep optical 
narrow-band  survey of the LMC, revealing a population of faint, weak-lined WN3 stars (which they 
characterize as WN3/O3 stars). Stars with these characteristics -- which would usually be classified 
as WN3ha according to the \citet{smith96} scheme -- comprise a negligible fraction of Galactic WR stars 
(e.g. WR3), a minority of the  moderately metal-poor  LMC WR population, and a majority of the more 
significantly metal deficient SMC WR population  \citep{hainich15}. 


It is anticipated that ongoing infrared searches using a combination of continuum 
methods and emission line characteristics will significantly improve the completeness of 
WR surveys in a sizeable volume of the Galaxy within the near future.

\section*{Acknowledgments}

We wish to thank Nicole Homeier, Bill Vacca and Frank Tramper for providing us with near-IR spectroscopic datasets, from NTT/SOFI, 
IRTF/SpeX and VLT/XSHOOTER respectively, for several WR stars. We appreciate useful discussions with Simon Goodwin and Richard 
Parker, and useful comments from the referee which helped improve the focus of the paper.
Financial support for CKR was provided by the UK Science and  Technology Facilities Council. PAC would like to thank ESO for arranging emergency 
dental care in La Serena immediately after the 2015 observing run.


\bibliographystyle{mnras}

\section*{Supplementary Information}

Additional information may be found in the online version of this article:

\noindent Appendix A: Near-IR spectral classification of WR stars

\noindent Appendix B: Spectroscopic observations of IR selected candidates

\noindent Please note: Oxford University Press is not responsible for the content or functionality of any supporting materials 
supplied by the authors. Any queries (other than missing material) should be directed to the corresponding author for this 
article.



\clearpage

\appendix


\section[]{Near-IR Spectral classification of WR stars}\label{classi}

\begin{figure*}
\begin{center}
\includegraphics[width=0.85\textwidth]{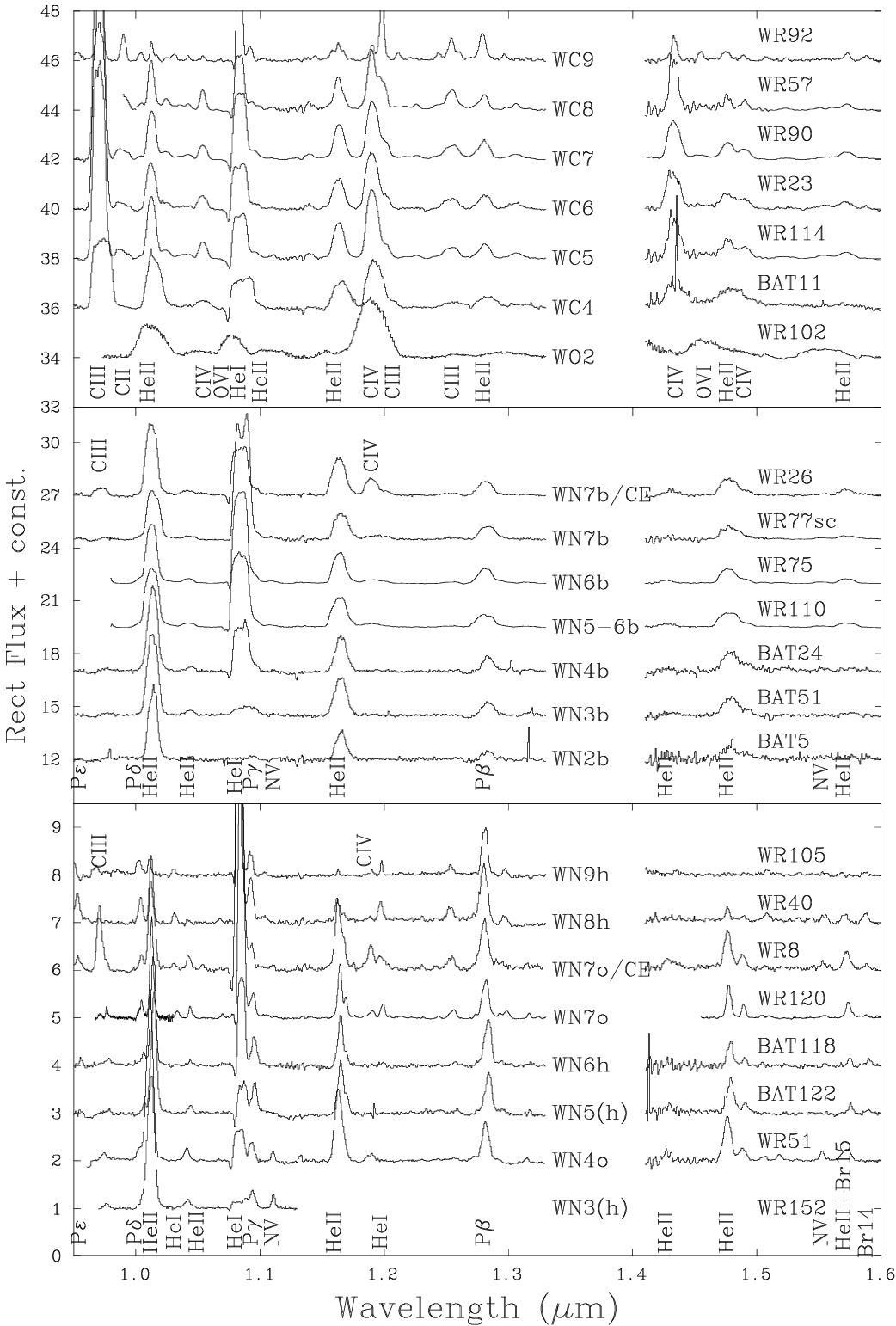}
\caption{
YJ-band spectroscopy of template Galactic and LMC Wolf-Rayet stars, from the present NTT/SOFI dataset, supplemented by 
archival spectroscopy from various telescopes/instruments. C\,{\sc iv}
1.19$\mu$m and C\,{\sc iii} 0.97$\mu$m are visible in the transition WN7/CE stars WR8 and WR26.}
\label{fig:old_wr_blue}
\end{center}
\end{figure*}

\begin{figure*}
\begin{center}
\includegraphics[width=0.85\textwidth]{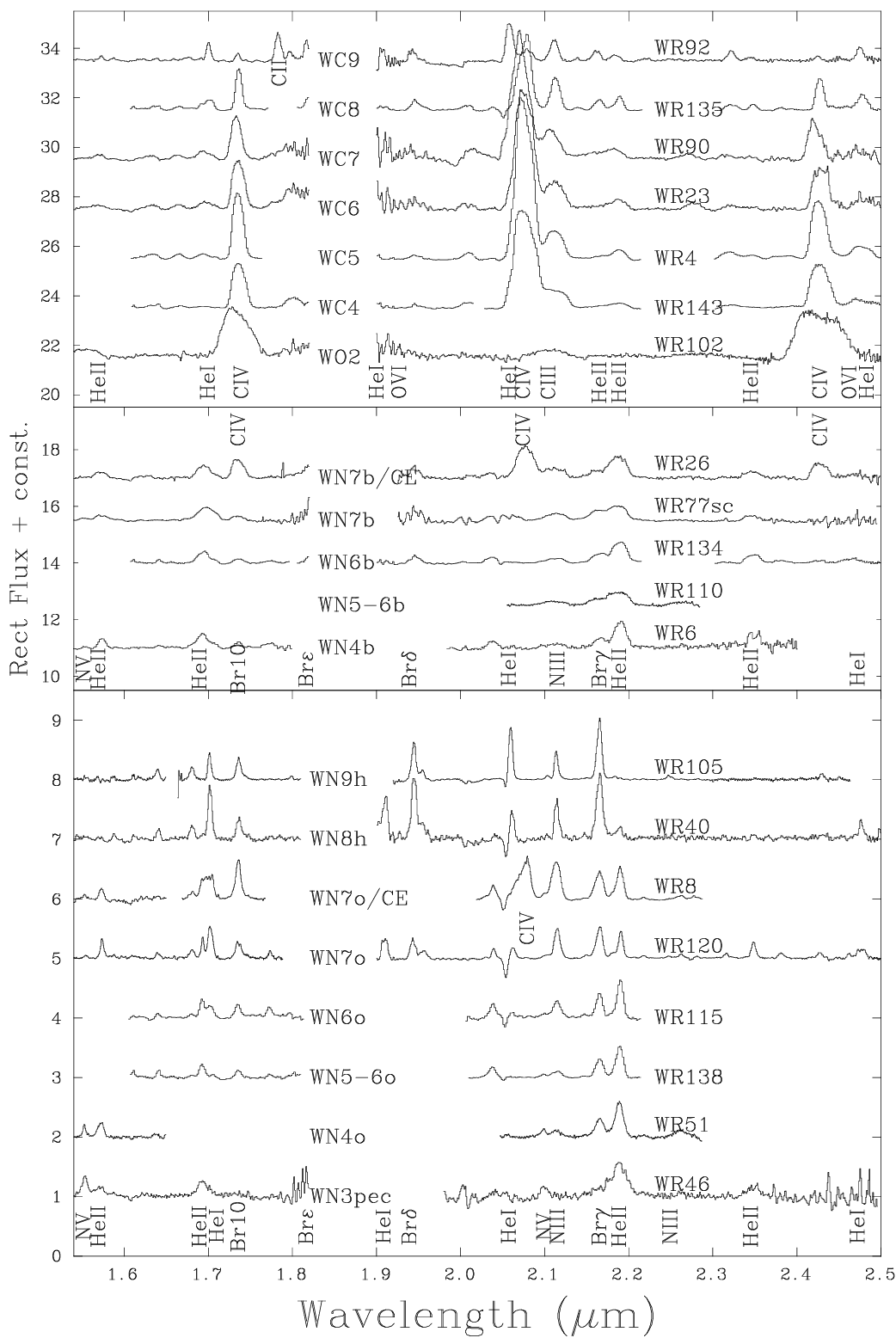}
\caption{
HK-band spectroscopy of template Galactic and LMC Wolf-Rayet stars, from the present NTT/SOFI dataset, supplemented by 
archival spectroscopy from various telescopes/instruments. C\,{\sc iv}
1.74$\mu$m, 2.07$\mu$m and 2.43$\mu$m are prominent in the transition WN7/CE stars WR8 and WR26.}
\label{fig:old_wr_red}
\end{center}
\end{figure*}

C06 utilised He\,{\sc ii}\,1.012\micron/He\,{\sc i}\,1.083\micron\ and C\,{\sc iv}\,1.19\micron/C\,{\sc iii}\,1.20\micron\ to classify
WN and WC stars from YJ-band spectroscopy, respectively, plus He\,{\sc ii}\,2.189\micron/Br$\gamma$ and 
C\,{\sc iv}\,2.08\micron/C\,{\sc iii} 2.11\micron\ from K-band spectroscopy. 
Subtypes for weak/narrow WN4--7 stars with a He\,{\sc ii} 1.01\micron\ 
FWHM $\leq$ 65\AA\
were  distinguished by the ratio of N\,{\sc v} 2.10\micron/(He\,{\sc i} + N\,{\sc iii} 2.11\micron), while 
WC8--9 stars were also classified from their
C\,{\sc iii} 0.97\micron/C\,{\sc ii} 0.99\micron\ ratio. However, separating broad-lined WN4--7 stars and WC5--7 stars from
near-IR spectroscopy proved to be extremely challenging. In addition, the use of He\,{\sc ii}\,2.189\micron/Br$\gamma$ as a classification
diagnostic adds a further complication owing to the sensitivity of this ratio on hydrogen content as well as ionization. 
Consequently, we revisit these issues in the following discussion.

\begin{figure}
\begin{center}
  \includegraphics[width=1\columnwidth]{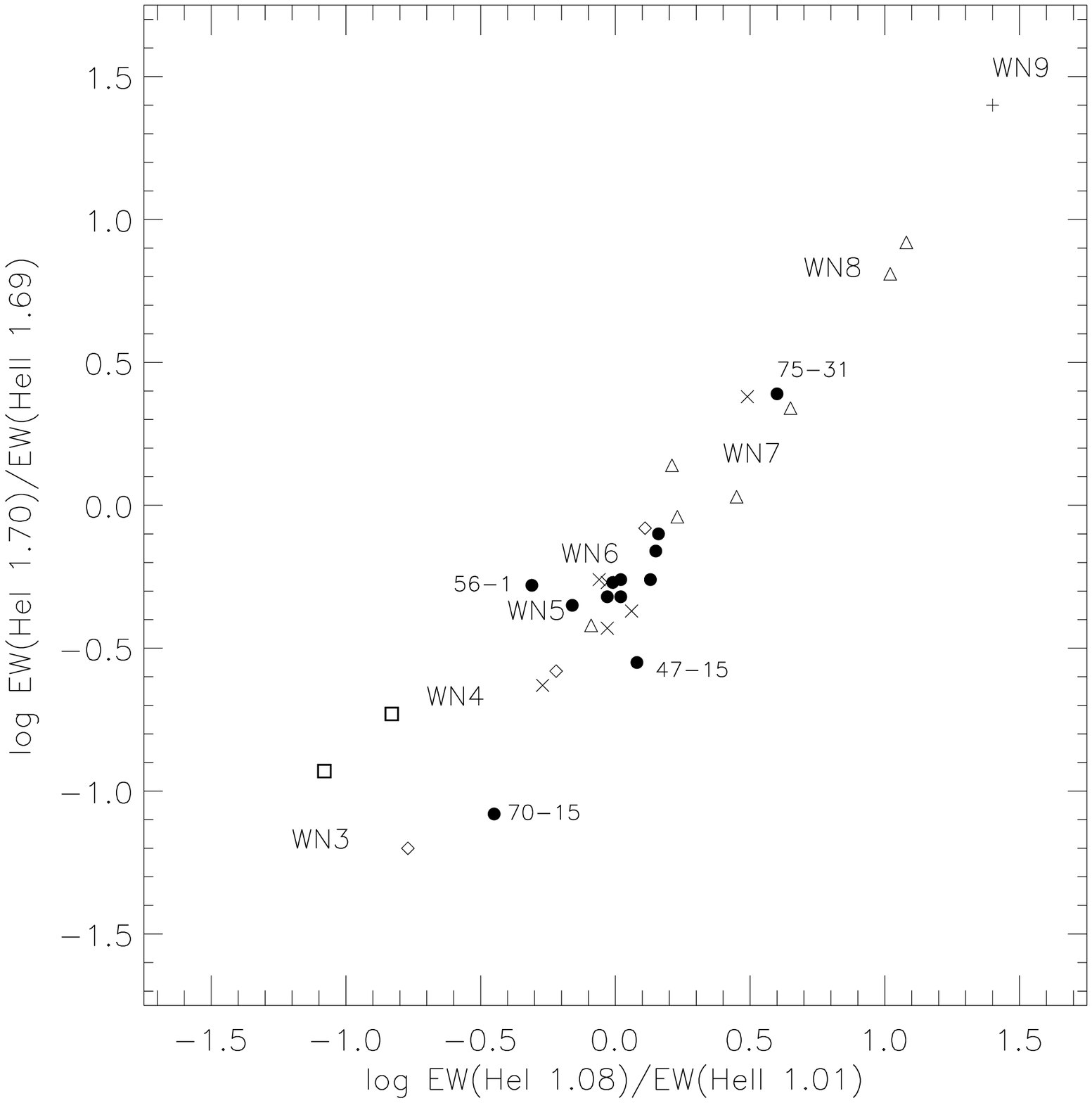}
\caption{Comparison between He\,{\sc i} 1.08$\mu$m/He\,{\sc ii} 1.01$\mu$m and He\,{\sc i} 
1.70$\mu$m/He\,{\sc ii} 1.69$\mu$m ratios for WN9 (plus symbol), WN7--8 (triangles), WN5--6 stars 
(diamonds) and  WN3--4 stars (squares), with newly identified WN stars indicated as filled circles, and 
strong-lined WN  stars shown as crosses (He\,{\sc i} 1.70 and He\,{\sc ii} 1.69 are severely blended).}
\label{fig:wn}  
\end{center}
\end{figure}

\begin{figure*}
\begin{center}
  \includegraphics[width=1.75\columnwidth]{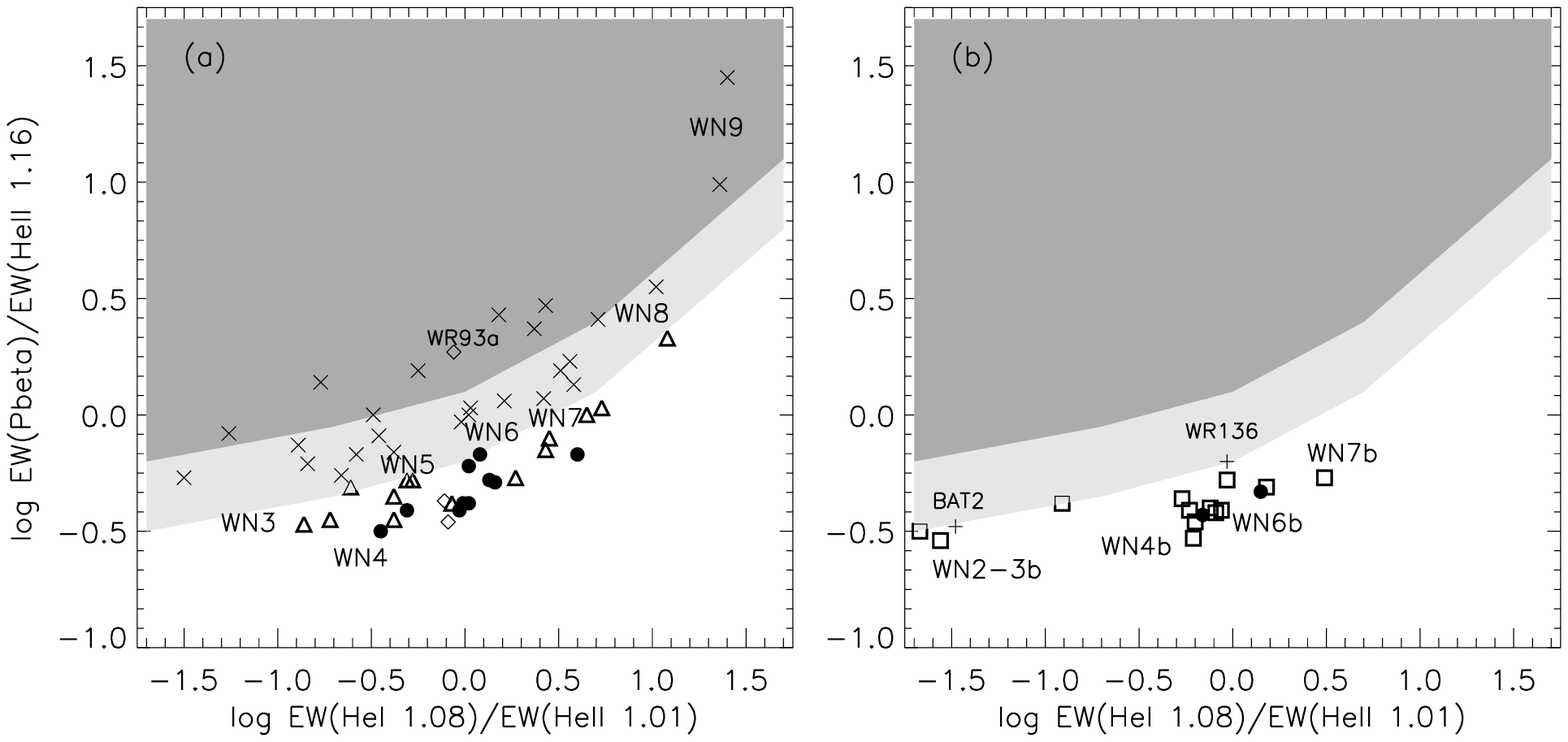}
  \includegraphics[width=1.75\columnwidth]{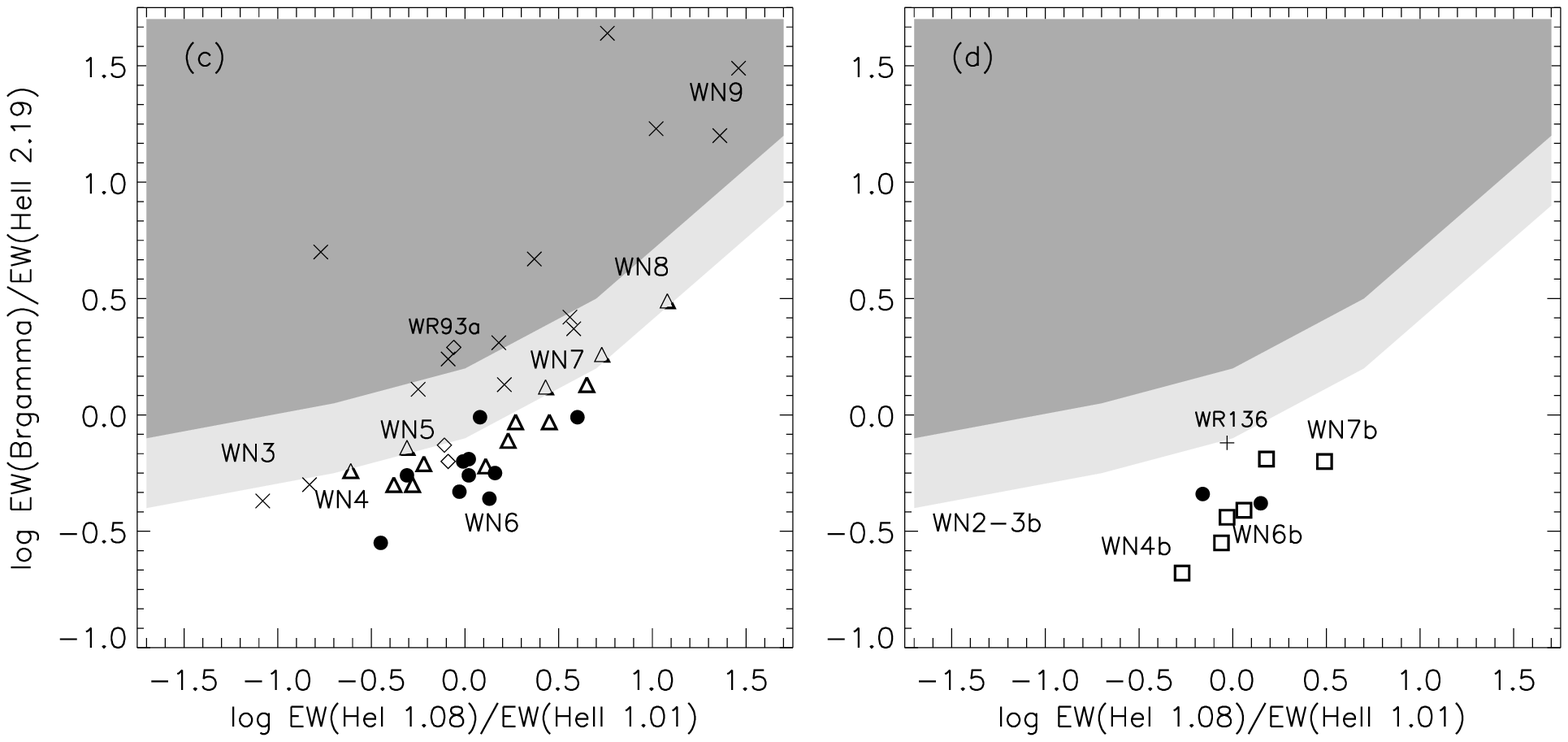}
  \includegraphics[width=1.75\columnwidth]{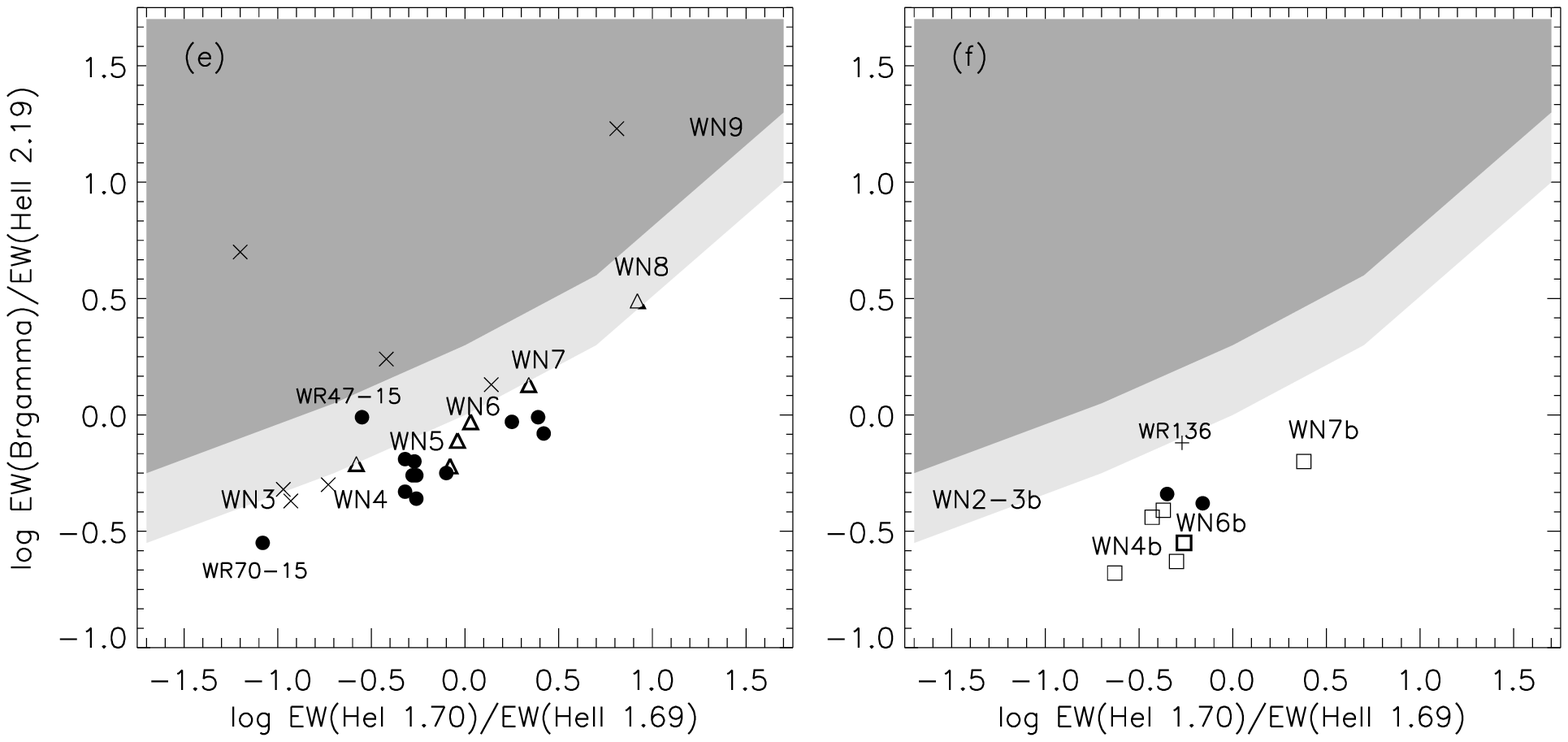}
  \caption{{\bf (a):} Comparison between He\,{\sc i} 1.08$\mu$m/He\,{\sc ii} 1.01$\mu$m and P$\beta$/He\,{\sc 
ii} 
1.16$\mu$m ratios for narrow-lined WN stars (hydrogen-free: triangles; hydrogen: crosses), with the pale 
(dark)  shaded regions indicating modest (high) hydrogen contents. Newly discovered WN stars are indicated 
by  filled circles while previously discovered WN stars lacking previous optical classifications (WR62a, 
68a, 93a) are indicated by diamonds; {\bf (b)} as (a) for broad-lined WN stars (hydrogen-free: squares; 
hydrogen:  pluses). {\bf (c)} Comparison betweeen He\,{\sc i} 1.08$\mu$m/He\,{\sc ii} 1.01$\mu$m 
and Br$\gamma$/He\,{\sc ii} 2.19$\mu$m ratios for narrow-lined WN stars, symbols as for (a); {\bf (d)} as (c) 
for broad-lined WN stars, symbols as in (b). {\bf (e)} Comparison between He\,{\sc i} 
1.70$\mu$m/He\,{\sc ii} 1.69$\mu$m and Br$\gamma$/He\,{\sc ii} 2.19$\mu$m ratios for narrow-lined WN 
stars, symbols as for (a); {\bf (f)} as (e) for broad-lined WN stars, symbols as for (b).}
\label{fig:hydrogen}
\end{center}
\end{figure*}

\begin{figure*}
\begin{center}
\includegraphics[width=1.75\columnwidth]{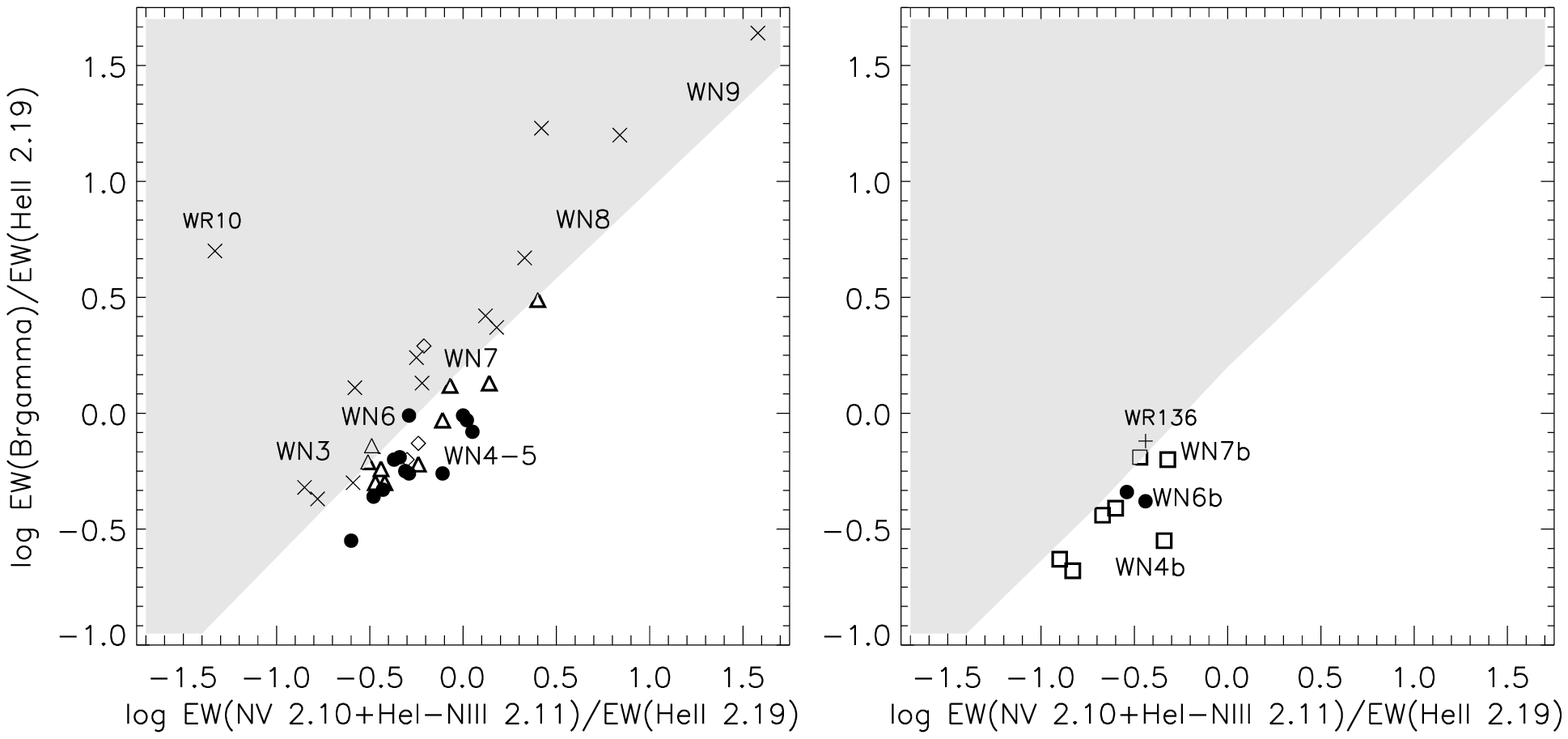}
\caption{{\bf Left panel} Comparison between (N\,{\sc v} 2.10$\mu$m + He\,{\sc i}+N\,{\sc iii} 
2.11$\mu$m)/He\,{\sc ii} 2.19$\mu$m and Br$\gamma$/He\,{\sc ii} 2.19$\mu$m ratios for narrow-lined WN 
stars,  symbols as in Fig.~\ref{fig:hydrogen}(a). The shaded region indicates the approximate domain 
of WN stars with hydrogen. WR10 is a WN5ha star with extremely weak 2.10-2.11$\mu$m emission.
{\bf Right panel} Comparison for broad-lined WN stars, symbols as in 
Fig.~\ref{fig:hydrogen}(b).}
\label{fig:hydrogen_k}  
\end{center}
\end{figure*}

\subsection{WN stars}

The primary classification diagnostics of WN stars involve helium (He\,{\sc ii} 5411/He\,{\sc i} 5876) and nitrogen (N\,{\sc iii} 4640, N\,{\sc 
iv} 4058, N\,{\sc v} 4610) lines, with strong, broad-lined WN stars (WNb or WN-s) further distinguished from weak 
narrow-lined stars (WN, WN-w). In common with C06, we identify strong/broad WN stars with $W_{\lambda}$(He\,{\sc ii} 5411) 
$\geq$ 
40\AA, and FWHM(He\,{\sc ii} 4686)$\geq$30\AA\ \citep{smith96} using the FWHM of prominent 
He\,{\sc 
ii} lines, 1.012$\mu$m and/or 2.189$\mu$m, namely $\geq$1900 km\,s$^{-1}$ for WNb stars, plus an equivalent width threshold of 
$W_{\lambda} \sim$ 250\AA\ for He\,{\sc ii} 1.0124$\mu$m, and/or $W_{\lambda}\sim$ 75\AA\ for He\,{\sc ii} 2.1885$\mu$m.

Central to the optical classification scheme are the ratios of N\,{\sc iii} 4636-41, N\,{\sc iv} 4068 and N\,{\sc v} 4603-20 lines. 
At near-IR wavelengths, nitrogen lines are scarse, weak, and often blended with helium lines. The most prominent nitrogen features arise 
from N\,{\sc v} (1.111, 1.552, 2.100$\mu$m) in early WN subtypes, although these recombination lines require high S/N spectroscopy 
for broad-lined stars. In addition, N\,{\sc iii} 2.116 is prominent in late WN stars, albeit severely blended with He\,{\sc i} 
2.112--2.113$\mu$m and/or C\,{\sc iii} 2.115$\mu$m for WN/C stars. 

Consequently we are largely reliant upon the relative strengths 
of helium lines for IR classification. In the YJ-band, He\,{\sc ii}\,1.012\micron/He\,{\sc i}\,1.083\micron\ represents an excellent classification 
diagnosic for most WN subtypes. He\,{\sc  i} 1.0830 is blended with P$\gamma$ (+ He\,{\sc ii} 1.093\micron) in early subtypes and broad-lined stars,
while He\,{\sc ii} 1.0124 is blended with P$\delta$ (+ 
He\,{\sc ii} 1.0045\micron) in broad-lined WN stars. Divisions between subtypes are assigned suing optically classified WN stars, although 
broad-lined WN stars cannot be distinguished from this diagnostic.

He\,{\sc ii} 1.692$\mu$m/He\,{\sc i} 1.700$\mu$m provides a H-band diagnostic in 
narrow-lined WN stars, especially for late subtypes, although blending  severely hinders its role in broad-lined stars. 
Figure~\ref{fig:wn} compares these ratios in WN stars, indicating that they are well correlated for most narrow-lined WN
subtypes. 

In the K-band, the relative strengths of N\,{\sc iii} 2.116$\mu$m (+ He\,{\sc i} 2.112-2.113) to N\,{\sc v} 2.100$\mu$m (or He\,{\sc 
  ii} 2.189) provide key diagnostics in weak-lined WN stars, although this is hindered at low spectral resolution and signal-to-noise
since these emission lines are generally weak. Other relevant pairs of diagnostics, 
  He\,{\sc ii} 1.162/P$\beta$ and He\,{\sc ii} 2.189/Br$\gamma$ provide a combination of ionization and 
hydrogen content information, so if classifications can be established from other diagnostics, these are helpful in
distinguishing hydrogen-rich from hydrogen-deficient WN stars.

\subsubsection{Hydrogen content}

\citet{conti83} and \citet{smith96} employed the Pickering-Balmer series to characterize the hydrogen content of WN stars, the 
latter introducing 'o', '(h)' and 'h' subtypes for narrow-lined WN stars based on the strength of H$\beta$ + He\,{\sc ii} 4859 
(8-4) with respect to adjacent Pickering series members He\,{\sc ii} $\lambda$5411 (7--4) and He\,{\sc ii} $\lambda$4541 (9--4), with 
'o' omitted for broad-lined WN stars. To date, no attempts have been made to develop similar criteria from near-IR series (e.g. 
Paschen, Brackett), largely because there are additional complications with respect to optical diagostics.

Firstly, leading He\,{\sc ii} near-IR lines are blended with other He\,{\sc ii} series (incl. $n$-7) whose contributions need 
to be taken into account (e.g. 1.488$\mu$m 14--7 with 1.476$\mu$m 9--6). Secondly, hydrogenic He\,{\sc i} emission contributes 
significantly to 
the hydrogen lines (e.g. He\,{\sc i} 5--3 at P$\beta$). The former issue can be overcome fairly easily, since medium resolution 
spectroscopy of late WN stars WR120 and WR123 (dataset U6 from Table~\ref{tab:log}) reveals $W_{\lambda}$(He\,{\sc ii} 
1.167)/$W_{\lambda}$(1.163) 
$\sim$ 0.35 and $W_{\lambda}$(He\,{\sc ii} 1.488)/$W_{\lambda}$(1.476) $\sim$ 0.35. However, contributions from neutral helium lines 
to hydrogen features are more problematic. By way of example, one would expect the observed equivalent width of 1.28$\mu$m to be 
$\sim$10\AA\ for the hydrogen-free WN8 star WR123 \citep{crowther95d} if He\,{\sc ii} (10-6) is the dominant contributor, since $W_{\lambda}$(He\,{\sc 
ii} 11-6) = 9.4$\pm$0.2\AA\ and $W_{\lambda}$(He\,{\sc ii} 9-6) = 11.5$\pm$0.6\AA, whereas $W_{\lambda}$ (1.281$\mu$m He\,{\sc ii} 
+ P$\beta$ + He\,{\sc i} 5-3) = 70$\pm$1\AA. Similar issues affect P$\gamma$ and its adjacent He\,{\sc ii} lines, 1.042$\mu$m 
(13-6) and 1.167$\mu$m (11-6). Therefore, it is not straightforward to determine the hydrogen content in low ionization WN stars from the Paschen
series and their adjacent He\,{\sc ii} lines.

Instead, we compare the equivalent width ratios of He\,{\sc i} 1.08$\mu$m/He\,{\sc ii} 1.01$\mu$m (ionization) to P$\beta$/He\,{\sc 
  ii} 1.16$\mu$m (ionization and hydrogen content) for narrow-lined WN stars in Fig.~\ref{fig:hydrogen}(a), and broad-lined WN
stars in Fig.~\ref{fig:hydrogen}(b). WN stars with optical (h) or 'h' classifications are indicated by cross (narrow-lined) or plus (broad-lined)
symbols. It is apparent that WN stars with hydrogen exhibit excess P$\beta$ strengths at a given ionization, such that we have overlaid
a pale shaded region for those stars with hydrogen classifications. A darker shaded region, vertically offset by 0.3 dex corresponds to stars
with significant hydrogen. Stars close to the boundary include 
WR66 (WN8(h)) for which \citet{hamann06} have obtained  5\% atmospheric hydrogen. We have adopted the same criteria for broad-lined
WN stars, for which WR136, the only Galactic WNb star with some atmospheric hydrogen in our sample \citep{crowther96} sits close to the threshold,
as does BAT2 (WN2b(h)), for which \citet{hainich14} obtained a {\it negligible} hydrogen content.

At longer wavelengths, the Brackett series and adjacent He\,{\sc ii} lines 
are not easily amenable to the determination of hydrogen content. However, we also consider the
equivalent width ratios of He\,{\sc i} 1.08$\mu$m/He\,{\sc ii} 1.01$\mu$m (ionization) to
Br$\gamma$/He\,{\sc   ii} 2.19$\mu$m (ionization and hydrogen content) for narrow-lined WN stars in Fig.~\ref{fig:hydrogen}(c), and broad-lined WN
stars in Fig.~\ref{fig:hydrogen}(d). These closely resemble the above comparisons, and we have attempted to indicate the regions of stars with atmospheric
hydrogen, although high ionization, weak-lined WN3--4 stars with modest hydrogen content \citep[e.g. WR128, WR152][]{crowther95b} are less
straightforward to interpret since they lie close to the boundary. Amongst broad-lined WN stars, WR136 again lies close to the threshold obtained from narrow-line
stars.

\begin{figure*}
\begin{center}
  \includegraphics[width=2\columnwidth]{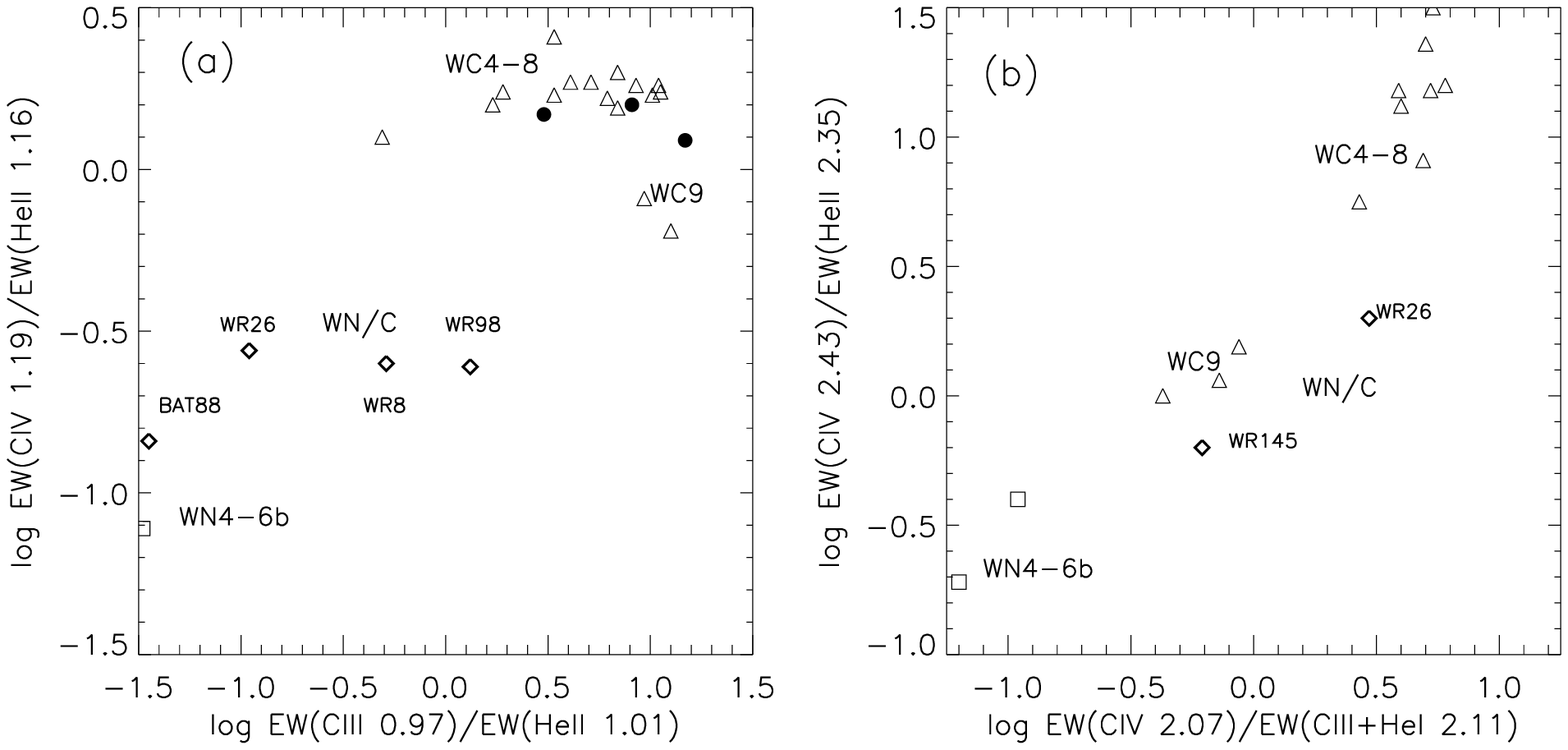}
\caption{{\bf (a)} Comparison between C\,{\sc iii} 0.97$\mu$m/He\,{\sc ii} 1.01$\mu$m and 
C\,{\sc iv} 1.19$\mu$m/He\,{\sc ii} 1.16$\mu$m ratios for WN/C stars 
(diamonds), WC (triangles) and broad-lined WN stars (squares) plus newly identified WC stars (filled circles). {\bf (b)} 
Comparison between 
C\,{\sc iv} 2.07$\mu$m/(C\,{\sc iii} + He\,{\sc i} + N\,{\sc iii} 2.11$\mu$m) ratios. Symbols are the same as panel (a).}
\label{fig:transition}  
\end{center}
\end{figure*}

We have also utilised the He\,{\sc i} 1.70$\mu$m/He\,{\sc ii} 1.69$\mu$m
ratio together with Br$\gamma$/2.189$\mu$m He\,{\sc ii}. In Fig.~\ref{fig:hydrogen}(e) and (f) we present a comparison between these equivalent width 
ratios for narrow-lined and broad-lined
WN stars, respectively. Fewer stars are available for this comparison owing to the lack of HK-band spectroscopy for LMC WN stars and
the difficulty in deblending He\,{\sc i} 1.70$\mu$m/He\,{\sc ii} 1.69$\mu$m, especially in broad-lined stars. Nevertheless,
WN(h) and WNh stars again exhibit an excess in Br$\gamma$ at a given ionization, once again with the notable exception of weak-lined WN3--4(h) stars.
The shaded region in Fig.~\ref{fig:hydrogen}(e) indicates regions for which hydrogen is present, which are duplicated for 
broad-lined stars in Fig.~\ref{fig:hydrogen}(f), with WR136 (WN6b(h)) again close to this threshold.

For WN stars with exceptionally high interstellar reddenings, establishing hydrogen contents of WN stars solely from K-band spectroscopt is extremely 
challenging. This is illustrated in Fig.~\ref{fig:hydrogen_k}, in which we compare the equivalent 
width ratios of (N\,{\sc v} 2.10 + He\,{\sc i} + N\,{\sc iii} 2.11)/He\,{\sc ii} 2.19 to Br$\gamma$/He\,{\sc ii} 2.19$\mu$m. These 
are ratios are well correlated, although a Br$\gamma$ excess is generally apparent in WN(h) and WNh stars (crosses and plus symbols), such that
we have indicated these in the shaded region, although no attempt has been made to distinguish between modest and high hydrogen contents from
these ratios.




\subsection{WN/C stars}

At visual wavelengths, intermediate WN/C stars usually closely resemble WN stars, albeit with prominent C\,{\sc iv} 5808\AA\ emission and usually weakly 
enhanced carbon recombination lines with respect to helium recombination lines (e.g. C\,{\sc iv} 5471/He\,{\sc ii} 5411). The most 
prominent carbon lines at near-IR wavelengths are C\,{\sc iv} 1.19, 1.74 2.075, 2.43$\mu$m, plus C\,{\sc iii} 0.971, 2.115$\mu$m.  
We have identified several pairs of carbon to helium lines amenable to the identification of WN/C stars at near-IR wavelengths, from 
spectroscopy of transition stars WR8, WR26, WR98, WR145 and BAT88. 

In Fig.~\ref{fig:transition}(a) we compare the ratios of  C\,{\sc iii} 0.97$\mu$m/He\,{\sc ii} 1.01$\mu$m to
C\,{\sc iv} 1.19$\mu$m/He\,{\sc ii} 1.16$\mu$m for WN (squares), WN/C (diamonds) and WC (triangles) stars. The intermediate
carbon-to-helium abundances of WN/C stars are reflected in their high C\,{\sc iv} 1.19$\mu$m/He\,{\sc ii} 1.16$\mu$m ratios, although
only late WN/C subtypes also exhibit elevated C\,{\sc iii} 0.97$\mu$m/He\,{\sc ii} 1.01$\mu$m ratios, since C\,{\sc iii} 0.97$\mu$m
is not observed in early-type WN/C stars (e.g. BAT88, WN4b/CE). Nevertheless,  C\,{\sc iv} 1.19/He\,{\sc ii} 1.16 is significantly larger
in BAT88 with respect to normal WN4b stars (e.g. WR6, WR18).


The only suitable pair of carbon-to-helium
recombination lines in the K-band is C\,{\sc iv} 2.43$\mu$m/He\,{\sc ii} 2.35$\mu$m, although C\,{\sc iv} 2.07$\mu$m is often also
pronounced in WN/C stars (recall WR8 and WR26 in Fig.~\ref{fig:old_wr_red}). We therefore consider the ratio of C\,{\sc iv} 
2.07$\mu$m/(C\,{\sc iii} + N\,{\sc iii} + He\,{\sc i} 2.11$\mu$m), although C\,{\sc iii} 2.115$\mu$m will also be enhanced at 
late subtypes. Fig.~\ref{fig:transition}(b) compares these ratios in WN/C stars with WN and WC
stars. WN/C stars are well separated from most normal subtypes with the notable exceptions of WC9 stars which have unusually weak
C\,{\sc iv} emission lines (and strong C\,{\sc iii} 2.11$\mu$m emission), but are otherwise
easily distinguished from transition WN/C stars.


Finally, C\,{\sc iv} 1.74$\mu$m/(He\,{\sc ii} 1.69 + He\,{\sc i} 1.70) is also a potential diagnostic, although this is 
complicated by the presence of Br10 + He\,{\sc ii} 1.735 in late WN stars such that it is most effective in hydrogen-free early 
subtypes. By way of example, C\,{\sc iv} 1.74$\mu$m is enhanced in WR8 (WN7o/CE) with respect to WR120 (WN7o) in
Fig.~\ref{fig:old_wr_red}.

\begin{figure}
\begin{center}
  \includegraphics[width=0.8\columnwidth]{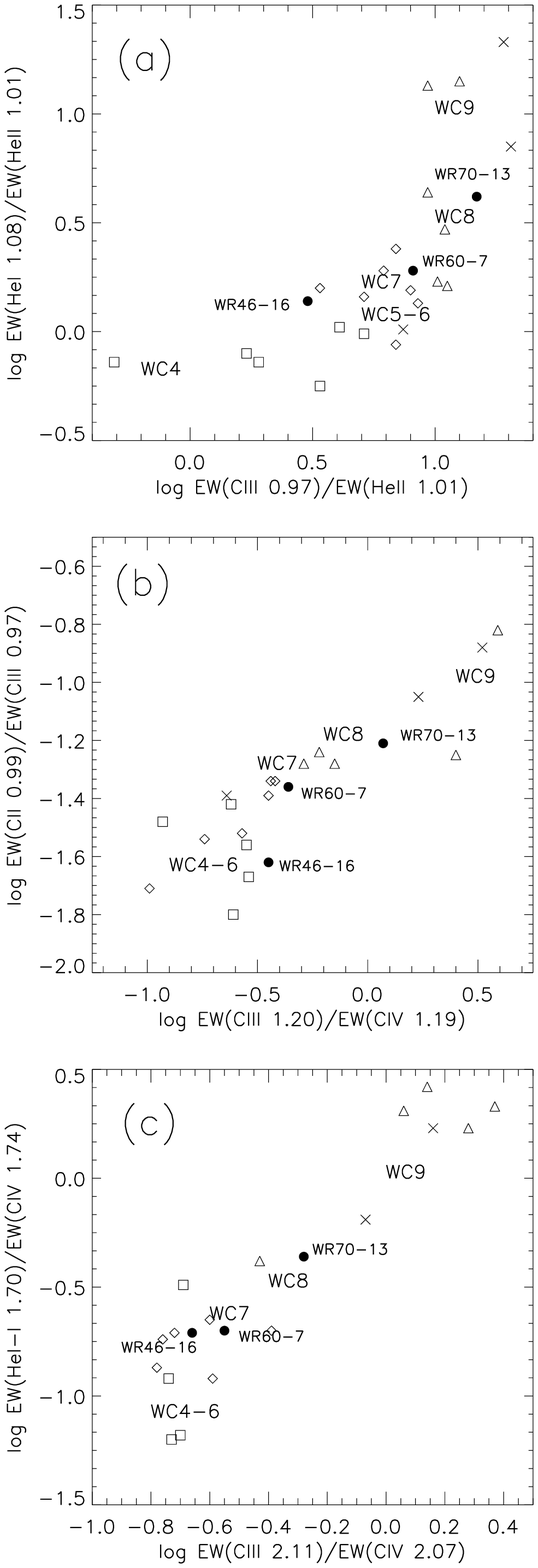}

\caption{{\bf (a)} Comparison between C\,{\sc iii} 1.20$\mu$m/C\,{\sc iv} 1.19$\mu$m 
and C\,{\sc ii} 0.99$\mu$m/C\,{\sc iii} 0.97$\mu$m ratios for WC8--9 stars (triangles), 
WC6--7 stars (diamonds), WC4--5 stars (squares) plus newly identified WC stars (filled circles) and newly 
classified WC stars WR75aa, WR75c and WR107a (crosses); {\bf (b)} Comparison between 
C\,{\sc iii} 0.97$\mu$m/He\,{\sc ii} 1.01$\mu$m and He\,{\sc i} 1.08$\mu$m/He\,{\sc 
ii} 1.01$\mu$m ratios for WC stars, symbols as above; {\bf (c)} comparison between 
C\,{\sc  iii} 2.11$\mu$m/C\,{\sc iv} 2.07$\mu$m and He\,{\sc i-ii} 1.70$\mu$m/C\,{\sc iv} 
1.74$\mu$m ratios for WC stars, symbols as above.}

\label{fig:carbon}  
\end{center}
\end{figure}

\subsection{WC stars}

For WC stars, the primary optical diagnostics involve the ratio of C\,{\sc iv} 5808\AA/C\,{\sc iii} 5696\AA, with other secondary 
criteria - O\,{\sc iii-v} 5592\AA\ for early WC stars according to \citet{smith90} and C\,{\sc ii} 4267\AA\ for late WC stars according to 
\citet{crowther98a}. In sharp contrast with WN stars, there are many carbon lines observed in the near-IR spectra of WC stars 
\citep{eenens91}, the strongest of which include C\,{\sc ii} 0.990, 1.785$\mu$m, C\,{\sc iii} 0.971, 1.199, 2.115$\mu$m, 
C\,{\sc iv} 1.190, 
1.430, 1.736, 2.075 and 2.427$\mu$m. Prominent helium lines include He\,{\sc i} 1.08, 1.70, and He\,{\sc ii} 1.01, 1.16$\mu$m, although C\,{\sc iv} contributes
significantly to lines which otherwise would be assumed to correspond to solely ionized helium, i.e. C\,{\sc iv} (10--8) to He\,{\sc ii} (5--4) at 1.01$\mu$m.

In the Y-band, we consider a range of diagnostic ratios to discriminate between WC subtypes. Fig.~\ref{fig:carbon}(a) compares
C\,{\sc iii} 1.20$\mu$m/C\,{\sc iv} 1.19$\mu$m  to C\,{\sc ii} 0.99$\mu$m/C\,{\sc iii} 0.97$\mu$m for late (WC8--9, triangles), mid (WC6--7, diamonds) and early (WC4--5, squares)
subtypes. The former cleanly separates WC8 and WC9 stars from earlier subtypes, but fails to discriminate between WC4--7 stars and is severely
hindered by severe blending in broad lined systems which are common at early subtypes. Similarly, the latter separates WC9 and WC7--8 from earlier subtypes, but C\,{\sc ii}
0.99$\mu$m is extremely weak at early subtypes (absent at WC4) and lies in the electron scattering wing of C\,{\sc iii}  
0.97$\mu$m for broad-lined stars. 

Helium lines are not generally considered when visually classifying WC stars, since the strongest features are severely blended with carbon lines,
e.g. He\,{\sc ii} 4686\AA\ (He\,{\sc i} 5876\AA) lies in the wing of C\,{\sc iii} 4650\AA\ (C\,{\sc iv} 5808\AA). However, in the Y-band, He\,{\sc ii} 1.01$\mu$m is relatively isolated 
-- albeit with a C\,{\sc iv} contribution - and He\,{\sc i} 1.08$\mu$m is prominent throughout the WC sequence. 
Therefore, Fig.~\ref{fig:carbon}(b) compares  C\,{\sc iii} 0.97$\mu$m/He\,{\sc ii} 1.01$\mu$m to
He\,{\sc i} 1.08$\mu$m/He\,{\sc ii} 1.01$\mu$m for late, mid and early subtypes. The former separates WC4 stars from later subtypes, owing to a ratio of $<$1, and also
discriminates between WC6--7 and WC8--9 stars at a ratio of $\sim$10. The latter ratio maps reasonably well onto the WC sequence, especially for WC9 ($>$10) and WC4 ($<$1) subtypes.


Fig.~\ref{fig:carbon}(c) compares the ratio of the strongest carbon
lines in the K-band, C\,{\sc iii} 2.11$\mu$m/C\,{\sc iv} 2.07$\mu$m, with the ratio of the (He\,{\sc ii} 1.69$\mu$m + 
He\,{\sc i} 1.70$\mu$m)/C\,{\sc iv} 1.74$\mu$m. For WC4--7 stars
the latter is effectively the ratio of the helium-to-carbon recombination lines (12--7 transition for helium, 9--8 transition for carbon), so will depend upon the C/He ratio as well
as ionization. In contrast, the 1.7$\mu$m helium feature is dominated by He\,{\sc i} 1.70$\mu$m in WC8 and especially WC9 
subtypes, so this represents a better ionization indicator for
late subtypes. The K-band diagnostic cleanly separates WC4--7 subtypes from WC8 and WC9 stars, but fails to distinguish between early subtypes, as has previously been shown
\citep{eenens91, figer97}. The H-band diagnostic also distingishes between WC8 and WC9 stars,
but also maps earlier WC subtypes reasonably well. C\,{\sc ii} 1.78$\mu$m is also prominent at late subtypes, such that C\,{\sc ii} 1.785$\mu$m/C\,{\sc iv} 1.736 $>$ 1 in WC9 subtypes.


\subsection{WO stars}

At visual wavelengths, strong O\,{\sc vi} 3818\AA\ emission discriminates WO stars from early WC stars, with ratios of O\,{\sc vi} to O\,{\sc v} 5590\AA\ 
and/or C\,{\sc vi} 5808\AA\ providing WO subclasses \citet{crowther98a}. The majority of known WO stars have been observed in the near-IR 
\citep[this work,][]{tramper15}.
Unfortunately, the sole ionization stage of oxygen present in the near-IR  spectrum of WO stars is O\,{\sc vi} (1.075$\mu$m, 1.458$\mu$m, 1.917$\mu$m, 
2.463$\mu$m), preventing discrimination between WO subtypes, plus the exceptionally broad lines of WO stars leads to severe blending in most spectral 
regions.

Nevertheless, beyond their line widths, WO stars present a highly distinctive near-IR appearance.
They can be distinguished from WC4 stars by the presence of O\,{\sc 
vi} features, with high O\,{\sc vi} 1.075$\mu$m/C\,{\sc iv} 1.190$\mu$m and 
O\,{\sc vi} 1.46$\mu$m/C\,{\sc iv} 1.40$\mu$m ratios,
although 1.46$\mu$m is severely blended with C\,{\sc iv} 1.43$\mu$m and He\,{\sc ii} 1.47$\mu$m in most WO stars, and 1.075$\mu$m
is blended with He\,{\sc i} 1.083$\mu$m. He\,{\sc iv} LH41-1042  \citet{neugent12, tramper15} is unambiguously characterized as a WO 
star  from the presence of O\,{\sc vi} features, although C\,{\sc iii} 0.97$\mu$m is prominent at this (WO4) subtype. Since O\,{\sc 
vi} lines are weak, WO stars also exhibit unusually weak helium lines  with respect to early  WC stars,  such that WO stars possess 
He\,{\sc ii}  1.16/C\,{\sc iv} 1.19 $\leq$ 0.2, versus $\sim$0.5--0.6 in WC4 subtypes (recall Fig.~\ref{fig:old_wr_blue}). In 
addition,  C\,{\sc iii} 0.97$\mu$m and C\,{\sc iv} 2.07$\mu$m are weak in WO3 stars and absent in WO2 stars.

\begin{landscape}
\begin{table}
  \begin{center}
    \begin{footnotesize}
      \caption{Near-IR equivalent width and FWHM measurements for optically classified Galactic WN stars. Equivalent widths (in \AA) are generally robust to $\pm$0.05 dex, except for weak lines $\pm$0.1 dex, 
while measured FWHM (in km\,s$^{-1}$) are generally reliable to $\pm$50 km\,s$^{-1}$ (approximate values are indicated with colons). The key to the spectroscopic datasets utilised is provided in 
Table~\ref{tab:log}.
      }\label{oldbies_wn}
\begin{tabular}{l@{\hspace{1mm}}l@{\hspace{1mm}}l@{\hspace{1mm}}c@{\hspace{1mm}}
    c@{\hspace{1mm}}c@{\hspace{1mm}}c@{\hspace{1mm}}c@{\hspace{1mm}}c@{\hspace{1mm}}
    c@{\hspace{1mm}}c@{\hspace{1mm}}c@{\hspace{1mm}}c@{\hspace{1mm}}c@{\hspace{1mm}}c@{\hspace{1mm}}
    c@{\hspace{1mm}}c@{\hspace{1mm}}c@{\hspace{1mm}}c@{\hspace{1mm}}c@{\hspace{1mm}}l}
\hline
WR & WN     & Ref  & \multicolumn{2}{c}{He\,{\sc ii} 1.01} & He\,{\sc i} 1.08 & 
P$\gamma$ & N\,{\sc v} 1.11   & He\,{\sc ii} 1.16 & P$\beta$          & He\,{\sc ii} 1.48 & N\,{\sc 
v} 1.55 & He\,{\sc ii} 1.69 & He\,{\sc i} 1.70 & He\,{\sc i} 2.06 & N\,{\sc iii-v} 
2.11 & Br$\gamma$ & \multicolumn{2}{c}{He\,{\sc ii} 2.19} & Note & Data \\
& SpType &      & FWHM & $\log W_{\lambda}$ & $\log W_{\lambda}$& $\log W_{\lambda}$  & $\log W_{\lambda}$ & $\log W_{\lambda}$
& $\log W_{\lambda}$ & $\log W_{\lambda}$  & $\log W_{\lambda}$ & $\log W_{\lambda}$ & $\log W_{\lambda}$ & $\log W_{\lambda}$&  
$\log W_{\lambda}$ & $\log W_{\lambda}$ & FWHM & $\log W_{\lambda}$   \\
\hline
WR3 & WN3(h)&S96 &2470 & 2.13 &$<$0  &0.9:& 0.8 &      &      &      &      & 1.23 & 0.2: &     & 0.90& 1.43& 2580 & 1.75& 2.11$\ll$2.10  &O1,U3 \\ 
WR6 & WN4b & S96 &2500 & 2.63 & \multicolumn{2}{c}{ --- 2.60 ---}
                                          &     & 2.30 & 2.02 & 2.18 & 0.8: & 1.43 & 1.46 &     & 1.53& 1.76& 2160 & 2.20& 2.11$\sim$2.10 & I1,M1\\
WR8 &WN7o/CE&S96 &1480 & 2.13 & 2.58 &1.43&     & 1.99 & 1.89 & 1.72 & 0.8  & 1.59 & 1.15 &2.0: & 1.92& 1.76& 1350 & 1.73& 2.11$\ll$2.07  &U2,N6 \\ 
WR10& WN5ha& S96 &1480 & 2.03 & 1.26 &1.2:&0.5: & 1.66 & 1.81 & 1.5: &      & 1.2: &$<$0.0&     & 0.3:& 1.48& 1250 & 1.63& 2.11$\ll$2.19  & N3   \\
WR18 &WN4b & S96 &3020 & 2.65 & 2.39 &1.6:&     & 2.34 & 1.99 & 2.00 & 0.9  & 1.75 & 1.1: &     & 1.4:& 1.56&3100  & 2.23& 2.11$\sim$2.10 &N6    \\ %
WR24&WN6ha & S96 &1510 & 1.72 & 1.48 &    &     & 1.33 & 1.52 &      &      &      &      &     & 0.88& 1.57 &1500:& 1.46& 2.11$\ll$2.19 & N3   \\
WR26&WN4-6b/CE&S96 &2920 & 2.66 & 2.59 &1.1 &     & 2.43 & 2.02 & 2.12 & 0.7  & 1.74 & 1.48 &      &1.90& 1.69& 3040 & 2.24& 2.11$\ll$2.07  & N6   \\ 
WR40& WN8h & S96 &1070 & 1.55 & 2.58 &1.66&     & 1.36 & 1.91 & 1.0  &      & 0.5: & 1.71 &1.41 & 1.58& 1.98 &1180 & 1.16& 2.11$\gg$2.19  & N6   \\
WR46&WN3b pec&S96&2750 & 2.41 &$<$0.6&    &     &      &      &      & 1.38 & 1.5: &$<$0.9&     & 1.20& 1.45 &2920 & 2.10& 2.11$\ll$2.10  &E1,N6 \\
WR51&WN4o  & S96 &1790 & 2.38 &1.77  &1.31&0.80 & 2.06 & 1.75 & 1.83 & 0.90 &      &      &     & 1.41& 1.61 &1600 & 1.85& 2.11$\sim$2.10  & N2  \\
WR54&WN5o  & S96 &1600 & 2.39 & 2.01 &1.45&0.95 & 2.13 & 1.78 & 1.88 & 0.78 &      &      &     & 1.41& 1.59 &1700 & 1.88& 2.11$\sim$2.10 & N2  \\
WR55&WN7o  &S96  &1270 & 2.03 & 2.47 &1.4 &     & 1.89 & 1.73 &      &      &      &      &0.6: & 1.56&1.76  & 1310&1.63 &2.11$\sim$2.19  & N2  \\
WR61&WN5o  &S96  &1640 & 2.41 & 2.13 &1.36&0.7: & 2.15 & 1.73 & 1.89 & 0.78  &      &      &     & 1.43&1.54  &1490 &1.85 &2.11$>$2.10     &N2   \\
WR66 &WN8h &S86  & 1330& 1.72 & 2.30 &1.36&     & 1.61 & 1.74 & 1.34 &      &      &      &0.9: & 1.45& 1.64 &1330 & 1.27&2.11$>$2.19     & N2  \\
WR75& WN6b & S96 &2780 & 2.56 & 2.83 &1.1:&1.1: & 2.30 & 2.06 &      &      &      &      & 0.8:& 1.66 &1.86 &2700 & 2.02&2.11$\gg$2.10   & N2  \\
WR77sc&WN7b& C06 &3100 & 2.48  &2.97 &1.3:&     & 2.29 & 2.02 & 2.01 & 0.7   & 1.5: & 1.9: &     & 1.76 &1.88 &2840 & 2.08& 2.11$<$2.19    & N5  \\
WR78&WN7h  & S96 &1100 & 1.80 & 2.01 &1.27&     &1.58  & 1.63 &      &      & 1.0: & 1.1: & 0.9:& 1.23& 1.58& 1630 & 1.45&2.11$\sim$2.19  &  N6   \\
WR82&WN7(h)& S96 &1120 & 1.90 & 2.46 &1.74&     &1.80  & 2.03 &1.46  &      &      &      & 0.6:& 1.48& 1.78 &860  &1.36 &2.11$\sim$2.19  & N2  \\
WR83&WN5o  & S96 &1540 & 2.28 & 1.96 &1.46&0.8: &2.07  & 1.79 &1.80  & 0.8  &      &      &     &1.36 & 1.72 &1520 &1.85 &2.11$>$2.10     & N2  \\
WR84&WN7o  & S96 &1240 & 2.13 & 2.40 &1.36&     &2.00  & 1.72 &1.73  &      &      &      & 0.7:& 1.52& 1.60 &1040 &1.63 &2.11$<$2.19     &  N2 \\
WR87&WN7h+a&S96 &1150 & 1.46 & 1.64 &1.0:&     &1.20   & 1.63 &0.99  &      &      &      &     &     &     &      &     &                &  N6  \\
WR89&WN8h+a&S96 & 960 & 1.3  & 1.67 &0.9:&     &1.15   & 1.52 &0.89  &      &      &      &     & 1.18& 1.52& 740  &0.85 &2.11$\sim$2.19  &  N6  \\
WR98&WN8o/C7&S96&1400 & 1.83 & 2.55 &1.45 &     &1.80  & 1.83 &1.53  &      &      &      &1.20 &1.78 & 1.74 &1080 &1.90 & 2.11$>$2.07    &  U3,N5 \\
WR105&WN9h & S96 &500 & 0.88  & 2.27 &1.38&1.30 &0.3:  &1.76  &      &      &$<$0.0& 1.36 &1.68 & 1.43& 1.84& 160  &--0.2:&2.11$\gg$2.19  &  C1,U5,U6\\
WR108&WN9ha& S96 &400 & 0.92  & 1.68 &0.83&     &      & 1.32 &      &      &      &      &     & 1.28 &1.34& 170  &--0.3:&2.11$\gg$2.19  &  C1,I1,U5\\
WR110&WN5--6b&S96&3500& 2.64  & 2.81 &1.3:&1.3: &2.37  & 2.06 & 2.13 & 1.0: &      &      &     & 1.55& 1.88& 3200 &2.02 &2.11$\gg$2.10   &  N2  \\
WR115&WN6o & S96 &1260& 2.48  & 2.51 &1.25&0.2: &      &      &      &      & 1.28 & 1.36 &     & 1.52& 1.54& 1160 & 1.76&2.11$<$2.19     &  I1,U3\\
WR116&WN8h &S96  & 830& 1.32  & 2.68 &1.78&     &1.18  & 2.08 & 0.7: &      &      &      &2.04 & 1.54& 1.90& 1100:&0.7  &2.11$\gg$2.19   & N2,U3,U4\\
WR120&WN7o & S96 & 970 & 1.86 & 2.51 &1.31&     & 1.76 & 1.76 & 1.50 &      &      &      &1.1: & 1.68&1.68 &1050  &1.54 & 2.11$\sim$2.19 &I1,U3,U4\\
WR123&WN8o & S96 & 930 & 1.60 & 2.68 &1.56&     & 1.52 & 1.85 & 1.06 &      &      &      &1.73 & 1.78& 1.87& 930  & 1.38& 2.11$\gg$2.19  &O1,U3,U4\\
WR128&WN4(h)&S96 &1900 & 2.24  &1.41 &1.32&1.04 &      & 1.75 &      &      & 1.36 & 0.6: &     & 1.28& 1.58& 1750 & 1.87&2.11$\ll$2.10   &  U3   \\
WR131&WN7h+a&S96&1000 &  1.68 & 1.59 &1.13&     &      & 1.48 &      &      & 0.9: & 0.5: &     & 0.95& 1.50 &1040 & 1.20&2.11$<$2.19     &  U3   \\
WR134&WN6b & S96 &2690& 2.60 & 2.54 & 1.64 &1.0:&      & 2.01 &      &      & 1.73 & 1.36: &     & 1.50& 1.69& 2560 & 2.10&2.11$>$2.10     & I1,U3 \\
WR136&WN6b(h)&S96&1920& 2.48 & 2.51 & 1.77 &0.7:&2.25* & 2.05 &      &      & 1.68 & 1.41: &     & 1.59& 1.91& 1800 & 2.03&2.11$\gg$2.10   & I1,U3 \\
WR138&WN5--6o&C11&1440& 2.03  & 1.79 &1.18 &0.2:&      & 1.54 &      &      & 1.28 & 0.7: &     & 1.25& 1.56& 1450 & 1.76& 2.11$>$2.10    &  I1,U3\\
WR145&WN7o/CE&S96&1520& 2.01 & 2.25  &1.1 &     &      & 1.58 &      &      & 1.34 & 1.30 &0.7: & 1.54& 1.59& 1510 & 1.71& 2.11$>$2.07    &  I1,U3\\
WR152&WN3(h)&S96 &1980 & 2.31  & 1.23 &1.35&0.97&      &      &      &      & 1.43 & 0.5: &     & 1.18& 1.59& 1950 & 1.96& 2.11$\ll$2.10  &  O1,U3\\
\hline 
\multicolumn{20}{l}{
  \begin{minipage}{\columnwidth} 
S90: \citep{smith90}; C06: \citep{crowther06}, C11: \citep{crowther11}, * W.D.Vacca (priv. comm.) \\
Note: 2.10 = N\,{\sc v} 2.100; 2.11 = He\,{\sc i} 2.112 + N\,{\sc iii} 2.116; 2.19 = He\,{\sc ii} 2.189 \\
\end{minipage}
}
\end{tabular}
\end{footnotesize}
\end{center}
\end{table}
\end{landscape}

\begin{landscape}
\begin{table}
  \begin{center}
    \begin{footnotesize}
\caption{Near-IR equivalent width and FWHM measurements for visually classified Magellanic Cloud WN stars (LMC: 
BAT\#, SMC: AB\#). Equivalent widths (in \AA) are generally robust to $\pm$0.05 dex, except for weak lines $\pm$0.1 dex, while measured FWHM (in km\,s$^{-1}$) are generally
reliable to $\pm$50 km\,s$^{-1}$ (approximate values are indicated with colons).
The key to the spectroscopic datasets utilised is provided in Table~\ref{tab:log}. Several sets of measurements are provided for the variable LBV + WN star AB5 (HD\,5980).}\label{oldbies_wn_mc}
\begin{tabular}{l@{\hspace{1mm}}l@{\hspace{1mm}}l@{\hspace{1mm}}c@{\hspace{1mm}}
    c@{\hspace{1mm}}c@{\hspace{1mm}}c@{\hspace{1mm}}c@{\hspace{1mm}}c@{\hspace{1mm}}
    c@{\hspace{1mm}}c@{\hspace{1mm}}c@{\hspace{1mm}}c@{\hspace{1mm}}l@{\hspace{1mm}}l}
\hline
WR & WN     & Ref  & \multicolumn{2}{c}{He\,{\sc ii} 1.01} & He\,{\sc i} 1.08 & 
P$\gamma$ & N\,{\sc v} 1.11   & He\,{\sc ii} 1.16 & P$\beta$          & He\,{\sc ii} 1.48 & N\,{\sc 
v} 1.55 & He\,{\sc ii} 1.57 & Note & Data \\
& SpType &      & FWHM & $\log W_{\lambda}$ & $\log W_{\lambda}$& $\log W_{\lambda}$  & $\log W_{\lambda}$ & $\log W_{\lambda}$
& $\log W_{\lambda}$ & $\log W_{\lambda}$  & $\log W_{\lambda}$ & $\log W_{\lambda}$ \\
\hline
BAT1  & WN3b & F03b &2050 & 2.75 & 1.08 &1.36 & 0.9: & 2.34 & 1.84 & 2.08 & 0.9: & 1.43&                        & N4 \\
BAT2 &WN2b(h)& F03b &2680 & 2.48 &$<$1.0&     &      & 2.13 & 1.65 & 1.9: &      &     &  He\,{\sc i} 1.08 absent& N4 \\
BAT3  & WN4b & F03b &1920 & 2.55 & 2.32 &1.5: &$<$0.8& 2.22 & 1.81 & 1.98 & 0.8: &     &                         & N4 \\
BAT5  & WN2b & F03b &2430 & 2.56 &$<$1.0&1.1: & 1.0: & 2.20 & 1.66 & 1.9  &      &     & He\,{\sc i} 1.08 absent & N4 \\
BAT7  & WN4b & F03b &3900 & 2.88 & 1.97 &1.60 &      & 2.44 & 2.06 & 2.18 &      &1.85 &                         & N4\\
BAT16 & WN8h & S96  & 950 & 1.67 & 2.38 &1.60 &      & 1.54 & 1.95 & 1.1: &      &     &                         &N3\\
BAT17 & WN4o & F03b &1780 & 2.29 & 2.22 &1.23 & 0.8: & 2.01 & 1.63 & 1.76 &0.8:  &1.1  & log (N\,{\sc v} 1.55/He\,{\sc ii} 
1.57) = 0 & N3\\
BAT18 & WN3(h)&F03b &1770 & 2.35 & 1.69 &1.47 & 0.5  & 2.06 & 1.80 & 1.87 &      &1.0: &                         & N4\\
BAT24 & WN4b & F03b &2610 & 2.68 & 2.56 &1.55 &$<$0.8& 2.34 & 1.94 & 2.12 & 0.8: &     &                         & N4  \\
BAT25 & WN4ha& F03b &1580 & 1.96 &$<$0.7&1.1  &      & 1.68 & 1.60 & 1.53 & 0.9: &     &                         & N4\\
BAT30 & WN6h & S96  & 970 & 1.91 & 1.93 &1.65 &      & 1.76 & 1.76 & 1.43 &      &0.9: &                         & N3\\
BAT31 & WN4b & F03b &1860 & 2.49 & 2.29 &1.4  & 0.6: & 2.22 & 1.76 & 1.98 &      & 1.3 &                        & N4\\
BAT37 & WN3o & F03b &1770 & 2.53 & 1.81 &1.33 & 0.7: & 2.25 &  1.80& 2.05 & 0.9: & 1.3 &                         & N4\\
BAT47 & WN3o\ddag & F03b &1640 & 2.61 & 1.75 & 1.49& 0.95 & 2.33 & 1.86 & 2.11 & 0.9: & 1.4 &                         & N4\\ 
BAT51 & WN3b & F03b &3020 &2.65  & 1.72 & 1.59& 0.5:  & 2.42 & 1.90 &2.14  &      &1.3  &                         & N3,N4\\
BAT60 &WN4(h)a&F03b &1840 &1.99  & 1.15 &1.03 & 0.3  & 1.77 & 1.56 &  1.56&      &     &                         & N4\\
BAT66 &WN3(h)& F03b &1840 &2.32  &$<$0.8&1.3: & 0.6  & 2.03 &  1.76& 1.83 &      &     &                         & N4\\
BAT67 & WN5ha& F03b &1720 &2.01  &1.43  & 1.2 &      & 1.70 & 1.53 &  1.43&      &     &                         & N4\\
BAT75 & WN4o & F03b &1730 &2.46  &2.08  &1.53 &      & 2.18 & 1.73 &  1.90&      & 1.1 &                         & N4\\
BAT76 & WN9h & C95  & 370 &0.66  & 2.12 & 1.30&      &      &      &      &      &     & log (Br$\gamma$/He\,{\sc ii} 2.189) = 
+1.5 & C1 \\
BAT77 & WN7ha & S08 & 890 & 1.12 & 1.55 & 1.4:&      & 1.08 & 1.55 &      &      &     &                         & N3\\
BAT78 & WN4(h)& W99 &1590 & 1.61 & 1.23 & 1.1:&      & 1.56 & 1.40 & 1.38 &      &     &                         & N3\\
BAT88 & WN4b/CE&F03b&2670 & 2.75 & 2.54 & 1.69&      & 2.46 & 1.93 & 2.20 &      & 1.54& log (C\,{\sc iv} 1.19/He\,{\sc ii} 1.16) 
= --0.9 & N4\\
BAT89 & WN7h & S96  &1040 & 1.93 & 2.35 & 1.60&      & 1.81 & 1.88 &  1.43&      & 1.0 &                         & N3\\
BAT92 & WN6+B1Ia&M86&1000 & 1.76 & 2.27 & 1.22 &      & 1.53 & 1.72 & 1.32 &      &     &                         & N4\\
BAT117 & WN5ha &F03b&2350 & 1.99 & 1.1: & 1.23 &      & 1.70 & 1.57 &  1.45&      &     &                         & N4\\
BAT118 & WN6h & S96 &1340 & 2.05 & 2.08 & 1.50&      & 1.78 &  1.81&  1.45&      & 0.8 &                         & N3\\
BAT119 & WN6(h)& S96&1460 & 2.01 & 1.99 & 1.34 &      & 1.74 & 1.71 & 1.43 &      &     &                         & N4\\
BAT122 & WN5(h)&F03b&1510 & 2.13 & 1.67 & 1.56&      & 1.85 &  1.76& 1.68 &      & 1.0 &                        & N3\\
BAT134 & WN4b &F03b &1960 & 2.62  &2.53 & 1.41&      & 2.32 & 1.90 & 2.05 &      & 1.48&                         & N4\\
\\
AB1   & WN3ha& F03a &1550 & 1.88 &$<$0.0& 0.78 &      & 1.53 & 1.38 & 1.4: &      &     & He\,{\sc i} 1.08 absent & N3,N4\\
AB2   & WN5ha& F03a &1020 & 1.56 & 0.9: & 0.7:&      &  1.1:& 1.28 &      &      &     &                        & N4\\
AB4   & WN6h & F03a &1070 & 1.94 & 1.45 &1.20 &      & 1.61 & 1.61 & 1.41 &      & 0.9 & log (Br$\gamma$/He\,{\sc ii} 2.189) = 
--0.08  & N1 \\
AB5   &      &      &1160 &2.00 & 2.31 & 1.58 &     &  1.82&  1.81& 1.63 &      & 1.1 & Sep 99                 & N1 \\
AB5   &      &      &2040 &2.13 & 2.10 & 1.45 &     &  1.89& 1.77 & 1.73 &      & 1.3 & Nov 03                 & N3 \\
AB5   &      &      &1500 &2.10 & 2.06 & 1.37 &     &  1.89& 1.76 & 1.65 &      & 1.0:& Nov 04                 & N4 \\
\hline 
\multicolumn{15}{l}{
  \begin{minipage}{0.8\columnwidth} 
M86: \citep{moffat86}, C95: \citep{crowther95a}, S96: \citep{smith96}, W99 \citep{walborn99}, F03a 
\citep{foellmi03a}, F03b \citep{foellmi03b}, S08: \citep{schnurr08}\\
\ddag We adopt WN3o for BAT47 on the owing to its modest FWHM ($\ll$ 1900 km/s) whereas \citet{foellmi03b} adopted WN3b.
\end{minipage}
}
\end{tabular}
\end{footnotesize}
\end{center}
\end{table}
\end{landscape}

\begin{landscape}
\begin{table}
  \begin{center}
    \begin{footnotesize}
      \caption{Near-IR equivalent width and FWHM measurements for visually classified WC and WO stars in the Milky Way and Magellanic Clouds.
Equivalent widths (in \AA) are generally robust to $\pm$0.05 dex, except for weak lines $\pm$0.1 dex, while measured FWHM (in km\,s$^{-1}$) are generally
reliable to $\pm$50 km\,s$^{-1}$ (approximate values are indicated with colons).        The key to the spectroscopic datasets utilised 
is provided in Table~\ref{tab:log}}
\label{tab:WCgalactic}
\begin{tabular}{l@{\hspace{1mm}}l@{\hspace{1mm}}l@{\hspace{1mm}}c@{\hspace{0mm}}
    c@{\hspace{0mm}}c@{\hspace{1mm}}c@{\hspace{1mm}}c@{\hspace{1mm}}c@{\hspace{1mm}}c@{\hspace{1mm}}
    c@{\hspace{1mm}}c@{\hspace{1mm}}c@{\hspace{0mm}}c@{\hspace{1mm}}c@{\hspace{0mm}}c@{\hspace{1mm}}
    c@{\hspace{1mm}}c@{\hspace{1mm}}c@{\hspace{1mm}}c@{\hspace{1mm}}l}
\hline
WR & WC & Ref & C\,{\sc iii} 0.97 & C\,{\sc ii} 0.99 & \multicolumn{2}{c}{He\,{\sc ii} 1.01} & 
He\,{\sc i} 1.08 & He\,{\sc ii} 1.16 &
C\,{\sc iv} 1.19 & C\,{\sc iii} 1.20 & C\,{\sc iv} 1.43 & He\,{\sc i-ii} 1.70 & 
\multicolumn{2}{c}{C\,{\sc iv} 1.74} & C\,{\sc ii} 1.78 & He\,{\sc i} 2.06 & C\,{\sc iv} 2.07 & 
C\,{\sc iii} 2.11 & C\,{\sc iv} 2.43 & Data \\ 
& SpType & & $\log W_{\lambda}$ & $\log W_{\lambda}$& FWHM & $\log W_{\lambda}$ & $\log W_{\lambda}$ & $\log W_{\lambda}$ & 
$\log W_{\lambda}$ & $\log W_{\lambda}$ & $\log W_{\lambda}$ & $\log W_{\lambda}$ & FWHM & $\log W_{\lambda}$ & $\log W_{\lambda}$ & 
$\log W_{\lambda}$& $\log W_{\lambda}$ & $\log W_{\lambda}$ & $\log W_{\lambda}$ &  \\
\hline
WR4  & WC5  & S90 &      &      &      &      & 2.20 &      &      &      &      & 1.70 & 2360 & 2.62 &      &      & 3.20 & 2.46 &          &U3\\
WR5  & WC6  & S90 &      &      &      &      & 2.38 &      &      &      &      & 1.76 & 1950 & 2.47 &      &      &      &      & 2.56     &U3\\
WR14 & WC7  & S90 & 2.94 & 1.42 & 1960 & 2.10 & 2.48 & 2.05 & 2.24 & 1.67 & 2.21 & 1.72 & 2150 & 2.43 & 1.6: &      & 3.01 & 2.29 & 2.47     &N3\\ 
WR15 & WC6  & S90 & 2.91 & 1.2: & 3300 & 2.38 & 2.58 & 2.21 & 2.44 & 1.45 & 2.05 & 1.53 & 3060 & 2.40 &      &      & 3.04 & 2.26 & 2.50     &N6\\ 
WR23 & WC6  & S90 & 2.94 & 1.40 & 2490 & 2.23 & 2.39 & 2.13 & 2.40 & 1.66 & 2.25 & 1.59 & 2620 & 2.51 &      &      & 3.09 & 2.50 & 2.57     &N6\\ 
WR52 & WC4--5&C98 & 2.84 & 1.36 & 2460 & 2.31 & 2.06 & 2.08 & 2.49 & 1.56 & 2.38:& 1.38 & 2560 & 2.58 &      &      & 3.22 & 2.49 & 2.70     &N2,N6\\ 
WR56 & WC7  & S90 & 3.06 & 1.72 & 1810 & 2.13 & 2.26 & 2.03 & 2.29 & 1.87 & 2.33 &      &      &      &      &      & 2.81 & 2.23 &          &N2\\ 
WR57 & WC8  & S90 & 3.10 & 1.82 & 1590 & 2.05 & 2.26 & 1.99 & 2.23 & 1.94 & 2.26 &      &      &      &      &      & 2.42 & 1.96 &          &N2\\ 
WR60 & WC8  & S90 & 3.09 & 1.85 & 2260 & 2.05 & 2.52 & 2.00 & 2.26 & 2.04 & 2.29 &      &      &      &      &      & 2.63 & 2.36 &          &N2\\ 
WR64 & WC7  & S90 & 3.08 & 1.74 & 1840 & 2.24 & 2.18 & 2.03 & 2.33 & 1.89 & 2.35 &      &      &      &      &      & 2.81 & 2.26 &          &N2\\ 
WR88 & WC9  & S90 & 2.45 & 1.20 & 900  & 1.48 & 2.61 & 1.49 & 1.40 & 1.80 & 1.56 & 1.77 & 1060 & 1.46 & 1.61 & 1.75 & 1.99 & 2.05 & 1.18     &N6\\ 
WR90 & WC7  & S90 & 2.96 & 1.57 & 2170 & 2.17 & 2.45 & 2.08 & 2.30 & 1.85 & 2.15:& 1.75 & 2310 & 2.40 &      &      & 3.07 & 2.47 & 2.44     &N6\\ 
WR92 & WC9  & S90 & 2.50 & 1.68 &  800 & 1.40 & 2.55 & 1.62 & 1.43 & 2.02 & 1.68 & 1.72 & 1050 & 1.30 & 1.96 & 2.22 & 1.85 & 1.99 & 1.11     &N6 \\ 
WR101& WC8  & L89 & 3.00 & 1.72 & 2230 & 1.99 & 2.22 & 2.03 & 2.26 & 2.11 & 2.38 &      &      &      &      &      &      &      &          &N5\\ 
WR103& WC9d & S90 &      &      &      &      & 2.49 &      &      &      &      & 1.51 &  900 & 1.28 & 1.88 & 1.94 & 1.50 & 1.78 &          &U3\\
WR111& WC5  & S90 & 2.98:& 1.41 & 2320 & 2.27 & 2.26 &      &      &      &      & 1.41 & 2250 & 2.59 & 1.2: &      & 3.23 & 2.53 & 2.66     &I1,U1,U3\\
WR114& WC5  & S90 & 2.94 & 1.52 & 2350 & 2.33 & 2.35 & 2.16 & 2.43 & 1.79 & 2.25 &      &      &      &      &      & 2.94 & 2.20 &          &N2\\ 
WR121& WC9d & S90 &      &      &      &      & 2.47 &      &      &      &      & 1.25 &  930 & 0.92 & 1.53 & 1.66 & 1.04 & 1.41 & 0.5:     &U3\\
WR135& WC8  & S90 &2.86  & 1.52 & 1270 & 1.89 & 2.53 &      &      &      &      & 1.80 & 1450 & 2.18 & 1.70 &      & 2.74 & 2.31 & 2.15     &I1,U1,U3\\
WR137& WC7d+O&S90 & 2.56 & 1.34 & 1660 &  1.66& 1.85 &      &      &      &      &      &      &      &      &      & 2.67 & 2.06 &          &I1,L1\\
WR140& WC7d+O&S90 &      &      &      &      &      &      &      &      &      &      &      &      &      &      & 2.69 & 2.30 &          &L1\\
WR143& WC4  & S90 &      &      &      &      &      &      &      &      &      &      &      &      &      &      & 3.10 & 2.37 & 2.65     &U3\\
WR146& WC4-5+O&C98&      &      &      &      & 2.23 &      &      &      &      & 1.61 & 3200 & 2.10 &      &      & 2.80 & 2.11 & 2.23     &U3\\
WR154& WC6  & S90 &      &      &      &      & 2.36 &      &      &      &      & 1.80 & 2360 & 2.54 &      &      & 3.12 & 2.36 &          &U3\\
\\
BAT9  &WC4  & S90 &2.27  &$<$0.6& 2890 & 2.58 & 2.44 & 2.33 & 2.43 & 1.89 & 2.05 &      &      &      &      &      &      &      &          &N4\\ 
BAT11&WC4&S90 &2.70  &$<$0.9& 3710 & 2.47 & 2.37 & 2.20 & 2.40 & 1.79 & 2.13 &      &      &      &      &      &      &      &          &N3\\ 
BAT121&WC4&B01&2.76  &$<$1.2& 3260 & 2.48 & 2.34 & 2.20 & 2.44 & 1.89 & 2.09 &      &      &      &      &      &      &      &          &N4\\ 
\hline
WR & WO     & Ref & C\,{\sc iii} 0.97 & \multicolumn{2}{c}{He\,{\sc ii} 1.01} & C\,{\sc iv} 1.05 & O\,{\sc vi} 1.07 & He\,{\sc i} 1.08 & He\,{\sc ii} 1.16 &
C\,{\sc iv} 1.19 & C\,{\sc iv} 1.40  & C\,{\sc iv} 1.43 & O\,{\sc vi} 1.46 & He\,{\sc ii} 1.48 &
\multicolumn{2}{c}{C\,{\sc iv} 1.74} & C\,{\sc iv} 2.07 & C\,{\sc iv} 2.10 & C\,{\sc iv} 2.43 & Data \\ 
   & SpType &     & $\log W_{\lambda}$ & FWHM  & $\log W_{\lambda}$ & $\log W_{\lambda}$ & $\log W_{\lambda}$ & 
$\log W_{\lambda}$ & $\log W_{\lambda}$ & $\log W_{\lambda}$ & $\log W_{\lambda}$ & $\log W_{\lambda}$ &  $\log W_{\lambda}$ & $\log W_{\lambda}$ & 
FWHM & $\log W_{\lambda}$              & $\log W_{\lambda}$ & $\log W_{\lambda}$ & $\log W_{\lambda}$ \\
\hline
WR93b &WO3 & C98  &    &  5800 & 2.69 & 1.98 & 2.04 &      & 2.04 & 2.86 & 2.58 & 1.9: & 1.9: &  1.8:   & 5060 &  2.92 &       & 2.21 & 3.20 &N5\\
WR102 &WO2 & C98  &    &  6600 & 2.50 & 1.74 & 2.22 &      & 1.5: & 2.79 & 2.65 & -- & \multicolumn{2}{c}{-- 2.32 -- } 
                                                                                                        & 5890 &  2.83 &       & 2.09 & 3.05 &N5\\
WR142 &WO2 & C98  &    &  7100 & 2.67 & 1.3  & 1.8  &      &      & 2.8  &      &      &      &         & 6700 &  2.86 &       & 1.95 & 3.0  & V1\\
\\
BAT123&WO3 & C98  &    &  4200 & 2.74 & 1.95 & 1.93 &      & 2.14 & 2.79 &      &      &\multicolumn{2}{c}{-- 2.0 -- } 
                                                                                                        & 3650 & 2.93  &       & 3.2  & 3.16 & V1 \\
LH41-1042&WO4&N12 & 2.05& 4200 & 2.71 & 1.78 & 2.03 & 1.7: & 2.13 & 2.88 &      & 2.2: & 2.1:  & 2.2:   & 3650 & 2.97  & 3.08   & 3.1  & 3.18 & V1 \\
\\
AB8 &WO3+O&C98&$\leq$1.3&3200 &1.78 & 1.2: & 1.3: &      & 1.81 & 2.34 &      &      &       &        & 3750 & 2.54  &       & 2.8:  & 2.9  & V1 \\
\hline
\multicolumn{20}{l}{
  \begin{minipage}{\columnwidth} 
L89: \citep{lundstrom89},  S90: \citep{smith90}; C98: \citep{crowther98a}; B01: \citep{bartzakos01}; N12: \citep{neugent12}
\end{minipage}
}
\end{tabular}
\end{footnotesize}
\end{center}
\end{table}
\end{landscape}

\bibliographystyle{mnras}

\begin{thebibliography}{}
\makeatletter
\relax
\def\mn@urlcharsother{\let\do\@makeother \do\$\do\&\do\#\do\^\do\_\do\%\do\~}
\def\mn@doi{\begingroup\mn@urlcharsother \@ifnextchar [ {\mn@doi@}
  {\mn@doi@[]}}
\def\mn@doi@[#1]#2{\def\@tempa{#1}\ifx\@tempa\@empty \href
  {http://dx.doi.org/#2} {doi:#2}\else \href {http://dx.doi.org/#2} {#1}\fi
  \endgroup}
\def\mn@eprint#1#2{\mn@eprint@#1:#2::\@nil}
\def\mn@eprint@arXiv#1{\href {http://arxiv.org/abs/#1} {{\tt arXiv:#1}}}
\def\mn@eprint@dblp#1{\href {http://dblp.uni-trier.de/rec/bibtex/#1.xml}
  {dblp:#1}}
\def\mn@eprint@#1:#2:#3:#4\@nil{\def\@tempa {#1}\def\@tempb {#2}\def\@tempc
  {#3}\ifx \@tempc \@empty \let \@tempc \@tempb \let \@tempb \@tempa \fi \ifx
  \@tempb \@empty \def\@tempb {arXiv}\fi \@ifundefined
  {mn@eprint@\@tempb}{\@tempb:\@tempc}{\expandafter \expandafter \csname
  mn@eprint@\@tempb\endcsname \expandafter{\@tempc}}}

\bibitem[\protect\citeauthoryear{{Acker} \& {Stenholm}}{{Acker} \&
  {Stenholm}}{1990}]{acker90}
{Acker} A.,  {Stenholm} B.,  1990, \aaps, \href
  {http://adsabs.harvard.edu/abs/1990A%26AS...86..219A} {86, 219}

\bibitem[\protect\citeauthoryear{{Anderson}, {Bania}, {Balser}, {Cunningham},
  {Wenger}, {Johnstone}  \& {Armentrout}}{{Anderson} et~al.}{2014}]{anderson14}
{Anderson} L.~D.,  {Bania} T.~M.,  {Balser} D.~S.,  {Cunningham} V.,  {Wenger}
  T.~V.,  {Johnstone} B.~M.,   {Armentrout} W.~P.,  2014, \mn@doi [\apjs]
  {10.1088/0067-0049/212/1/1}, \href
  {http://adsabs.harvard.edu/abs/2014ApJS..212....1A} {212, 1}

\bibitem[\protect\citeauthoryear{{Benjamin} et~al.,}{{Benjamin}
  et~al.}{2003}]{benjamin03}
{Benjamin} R.~A.,  et~al., 2003, PASP, 115, 953

\bibitem[\protect\citeauthoryear{{Bibby}, {Crowther}, {Furness}  \&
  {Clark}}{{Bibby} et~al.}{2008}]{bibby08}
{Bibby} J.~L.,  {Crowther} P.~A.,  {Furness} J.~P.,   {Clark} J.~S.,  2008,
  \mn@doi [\mnras] {10.1111/j.1745-3933.2008.00453.x}, \href
  {http://adsabs.harvard.edu/abs/2008MNRAS.386L..23B} {386, L23}

\bibitem[\protect\citeauthoryear{{Bohannan}}{{Bohannan}}{1997}]{bohannan97}
{Bohannan} B.,  1997, in {Nota} A.,  {Lamers} H.,  eds,  ASP Conf Ser Vol. 120,
  Luminous Blue Variables: Massive Stars in Transition. p.~3

\bibitem[\protect\citeauthoryear{{Bohannan} \& {Crowther}}{{Bohannan} \&
  {Crowther}}{1999}]{bohannan99}
{Bohannan} B.,  {Crowther} P.~A.,  1999, \mn@doi [\apj] {10.1086/306647}, \href
  {http://adsabs.harvard.edu/abs/1999ApJ...511..374B} {511, 374}

\bibitem[\protect\citeauthoryear{{Borissova} et~al.,}{{Borissova}
  et~al.}{2011}]{borissova11}
{Borissova} J.,  et~al., 2011, \mn@doi [\aap] {10.1051/0004-6361/201116662},
  \href {http://adsabs.harvard.edu/abs/2011A%26A...532A.131B} {532, A131}

\bibitem[\protect\citeauthoryear{{Bressert} et~al.,}{{Bressert}
  et~al.}{2010}]{bressert10}
{Bressert} E.,  et~al., 2010, \mn@doi [\mnras]
  {10.1111/j.1745-3933.2010.00946.x}, \href
  {http://adsabs.harvard.edu/abs/2010MNRAS.409L..54B} {409, L54}

\bibitem[\protect\citeauthoryear{{Chen{\'e}} et~al.,}{{Chen{\'e}}
  et~al.}{2013}]{chene13}
{Chen{\'e}} A.-N.,  et~al., 2013, \mn@doi [\aap] {10.1051/0004-6361/201220107},
  \href {http://adsabs.harvard.edu/abs/2013A%26A...549A..98C} {549, A98}

\bibitem[\protect\citeauthoryear{{Chen{\'e}} et~al.,}{{Chen{\'e}}
  et~al.}{2015}]{chene15}
{Chen{\'e}} A.-N.,  et~al., 2015, \mn@doi [\aap] {10.1051/0004-6361/201525958},
  \href {http://adsabs.harvard.edu/abs/2015A%26A...584A..31C} {584, A31}

\bibitem[\protect\citeauthoryear{{Churchwell} et~al.,}{{Churchwell}
  et~al.}{2006}]{churchwell06}
{Churchwell} E.,  et~al., 2006, \mn@doi [\apj] {10.1086/507015}, \href
  {http://adsabs.harvard.edu/abs/2006ApJ...649..759C} {649, 759}

\bibitem[\protect\citeauthoryear{{Clark}, {Negueruela}, {Crowther}  \&
  {Goodwin}}{{Clark} et~al.}{2005}]{clark05}
{Clark} J.~S.,  {Negueruela} I.,  {Crowther} P.~A.,   {Goodwin} S.~P.,  2005,
  \mn@doi [\aap] {10.1051/0004-6361:20042413}, \href
  {http://adsabs.harvard.edu/abs/2005A%26A...434..949C} {434, 949}

\bibitem[\protect\citeauthoryear{{Clark} et~al.,}{{Clark}
  et~al.}{2009}]{clark09}
{Clark} J.~S.,  et~al., 2009, \mn@doi [\aap] {10.1051/0004-6361/200911945},
  \href {http://adsabs.harvard.edu/abs/2009A%26A...498..109C} {498, 109}

\bibitem[\protect\citeauthoryear{{Cooper} et~al.,}{{Cooper}
  et~al.}{2013}]{cooper13}
{Cooper} H.~D.~B.,  et~al., 2013, \mn@doi [\mnras] {10.1093/mnras/sts681},
  \href {http://adsabs.harvard.edu/abs/2013MNRAS.430.1125C} {430, 1125}

\bibitem[\protect\citeauthoryear{{Crowther}}{{Crowther}}{2000}]{crowther00}
{Crowther} P.~A.,  2000, \aap, \href
  {http://adsabs.harvard.edu/abs/2000A%26A...356..191C} {356, 191}

\bibitem[\protect\citeauthoryear{{Crowther}}{{Crowther}}{2007}]{crowther07}
{Crowther} P.~A.,  2007, ARA\&A, 45, 177

\bibitem[\protect\citeauthoryear{{Crowther}}{{Crowther}}{2015}]{crowther15}
{Crowther} P.~A.,  2015, in {Hamann} W.-R.,  {Sander} A.,   {Todt} H.,  eds,
  Wolf-Rayet Stars: Proceedings of an International Workshop,
  Universit{\"a}tsverlag Potsdam, p.21-26. pp 21--26 (\mn@eprint {arXiv}
  {1509.00495})

\bibitem[\protect\citeauthoryear{{Crowther} \& {Smith}}{{Crowther} \&
  {Smith}}{1996}]{crowther96}
{Crowther} P.~A.,  {Smith} L.~J.,  1996, \aap, \href
  {http://adsabs.harvard.edu/abs/1996A%26A...305..541C} {305, 541}

\bibitem[\protect\citeauthoryear{{Crowther} \& {Walborn}}{{Crowther} \&
  {Walborn}}{2011}]{crowther11}
{Crowther} P.~A.,  {Walborn} N.~R.,  2011, \mn@doi [\mnras]
  {10.1111/j.1365-2966.2011.19129.x}, \href
  {http://adsabs.harvard.edu/abs/2011MNRAS.416.1311C} {416, 1311}

\bibitem[\protect\citeauthoryear{{Crowther}, {Smith}  \& {Willis}}{{Crowther}
  et~al.}{1995}]{crowther95c}
{Crowther} P.~A.,  {Smith} L.~J.,   {Willis} A.~J.,  1995, \aap, \href
  {http://adsabs.harvard.edu/abs/1995A%26A...304..269C} {304, 269}

\bibitem[\protect\citeauthoryear{{Crowther}, {Hadfield}, {Clark}, {Negueruela}
  \& {Vacca}}{{Crowther} et~al.}{2006}]{crowther06}
{Crowther} P.~A.,  {Hadfield} L.~J.,  {Clark} J.~S.,  {Negueruela} I.,
  {Vacca} W.~D.,  2006, \mn@doi [\mnras] {10.1111/j.1365-2966.2006.10952.x},
  \href {http://adsabs.harvard.edu/abs/2006MNRAS.372.1407C} {372, 1407}

\bibitem[\protect\citeauthoryear{{Dalton} et~al.,}{{Dalton}
  et~al.}{2016}]{weave}
{Dalton} G.,  et~al., 2016, in Ground-based and Airborne Instrumentation for
  Astronomy VI. p. 99081G, \mn@doi{10.1117/12.2231078}

\bibitem[\protect\citeauthoryear{{Davies}, {Figer}, {Kudritzki}, {MacKenty},
  {Najarro}  \& {Herrero}}{{Davies} et~al.}{2007}]{davies07}
{Davies} B.,  {Figer} D.~F.,  {Kudritzki} R.-P.,  {MacKenty} J.,  {Najarro} F.,
    {Herrero} A.,  2007, \mn@doi [\apj] {10.1086/522224}, \href
  {http://adsabs.harvard.edu/abs/2007ApJ...671..781D} {671, 781}

\bibitem[\protect\citeauthoryear{{Davies}, {de La Fuente}, {Najarro}, {Hinton},
  {Trombley}, {Figer}  \& {Puga}}{{Davies} et~al.}{2012a}]{davies12b}
{Davies} B.,  {de La Fuente} D.,  {Najarro} F.,  {Hinton} J.~A.,  {Trombley}
  C.,  {Figer} D.~F.,   {Puga} E.,  2012a, \mnras, 419, 1860

\bibitem[\protect\citeauthoryear{{Davies} et~al.,}{{Davies}
  et~al.}{2012b}]{davies12a}
{Davies} B.,  et~al., 2012b, \mn@doi [\mnras]
  {10.1111/j.1365-2966.2011.19736.x}, \href
  {http://adsabs.harvard.edu/abs/2012MNRAS.419.1871D} {419, 1871}

\bibitem[\protect\citeauthoryear{{Dias}, {Alessi}, {Moitinho}  \&
  {L{\'e}pine}}{{Dias} et~al.}{2002}]{dias02}
{Dias} W.~S.,  {Alessi} B.~S.,  {Moitinho} A.,   {L{\'e}pine} J.~R.~D.,  2002,
  \mn@doi [\aap] {10.1051/0004-6361:20020668}, \href
  {http://adsabs.harvard.edu/abs/2002A%26A...389..871D} {389, 871}

\bibitem[\protect\citeauthoryear{{Dutra}, {Bica}, {Soares}  \&
  {Barbuy}}{{Dutra} et~al.}{2003}]{dutra03}
{Dutra} C.~M.,  {Bica} E.,  {Soares} J.,   {Barbuy} B.,  2003, \aap, 400, 533

\bibitem[\protect\citeauthoryear{{Eenens} \& {Williams}}{{Eenens} \&
  {Williams}}{1994}]{eenens94}
{Eenens} P.~R.~J.,  {Williams} P.~M.,  1994, \mn@doi [\mnras]
  {10.1093/mnras/269.4.1082}, \href
  {http://adsabs.harvard.edu/abs/1994MNRAS.269.1082E} {269, 1082}

\bibitem[\protect\citeauthoryear{{Eenens}, {Williams}  \& {Wade}}{{Eenens}
  et~al.}{1991}]{eenens91}
{Eenens} P.~R.~J.,  {Williams} P.~M.,   {Wade} R.,  1991, \mn@doi [\mnras]
  {10.1093/mnras/252.2.300}, \href
  {http://adsabs.harvard.edu/abs/1991MNRAS.252..300E} {252, 300}

\bibitem[\protect\citeauthoryear{{Eikenberry} et~al.,}{{Eikenberry}
  et~al.}{2004}]{eikenberry04}
{Eikenberry} S.~S.,  et~al., 2004, \mn@doi [\apj] {10.1086/422180}, \href
  {http://adsabs.harvard.edu/abs/2004ApJ...616..506E} {616, 506}

\bibitem[\protect\citeauthoryear{{Eldridge}, {Langer}  \& {Tout}}{{Eldridge}
  et~al.}{2011}]{eldridge11}
{Eldridge} J.~J.,  {Langer} N.,   {Tout} C.~A.,  2011, \mn@doi [\mnras]
  {10.1111/j.1365-2966.2011.18650.x}, \href
  {http://adsabs.harvard.edu/abs/2011MNRAS.414.3501E} {414, 3501}

\bibitem[\protect\citeauthoryear{{Faherty}, {Shara}, {Zurek}, {Kanarek}  \&
  {Moffat}}{{Faherty} et~al.}{2014}]{faherty14}
{Faherty} J.~K.,  {Shara} M.~M.,  {Zurek} D.,  {Kanarek} G.,   {Moffat}
  A.~F.~J.,  2014, \aj, 147, 115

\bibitem[\protect\citeauthoryear{{Figer}, {McLean}  \& {Najarro}}{{Figer}
  et~al.}{1997}]{figer97}
{Figer} D.~F.,  {McLean} I.~S.,   {Najarro} F.,  1997, \mn@doi [\apj]
  {10.1086/304488}, \href {http://adsabs.harvard.edu/abs/1997ApJ...486..420F}
  {486, 420}

\bibitem[\protect\citeauthoryear{{Figer}, {Najarro}, {Morris}, {McLean},
  {Geballe}, {Ghez}  \& {Langer}}{{Figer} et~al.}{1998}]{figer98}
{Figer} D.~F.,  {Najarro} F.,  {Morris} M.,  {McLean} I.~S.,  {Geballe} T.~R.,
  {Ghez} A.~M.,   {Langer} N.,  1998, \mn@doi [\apj] {10.1086/306237}, \href
  {http://adsabs.harvard.edu/abs/1998ApJ...506..384F} {506, 384}

\bibitem[\protect\citeauthoryear{{Figer}, {Najarro}  \& {Kudritzki}}{{Figer}
  et~al.}{2004}]{figer04}
{Figer} D.~F.,  {Najarro} F.,   {Kudritzki} R.~P.,  2004, \mn@doi [\apjl]
  {10.1086/423306}, \href {http://adsabs.harvard.edu/abs/2004ApJ...610L.109F}
  {610, L109}

\bibitem[\protect\citeauthoryear{{Figer}, {MacKenty}, {Robberto}, {Smith},
  {Najarro}, {Kudritzki}  \& {Herrero}}{{Figer} et~al.}{2006}]{figer06}
{Figer} D.~F.,  {MacKenty} J.~W.,  {Robberto} M.,  {Smith} K.,  {Najarro} F.,
  {Kudritzki} R.~P.,   {Herrero} A.,  2006, \mn@doi [\apj] {10.1086/503275},
  \href {http://adsabs.harvard.edu/abs/2006ApJ...643.1166F} {643, 1166}

\bibitem[\protect\citeauthoryear{{Fujii} \& {Portegies Zwart}}{{Fujii} \&
  {Portegies Zwart}}{2011}]{fujii11}
{Fujii} M.~S.,  {Portegies Zwart} S.,  2011, \mn@doi [Science]
  {10.1126/science.1211927}, \href
  {http://adsabs.harvard.edu/abs/2011Sci...334.1380F} {334, 1380}

\bibitem[\protect\citeauthoryear{{Gaia Collaboration} et~al.,}{{Gaia
  Collaboration} et~al.}{2016}]{GAIA-DR1}
{Gaia Collaboration} et~al., 2016, \mn@doi [\aap]
  {10.1051/0004-6361/201629512}, \href
  {http://adsabs.harvard.edu/abs/2016A%26A...595A...2G} {595, A2}

\bibitem[\protect\citeauthoryear{{Geballe}, {Najarro}  \& {Figer}}{{Geballe}
  et~al.}{2000}]{geballe00}
{Geballe} T.~R.,  {Najarro} F.,   {Figer} D.~F.,  2000, \mn@doi [\apjl]
  {10.1086/312501}, \href {http://adsabs.harvard.edu/abs/2000ApJ...530L..97G}
  {530, L97}

\bibitem[\protect\citeauthoryear{{Georgelin} \& {Georgelin}}{{Georgelin} \&
  {Georgelin}}{1976}]{georgelin76}
{Georgelin} Y.~M.,  {Georgelin} Y.~P.,  1976, \aap, 49, 57

\bibitem[\protect\citeauthoryear{{Gvaramadze}, {Kniazev}  \&
  {Berdnikov}}{{Gvaramadze} et~al.}{2015}]{gvaramadze15}
{Gvaramadze} V.~V.,  {Kniazev} A.~Y.,   {Berdnikov} L.~N.,  2015, \mn@doi
  [\mnras] {10.1093/mnras/stv2278}, \href
  {http://adsabs.harvard.edu/abs/2015MNRAS.454.3710G} {454, 3710}

\bibitem[\protect\citeauthoryear{{Hadfield}, {van Dyk}, {Morris}, {Smith},
  {Marston}  \& {Peterson}}{{Hadfield} et~al.}{2007}]{hadfield07}
{Hadfield} L.~J.,  {van Dyk} S.~D.,  {Morris} P.~W.,  {Smith} J.~D.,  {Marston}
  A.~P.,   {Peterson} D.~E.,  2007, \mnras, 376, 248

\bibitem[\protect\citeauthoryear{{Hainich}, {Pasemann}, {Todt}, {Shenar},
  {Sander}  \& {Hamann}}{{Hainich} et~al.}{2015}]{hainich15}
{Hainich} R.,  {Pasemann} D.,  {Todt} H.,  {Shenar} T.,  {Sander} A.,
  {Hamann} W.-R.,  2015, \mn@doi [\aap] {10.1051/0004-6361/201526241}, \href
  {http://adsabs.harvard.edu/abs/2015A%26A...581A..21H} {581, A21}

\bibitem[\protect\citeauthoryear{{H{\'e}nault-Brunet}
  et~al.,}{{H{\'e}nault-Brunet} et~al.}{2012}]{henault12}
{H{\'e}nault-Brunet} V.,  et~al., 2012, \mn@doi [\aap]
  {10.1051/0004-6361/201219471}, \href
  {http://adsabs.harvard.edu/abs/2012A%26A...546A..73H} {546, A73}

\bibitem[\protect\citeauthoryear{{Hillier}, {Jones}  \& {Hyland}}{{Hillier}
  et~al.}{1983}]{hillier83}
{Hillier} D.~J.,  {Jones} T.~J.,   {Hyland} A.~R.,  1983, \mn@doi [\apj]
  {10.1086/161189}, \href {http://adsabs.harvard.edu/abs/1983ApJ...271..221H}
  {271, 221}

\bibitem[\protect\citeauthoryear{{Homeier}, {Blum}, {Pasquali}, {Conti}  \&
  {Damineli}}{{Homeier} et~al.}{2003}]{homeier03}
{Homeier} N.~L.,  {Blum} R.~D.,  {Pasquali} A.,  {Conti} P.~S.,   {Damineli}
  A.,  2003, \aap, 408, 153

\bibitem[\protect\citeauthoryear{{Hopewell} et~al.,}{{Hopewell}
  et~al.}{2005}]{hopewell05}
{Hopewell} E.~C.,  et~al., 2005, \mn@doi [\mnras]
  {10.1111/j.1365-2966.2005.09487.x}, \href
  {http://adsabs.harvard.edu/abs/2005MNRAS.363..857H} {363, 857}

\bibitem[\protect\citeauthoryear{{Howarth} \& {Schmutz}}{{Howarth} \&
  {Schmutz}}{1992}]{howarth92}
{Howarth} I.~D.,  {Schmutz} W.,  1992, \aap, \href
  {http://adsabs.harvard.edu/abs/1992A%26A...261..503H} {261, 503}

\bibitem[\protect\citeauthoryear{{Humphreys}, {Weis}, {Davidson}  \&
  {Gordon}}{{Humphreys} et~al.}{2016}]{humphreys16}
{Humphreys} R.~M.,  {Weis} K.,  {Davidson} K.,   {Gordon} M.~S.,  2016, \mn@doi
  [\apj] {10.3847/0004-637X/825/1/64}, \href
  {http://adsabs.harvard.edu/abs/2016ApJ...825...64H} {825, 64}

\bibitem[\protect\citeauthoryear{{Jeffries} et~al.,}{{Jeffries}
  et~al.}{2014}]{jeffries14}
{Jeffries} R.~D.,  et~al., 2014, \mn@doi [\aap] {10.1051/0004-6361/201323288},
  \href {http://adsabs.harvard.edu/abs/2014A%26A...563A..94J} {563, A94}

\bibitem[\protect\citeauthoryear{{Kanarek}, {Shara}, {Faherty}, {Zurek}  \&
  {Moffat}}{{Kanarek} et~al.}{2015}]{kanarek15}
{Kanarek} G.,  {Shara} M.,  {Faherty} J.,  {Zurek} D.,   {Moffat} A.,  2015,
  \mn@doi [\mnras] {10.1093/mnras/stv1342}, \href
  {http://adsabs.harvard.edu/abs/2015MNRAS.452.2858K} {452, 2858}

\bibitem[\protect\citeauthoryear{{Kniazev}, {Gvaramadze}  \&
  {Berdnikov}}{{Kniazev} et~al.}{2015}]{kniazev15}
{Kniazev} A.~Y.,  {Gvaramadze} V.~V.,   {Berdnikov} L.~N.,  2015, \mn@doi
  [\mnras] {10.1093/mnrasl/slv023}, \href
  {http://adsabs.harvard.edu/abs/2015MNRAS.449L..60K} {449, L60}

\bibitem[\protect\citeauthoryear{{Kniazev}, {Gvaramadze}  \&
  {Berdnikov}}{{Kniazev} et~al.}{2016}]{kniazev16}
{Kniazev} A.~Y.,  {Gvaramadze} V.~V.,   {Berdnikov} L.~N.,  2016, \mn@doi
  [\mnras] {10.1093/mnras/stw889}, \href
  {http://adsabs.harvard.edu/abs/2016MNRAS.459.3068K} {459, 3068}

\bibitem[\protect\citeauthoryear{{Kothes} \& {Dougherty}}{{Kothes} \&
  {Dougherty}}{2007}]{kothes07}
{Kothes} R.,  {Dougherty} S.~M.,  2007, \mn@doi [\aap]
  {10.1051/0004-6361:20077309}, \href
  {http://adsabs.harvard.edu/abs/2007A%26A...468..993K} {468, 993}

\bibitem[\protect\citeauthoryear{{Koumpia} \& {Bonanos}}{{Koumpia} \&
  {Bonanos}}{2012}]{koumpia12}
{Koumpia} E.,  {Bonanos} A.~Z.,  2012, \mn@doi [\aap]
  {10.1051/0004-6361/201219465}, \href
  {http://adsabs.harvard.edu/abs/2012A%26A...547A..30K} {547, A30}

\bibitem[\protect\citeauthoryear{{Kurtev}, {Borissova}, {Georgiev}, {Ortolani}
  \& {Ivanov}}{{Kurtev} et~al.}{2007}]{kurtev07}
{Kurtev} R.,  {Borissova} J.,  {Georgiev} L.,  {Ortolani} S.,   {Ivanov} V.~D.,
   2007, \mn@doi [\aap] {10.1051/0004-6361:20066706}, \href
  {http://adsabs.harvard.edu/abs/2007A%26A...475..209K} {475, 209}

\bibitem[\protect\citeauthoryear{{Langer}}{{Langer}}{2012}]{langer12}
{Langer} N.,  2012, \mn@doi [\araa] {10.1146/annurev-astro-081811-125534},
  \href {http://adsabs.harvard.edu/abs/2012ARA%26A..50..107L} {50, 107}

\bibitem[\protect\citeauthoryear{{Larson}}{{Larson}}{2003}]{larson03}
{Larson} R.~B.,  2003, in {De Buizer} J.~M.,  {van der Bliek} N.~S.,  eds,  ASP
  Conf Ser Vol. 287, Galactic Star Formation Across the Stellar Mass Spectrum.
  pp 65--80 (\mn@eprint {} {astro-ph/0205466})

\bibitem[\protect\citeauthoryear{{Liermann}, {Hamann}  \&
  {Oskinova}}{{Liermann} et~al.}{2009}]{liermann09}
{Liermann} A.,  {Hamann} W.-R.,   {Oskinova} L.~M.,  2009, \mn@doi [\aap]
  {10.1051/0004-6361:200810371}, \href
  {http://adsabs.harvard.edu/abs/2009A%26A...494.1137L} {494, 1137}

\bibitem[\protect\citeauthoryear{{Liermann}, {Hamann}  \&
  {Oskinova}}{{Liermann} et~al.}{2012}]{liermann12}
{Liermann} A.,  {Hamann} W.-R.,   {Oskinova} L.~M.,  2012, \mn@doi [\aap]
  {10.1051/0004-6361/201117534}, \href
  {http://adsabs.harvard.edu/abs/2012A%26A...540A..14L} {540, A14}

\bibitem[\protect\citeauthoryear{{Lundstrom} \& {Stenholm}}{{Lundstrom} \&
  {Stenholm}}{1984}]{lundstrom84}
{Lundstrom} I.,  {Stenholm} B.,  1984, \aaps, \href
  {http://adsabs.harvard.edu/abs/1984A%26AS...58..163L} {58, 163}

\bibitem[\protect\citeauthoryear{{Ma{\'{\i}}z Apell{\'a}niz}
  et~al.,}{{Ma{\'{\i}}z Apell{\'a}niz} et~al.}{2013}]{gosc13}
{Ma{\'{\i}}z Apell{\'a}niz} J.,  et~al., 2013, in Massive Stars: From alpha to
  Omega. p.~198 (\mn@eprint {arXiv} {1306.6417})

\bibitem[\protect\citeauthoryear{{Marston}, {Mauerhan}, {Van Dyk}, {Cohen}  \&
  {Morris}}{{Marston} et~al.}{2013}]{marston13}
{Marston} A.,  {Mauerhan} J.~C.,  {Van Dyk} S.,  {Cohen} M.,   {Morris} P.,
  2013, in Massive Stars: From alpha to Omega. p.~167 (\mn@eprint {arXiv}
  {1309.1584})

\bibitem[\protect\citeauthoryear{{Massey} \& {Johnson}}{{Massey} \&
  {Johnson}}{1993}]{massey93}
{Massey} P.,  {Johnson} J.,  1993, \mn@doi [\aj] {10.1086/116487}, \href
  {http://adsabs.harvard.edu/abs/1993AJ....105..980M} {105, 980}

\bibitem[\protect\citeauthoryear{{Massey}, {Neugent}, {Morrell}  \&
  {Hillier}}{{Massey} et~al.}{2014}]{massey14}
{Massey} P.,  {Neugent} K.~F.,  {Morrell} N.,   {Hillier} D.~J.,  2014, \mn@doi
  [\apj] {10.1088/0004-637X/788/1/83}, \href
  {http://adsabs.harvard.edu/abs/2014ApJ...788...83M} {788, 83}

\bibitem[\protect\citeauthoryear{{Massey}, {Neugent}  \& {Morrell}}{{Massey}
  et~al.}{2015a}]{Massey15b}
{Massey} P.,  {Neugent} K.~F.,   {Morrell} N.~I.,  2015a, in {Hamann} W.-R.,
  {Sander} A.,   {Todt} H.,  eds, Wolf-Rayet Stars: Proceedings of an
  International Workshop, Universit{\"a}tsverlag Potsdam, p.35-42. pp 35--42
  (\mn@eprint {arXiv} {1507.07297})

\bibitem[\protect\citeauthoryear{{Massey}, {Neugent}  \& {Morrell}}{{Massey}
  et~al.}{2015b}]{Massey15a}
{Massey} P.,  {Neugent} K.~F.,   {Morrell} N.,  2015b, \mn@doi [\apj]
  {10.1088/0004-637X/807/1/81}, \href
  {http://adsabs.harvard.edu/abs/2015ApJ...807...81M} {807, 81}

\bibitem[\protect\citeauthoryear{{Mauerhan}, {Van Dyk}  \& {Morris}}{{Mauerhan}
  et~al.}{2009}]{mauerhan09}
{Mauerhan} J.~C.,  {Van Dyk} S.~D.,   {Morris} P.~W.,  2009, \mn@doi [\pasp]
  {10.1086/603544}, \href {http://adsabs.harvard.edu/abs/2009PASP..121..591M}
  {121, 591}

\bibitem[\protect\citeauthoryear{{Mauerhan}, {Muno}, {Morris}, {Stolovy}  \&
  {Cotera}}{{Mauerhan} et~al.}{2010}]{mauerhan10}
{Mauerhan} J.~C.,  {Muno} M.~P.,  {Morris} M.~R.,  {Stolovy} S.~R.,   {Cotera}
  A.,  2010, \apj, 710, 706

\bibitem[\protect\citeauthoryear{{Mauerhan}, {Van Dyk}  \& {Morris}}{{Mauerhan}
  et~al.}{2011}]{mauerhan11}
{Mauerhan} J.~C.,  {Van Dyk} S.~D.,   {Morris} P.~W.,  2011, \aj, 142, 40

\bibitem[\protect\citeauthoryear{{Mercer} et~al.,}{{Mercer}
  et~al.}{2005}]{mercer05}
{Mercer} E.~P.,  et~al., 2005, \apj, 635, 560

\bibitem[\protect\citeauthoryear{{Messineo}, {Menten}, {Churchwell}  \&
  {Habing}}{{Messineo} et~al.}{2012}]{messineo12}
{Messineo} M.,  {Menten} K.~M.,  {Churchwell} E.,   {Habing} H.,  2012, \aap,
  537, A10

\bibitem[\protect\citeauthoryear{{Meynet} \& {Maeder}}{{Meynet} \&
  {Maeder}}{2003}]{meynet03}
{Meynet} G.,  {Maeder} A.,  2003, \mn@doi [\aap] {10.1051/0004-6361:20030512},
  \href {http://adsabs.harvard.edu/abs/2003A%26A...404..975M} {404, 975}

\bibitem[\protect\citeauthoryear{{Miszalski}, {Miko{\l}ajewska}  \&
  {Udalski}}{{Miszalski} et~al.}{2013}]{miszalski13}
{Miszalski} B.,  {Miko{\l}ajewska} J.,   {Udalski} A.,  2013, \mn@doi [\mnras]
  {10.1093/mnras/stt673}, \href
  {http://adsabs.harvard.edu/abs/2013MNRAS.432.3186M} {432, 3186}

\bibitem[\protect\citeauthoryear{{Najarro}, {de la Fuente}, {Geballe}, {Figer}
  \& {Hillier}}{{Najarro} et~al.}{2015}]{najarro15}
{Najarro} F.,  {de la Fuente} D.,  {Geballe} T.~R.,  {Figer} D.~F.,   {Hillier}
  D.~J.,  2015, in {Hamann} W.-R.,  {Sander} A.,   {Todt} H.,  eds, Wolf-Rayet
  Stars: Proceedings of an International Workshop, Universit{\"a}tsverlag
  Potsdam, p.113-116. pp 113--116

\bibitem[\protect\citeauthoryear{{Nebot G{\'o}mez-Mor{\'a}n}, {Motch},
  {Pineau}, {Carrera}, {Pakull}  \& {Riddick}}{{Nebot G{\'o}mez-Mor{\'a}n}
  et~al.}{2015}]{nebot15}
{Nebot G{\'o}mez-Mor{\'a}n} A.,  {Motch} C.,  {Pineau} F.-X.,  {Carrera} F.~J.,
   {Pakull} M.~W.,   {Riddick} F.,  2015, \mn@doi [\mnras]
  {10.1093/mnras/stv1020}, \href
  {http://adsabs.harvard.edu/abs/2015MNRAS.452..884N} {452, 884}

\bibitem[\protect\citeauthoryear{{O'Connor} \& {Ott}}{{O'Connor} \&
  {Ott}}{2011}]{oConnor11}
{O'Connor} E.,  {Ott} C.~D.,  2011, \mn@doi [\apj]
  {10.1088/0004-637X/730/2/70}, \href
  {http://adsabs.harvard.edu/abs/2011ApJ...730...70O} {730, 70}

\bibitem[\protect\citeauthoryear{{Parker} \& {Dale}}{{Parker} \&
  {Dale}}{2017}]{parker17}
{Parker} R.~J.,  {Dale} J.~E.,  2017, preprint, \href
  {http://adsabs.harvard.edu/abs/2017arXiv170504686P} {} (\mn@eprint {arXiv}
  {1705.04686})

\bibitem[\protect\citeauthoryear{{Parker} \& {Goodwin}}{{Parker} \&
  {Goodwin}}{2007}]{parker07}
{Parker} R.~J.,  {Goodwin} S.~P.,  2007, \mn@doi [\mnras]
  {10.1111/j.1365-2966.2007.12179.x}, \href
  {http://adsabs.harvard.edu/abs/2007MNRAS.380.1271P} {380, 1271}

\bibitem[\protect\citeauthoryear{{Porter}, {Drew}  \& {Lumsden}}{{Porter}
  et~al.}{1998}]{porter98}
{Porter} J.~M.,  {Drew} J.~E.,   {Lumsden} S.~L.,  1998, \aap, \href
  {http://adsabs.harvard.edu/abs/1998A%26A...332..999P} {332, 999}

\bibitem[\protect\citeauthoryear{{Rahman}, {Moon}  \& {Matzner}}{{Rahman}
  et~al.}{2011}]{rahman11}
{Rahman} M.,  {Moon} D.-S.,   {Matzner} C.~D.,  2011, \mn@doi [\apjl]
  {10.1088/2041-8205/743/2/L28}, \href
  {http://adsabs.harvard.edu/abs/2011ApJ...743L..28R} {743, L28}

\bibitem[\protect\citeauthoryear{{Roman-Lopes}}{{Roman-Lopes}}{2011}]{roman11}
{Roman-Lopes} A.,  2011, \mn@doi [\mnras] {10.1111/j.1365-2966.2010.17431.x},
  \href {http://adsabs.harvard.edu/abs/2011MNRAS.410..161R} {410, 161}

\bibitem[\protect\citeauthoryear{{Rosslowe} \& {Crowther}}{{Rosslowe} \&
  {Crowther}}{2015a}]{rosslowe15a}
{Rosslowe} C.~K.,  {Crowther} P.~A.,  2015a, \mnras, 447, 2322

\bibitem[\protect\citeauthoryear{{Rosslowe} \& {Crowther}}{{Rosslowe} \&
  {Crowther}}{2015b}]{rosslowe15b}
{Rosslowe} C.~K.,  {Crowther} P.~A.,  2015b, \mn@doi [\mnras]
  {10.1093/mnras/stv502}, \href
  {http://adsabs.harvard.edu/abs/2015MNRAS.449.2436R} {449, 2436}

\bibitem[\protect\citeauthoryear{{Rousselot}, {Lidman}, {Cuby}, {Moreels}  \&
  {Monnet}}{{Rousselot} et~al.}{2000}]{rousselot00}
{Rousselot} P.,  {Lidman} C.,  {Cuby} J.-G.,  {Moreels} G.,   {Monnet} G.,
  2000, \aap, \href {http://adsabs.harvard.edu/abs/2000A%26A...354.1134R} {354,
  1134}

\bibitem[\protect\citeauthoryear{{Russeil}}{{Russeil}}{2003}]{russeil03}
{Russeil} D.,  2003, \mn@doi [\aap] {10.1051/0004-6361:20021504}, \href
  {http://adsabs.harvard.edu/abs/2003A%26A...397..133R} {397, 133}

\bibitem[\protect\citeauthoryear{{Russeil}, {Adami}, {Amram}, {Le Coarer},
  {Georgelin}, {Marcelin}  \& {Parker}}{{Russeil} et~al.}{2005}]{Russeil05}
{Russeil} D.,  {Adami} C.,  {Amram} P.,  {Le Coarer} E.,  {Georgelin} Y.~M.,
  {Marcelin} M.,   {Parker} Q.,  2005, \mn@doi [\aap]
  {10.1051/0004-6361:20048090}, \href
  {http://adsabs.harvard.edu/abs/2005A%26A...429..497R} {429, 497}

\bibitem[\protect\citeauthoryear{{Sagar}, {Munari}  \& {de Boer}}{{Sagar}
  et~al.}{2001}]{sagar01}
{Sagar} R.,  {Munari} U.,   {de Boer} K.~S.,  2001, \mn@doi [\mnras]
  {10.1046/j.1365-8711.2001.04438.x}, \href
  {http://adsabs.harvard.edu/abs/2001MNRAS.327...23S} {327, 23}

\bibitem[\protect\citeauthoryear{{Sana} et~al.,}{{Sana} et~al.}{2012}]{sana12}
{Sana} H.,  et~al., 2012, \mn@doi [Science] {10.1126/science.1223344}, \href
  {http://adsabs.harvard.edu/abs/2012Sci...337..444S} {337, 444}

\bibitem[\protect\citeauthoryear{{Sander}, {Hamann}  \& {Todt}}{{Sander}
  et~al.}{2012}]{sander12}
{Sander} A.,  {Hamann} W.-R.,   {Todt} H.,  2012, \aap, 540, A144

\bibitem[\protect\citeauthoryear{{Schneider} et~al.,}{{Schneider}
  et~al.}{2014}]{schneider14}
{Schneider} F.~R.~N.,  et~al., 2014, \mn@doi [\apj]
  {10.1088/0004-637X/780/2/117}, \href
  {http://adsabs.harvard.edu/abs/2014ApJ...780..117S} {780, 117}

\bibitem[\protect\citeauthoryear{{Shara}, {Moffat}, {Smith}, {Niemela},
  {Potter}  \& {Lamontagne}}{{Shara} et~al.}{1999}]{shara99}
{Shara} M.~M.,  {Moffat} A.~F.~J.,  {Smith} L.~F.,  {Niemela} V.~S.,  {Potter}
  M.,   {Lamontagne} R.,  1999, \mn@doi [\aj] {10.1086/300908}, \href
  {http://adsabs.harvard.edu/abs/1999AJ....118..390S} {118, 390}

\bibitem[\protect\citeauthoryear{{Shara} et~al.,}{{Shara}
  et~al.}{2009}]{shara09}
{Shara} M.~M.,  et~al., 2009, \aj, 138, 402

\bibitem[\protect\citeauthoryear{{Shara}, {Faherty}, {Zurek}, {Moffat},
  {Gerke}, {Doyon}, {Artigau}  \& {Drissen}}{{Shara} et~al.}{2012}]{shara12}
{Shara} M.~M.,  {Faherty} J.~K.,  {Zurek} D.,  {Moffat} A.~F.~J.,  {Gerke} J.,
  {Doyon} R.,  {Artigau} E.,   {Drissen} L.,  2012, \aj, 143, 149

\bibitem[\protect\citeauthoryear{{Simpson} et~al.,}{{Simpson}
  et~al.}{2012}]{simpson12}
{Simpson} R.~J.,  et~al., 2012, \mn@doi [\mnras]
  {10.1111/j.1365-2966.2012.20770.x}, \href
  {http://adsabs.harvard.edu/abs/2012MNRAS.424.2442S} {424, 2442}

\bibitem[\protect\citeauthoryear{{Skrutskie}, {Cutri}, {Stiening}, {Weinberg},
  {Schneider}, {Carpenter}, {Beichman}  \& {Capps}}{{Skrutskie}
  et~al.}{2006}]{skrutskie06}
{Skrutskie} M.~F.,  {Cutri} R.~M.,  {Stiening} R.,  {Weinberg} M.~D.,
  {Schneider} S.,  {Carpenter} J.~M.,  {Beichman} C.,   {Capps} R. e.~a.,
  2006, \aj, 131, 1163

\bibitem[\protect\citeauthoryear{{Smith}}{{Smith}}{2016}]{smith16}
{Smith} N.,  2016, \mn@doi [\mnras] {10.1093/mnras/stw1533}, \href
  {http://adsabs.harvard.edu/abs/2016MNRAS.461.3353S} {461, 3353}

\bibitem[\protect\citeauthoryear{{Smith} \& {Tombleson}}{{Smith} \&
  {Tombleson}}{2015}]{smith15}
{Smith} N.,  {Tombleson} R.,  2015, \mnras, 447, 598

\bibitem[\protect\citeauthoryear{{Smith}, {Shara}  \& {Moffat}}{{Smith}
  et~al.}{1990}]{smith90}
{Smith} L.~F.,  {Shara} M.~M.,   {Moffat} A.~F.~J.,  1990, \mn@doi [\apj]
  {10.1086/168978}, \href {http://adsabs.harvard.edu/abs/1990ApJ...358..229S}
  {358, 229}

\bibitem[\protect\citeauthoryear{{Smith}, {Crowther}  \& {Prinja}}{{Smith}
  et~al.}{1994}]{smith94}
{Smith} L.~J.,  {Crowther} P.~A.,   {Prinja} R.~K.,  1994, \aap, \href
  {http://adsabs.harvard.edu/abs/1994A%26A...281..833S} {281, 833}

\bibitem[\protect\citeauthoryear{{Smith}, {Shara}  \& {Moffat}}{{Smith}
  et~al.}{1996}]{smith96}
{Smith} L.~F.,  {Shara} M.~M.,   {Moffat} A.~F.~J.,  1996, \mnras, 281, 163

\bibitem[\protect\citeauthoryear{{Smith}, {Vink}  \& {de Koter}}{{Smith}
  et~al.}{2004}]{smith04}
{Smith} N.,  {Vink} J.~S.,   {de Koter} A.,  2004, \mn@doi [\apj]
  {10.1086/424030}, \href {http://adsabs.harvard.edu/abs/2004ApJ...615..475S}
  {615, 475}

\bibitem[\protect\citeauthoryear{{Stead} \& {Hoare}}{{Stead} \&
  {Hoare}}{2009}]{stead09}
{Stead} J.~J.,  {Hoare} M.~G.,  2009, \mnras, 400, 731

\bibitem[\protect\citeauthoryear{{Toal{\'a}}, {Guerrero}, {Ramos-Larios}  \&
  {Guzm{\'a}n}}{{Toal{\'a}} et~al.}{2015}]{toala15}
{Toal{\'a}} J.~A.,  {Guerrero} M.~A.,  {Ramos-Larios} G.,   {Guzm{\'a}n} V.,
  2015, \mn@doi [\aap] {10.1051/0004-6361/201525706}, \href
  {http://adsabs.harvard.edu/abs/2015A%26A...578A..66T} {578, A66}

\bibitem[\protect\citeauthoryear{{Tramper} et~al.,}{{Tramper}
  et~al.}{2015}]{tramper15}
{Tramper} F.,  et~al., 2015, \mn@doi [\aap] {10.1051/0004-6361/201425390},
  \href {http://adsabs.harvard.edu/abs/2015A%26A...581A.110T} {581, A110}

\bibitem[\protect\citeauthoryear{{Vall{\'e}e}}{{Vall{\'e}e}}{2014}]{vallee14}
{Vall{\'e}e} J.~P.,  2014, \mn@doi [\apjs] {10.1088/0067-0049/215/1/1}, \href
  {http://adsabs.harvard.edu/abs/2014ApJS..215....1V} {215, 1}

\bibitem[\protect\citeauthoryear{{Vall{\'e}e}}{{Vall{\'e}e}}{2015}]{vallee15}
{Vall{\'e}e} J.~P.,  2015, \mn@doi [\mnras] {10.1093/mnras/stv862}, \href
  {http://adsabs.harvard.edu/abs/2015MNRAS.450.4277V} {450, 4277}

\bibitem[\protect\citeauthoryear{{Vreux}, {Dennefeld}  \& {Andrillat}}{{Vreux}
  et~al.}{1983}]{vreux83}
{Vreux} J.~M.,  {Dennefeld} M.,   {Andrillat} Y.,  1983, \aaps, \href
  {http://adsabs.harvard.edu/abs/1983A%26AS...54..437V} {54, 437}

\bibitem[\protect\citeauthoryear{{Vreux}, {Andrillat}  \& {Biemont}}{{Vreux}
  et~al.}{1990}]{vreux90}
{Vreux} J.-M.,  {Andrillat} Y.,   {Biemont} E.,  1990, \aap, \href
  {http://adsabs.harvard.edu/abs/1990A%26A...238..207V} {238, 207}

\bibitem[\protect\citeauthoryear{{Wachter}, {Mauerhan}, {Van Dyk}, {Hoard},
  {Kafka}  \& {Morris}}{{Wachter} et~al.}{2010}]{wachter10}
{Wachter} S.,  {Mauerhan} J.~C.,  {Van Dyk} S.~D.,  {Hoard} D.~W.,  {Kafka} S.,
    {Morris} P.~W.,  2010, \aj, 139, 2330

\bibitem[\protect\citeauthoryear{{Walborn}, {Gamen}, {Morrell}, {Barb{\'a}},
  {Fern{\'a}ndez Laj{\'u}s}  \& {Angeloni}}{{Walborn} et~al.}{2017}]{walborn17}
{Walborn} N.~R.,  {Gamen} R.~C.,  {Morrell} N.~I.,  {Barb{\'a}} R.~H.,
  {Fern{\'a}ndez Laj{\'u}s} E.,   {Angeloni} R.,  2017, \mn@doi [\aj]
  {10.3847/1538-3881/aa6195}, \href
  {http://adsabs.harvard.edu/abs/2017AJ....154...15W} {154, 15}

\bibitem[\protect\citeauthoryear{{Weidner}, {Kroupa}  \& {Bonnell}}{{Weidner}
  et~al.}{2010}]{weidner10}
{Weidner} C.,  {Kroupa} P.,   {Bonnell} I.~A.~D.,  2010, \mn@doi [\mnras]
  {10.1111/j.1365-2966.2009.15633.x}, \href
  {http://adsabs.harvard.edu/abs/2010MNRAS.401..275W} {401, 275}

\bibitem[\protect\citeauthoryear{{Wright}, {Bouy}, {Drew}, {Sarro}, {Bertin},
  {Cuillandre}  \& {Barrado}}{{Wright} et~al.}{2016}]{wright16}
{Wright} N.~J.,  {Bouy} H.,  {Drew} J.~E.,  {Sarro} L.~M.,  {Bertin} E.,
  {Cuillandre} J.-C.,   {Barrado} D.,  2016, \mn@doi [\mnras]
  {10.1093/mnras/stw1148}, \href
  {http://adsabs.harvard.edu/abs/2016MNRAS.460.2593W} {460, 2593}

\bibitem[\protect\citeauthoryear{{de Jong} et~al.,}{{de Jong}
  et~al.}{2016}]{4most}
{de Jong} R.~S.,  et~al., 2016, in Ground-based and Airborne Instrumentation
  for Astronomy VI. p. 99081O, \mn@doi{10.1117/12.2232832}

\bibitem[\protect\citeauthoryear{{de Mink}, {Sana}, {Langer}, {Izzard}  \&
  {Schneider}}{{de Mink} et~al.}{2014}]{demink14}
{de Mink} S.~E.,  {Sana} H.,  {Langer} N.,  {Izzard} R.~G.,   {Schneider}
  F.~R.~N.,  2014, \mn@doi [\apj] {10.1088/0004-637X/782/1/7}, \href
  {http://adsabs.harvard.edu/abs/2014ApJ...782....7D} {782, 7}

\bibitem[\protect\citeauthoryear{{de la Fuente}, {Najarro}, {Trombley},
  {Davies}  \& {Figer}}{{de la Fuente} et~al.}{2015}]{delaFuente15}
{de la Fuente} D.,  {Najarro} F.,  {Trombley} C.,  {Davies} B.,   {Figer}
  D.~F.,  2015, \mn@doi [\aap] {10.1051/0004-6361/201425371}, \href
  {http://adsabs.harvard.edu/abs/2015A%26A...575A..10D} {575, A10}

\bibitem[\protect\citeauthoryear{{van der Hucht}}{{van der
  Hucht}}{2001}]{vdh01}
{van der Hucht} K.~A.,  2001, \nar, 45, 135

\makeatother
\end{thebibliography}

\begin{thebibliography}{}
\makeatletter
\relax
\def\mn@urlcharsother{\let\do\@makeother \do\$\do\&\do\#\do\^\do\_\do\%\do\~}
\def\mn@doi{\begingroup\mn@urlcharsother \@ifnextchar [ {\mn@doi@}
  {\mn@doi@[]}}
\def\mn@doi@[#1]#2{\def\@tempa{#1}\ifx\@tempa\@empty \href
  {http://dx.doi.org/#2} {doi:#2}\else \href {http://dx.doi.org/#2} {#1}\fi
  \endgroup}
\def\mn@eprint#1#2{\mn@eprint@#1:#2::\@nil}
\def\mn@eprint@arXiv#1{\href {http://arxiv.org/abs/#1} {{\tt arXiv:#1}}}
\def\mn@eprint@dblp#1{\href {http://dblp.uni-trier.de/rec/bibtex/#1.xml}
  {dblp:#1}}
\def\mn@eprint@#1:#2:#3:#4\@nil{\def\@tempa {#1}\def\@tempb {#2}\def\@tempc
  {#3}\ifx \@tempc \@empty \let \@tempc \@tempb \let \@tempb \@tempa \fi \ifx
  \@tempb \@empty \def\@tempb {arXiv}\fi \@ifundefined
  {mn@eprint@\@tempb}{\@tempb:\@tempc}{\expandafter \expandafter \csname
  mn@eprint@\@tempb\endcsname \expandafter{\@tempc}}}

\bibitem[\protect\citeauthoryear{{Bartzakos}, {Moffat}  \&
  {Niemela}}{{Bartzakos} et~al.}{2001}]{bartzakos01}
{Bartzakos} P.,  {Moffat} A.~F.~J.,   {Niemela} V.~S.,  2001, \mn@doi [\mnras]
  {10.1046/j.1365-8711.2001.04126.x}, \href
  {http://adsabs.harvard.edu/abs/2001MNRAS.324...18B} {324, 18}

\bibitem[\protect\citeauthoryear{{Conti}, {Leep}  \& {Perry}}{{Conti}
  et~al.}{1983}]{conti83}
{Conti} P.~S.,  {Leep} M.~E.,   {Perry} D.~N.,  1983, \mn@doi [\apj]
  {10.1086/160948}, \href {http://adsabs.harvard.edu/abs/1983ApJ...268..228C}
  {268, 228}

\bibitem[\protect\citeauthoryear{{Crowther} \& {Smith}}{{Crowther} \&
  {Smith}}{1996}]{crowther96}
{Crowther} P.~A.,  {Smith} L.~J.,  1996, \aap, \href
  {http://adsabs.harvard.edu/abs/1996A%26A...305..541C} {305, 541}

\bibitem[\protect\citeauthoryear{{Crowther} \& {Walborn}}{{Crowther} \&
  {Walborn}}{2011}]{crowther11}
{Crowther} P.~A.,  {Walborn} N.~R.,  2011, \mn@doi [\mnras]
  {10.1111/j.1365-2966.2011.19129.x}, \href
  {http://adsabs.harvard.edu/abs/2011MNRAS.416.1311C} {416, 1311}

\bibitem[\protect\citeauthoryear{{Crowther}, {Hillier}  \& {Smith}}{{Crowther}
  et~al.}{1995a}]{crowther95a}
{Crowther} P.~A.,  {Hillier} D.~J.,   {Smith} L.~J.,  1995a, \aap, \href
  {http://adsabs.harvard.edu/abs/1995A%26A...293..172C} {293, 172}

\bibitem[\protect\citeauthoryear{{Crowther}, {Hillier}  \& {Smith}}{{Crowther}
  et~al.}{1995b}]{crowther95d}
{Crowther} P.~A.,  {Hillier} D.~J.,   {Smith} L.~J.,  1995b, \aap, \href
  {http://adsabs.harvard.edu/abs/1995A%26A...293..403C} {293, 403}

\bibitem[\protect\citeauthoryear{{Crowther}, {Smith}  \& {Hillier}}{{Crowther}
  et~al.}{1995c}]{crowther95b}
{Crowther} P.~A.,  {Smith} L.~J.,   {Hillier} D.~J.,  1995c, \aap, \href
  {http://adsabs.harvard.edu/abs/1995A%26A...302..457C} {302, 457}

\bibitem[\protect\citeauthoryear{{Crowther}, {De Marco}  \&
  {Barlow}}{{Crowther} et~al.}{1998}]{crowther98a}
{Crowther} P.~A.,  {De Marco} O.,   {Barlow} M.~J.,  1998, \mnras, 296, 367

\bibitem[\protect\citeauthoryear{{Crowther}, {Hadfield}, {Clark}, {Negueruela}
  \& {Vacca}}{{Crowther} et~al.}{2006}]{crowther06}
{Crowther} P.~A.,  {Hadfield} L.~J.,  {Clark} J.~S.,  {Negueruela} I.,
  {Vacca} W.~D.,  2006, \mn@doi [\mnras] {10.1111/j.1365-2966.2006.10952.x},
  \href {http://adsabs.harvard.edu/abs/2006MNRAS.372.1407C} {372, 1407}

\bibitem[\protect\citeauthoryear{{Eenens}, {Williams}  \& {Wade}}{{Eenens}
  et~al.}{1991}]{eenens91}
{Eenens} P.~R.~J.,  {Williams} P.~M.,   {Wade} R.,  1991, \mn@doi [\mnras]
  {10.1093/mnras/252.2.300}, \href
  {http://adsabs.harvard.edu/abs/1991MNRAS.252..300E} {252, 300}

\bibitem[\protect\citeauthoryear{{Figer}, {McLean}  \& {Najarro}}{{Figer}
  et~al.}{1997}]{figer97}
{Figer} D.~F.,  {McLean} I.~S.,   {Najarro} F.,  1997, \mn@doi [\apj]
  {10.1086/304488}, \href {http://adsabs.harvard.edu/abs/1997ApJ...486..420F}
  {486, 420}

\bibitem[\protect\citeauthoryear{{Foellmi}, {Moffat}  \& {Guerrero}}{{Foellmi}
  et~al.}{2003a}]{foellmi03a}
{Foellmi} C.,  {Moffat} A.~F.~J.,   {Guerrero} M.~A.,  2003a, \mn@doi [\mnras]
  {10.1046/j.1365-8711.2003.06052.x}, \href
  {http://adsabs.harvard.edu/abs/2003MNRAS.338..360F} {338, 360}

\bibitem[\protect\citeauthoryear{{Foellmi}, {Moffat}  \& {Guerrero}}{{Foellmi}
  et~al.}{2003b}]{foellmi03b}
{Foellmi} C.,  {Moffat} A.~F.~J.,   {Guerrero} M.~A.,  2003b, \mn@doi [\mnras]
  {10.1046/j.1365-8711.2003.06161.x}, \href
  {http://adsabs.harvard.edu/abs/2003MNRAS.338.1025F} {338, 1025}

\bibitem[\protect\citeauthoryear{{Hainich} et~al.,}{{Hainich}
  et~al.}{2014}]{hainich14}
{Hainich} R.,  et~al., 2014, \mn@doi [\aap] {10.1051/0004-6361/201322696},
  \href {http://adsabs.harvard.edu/abs/2014A%26A...565A..27H} {565, A27}

\bibitem[\protect\citeauthoryear{{Hamann}, {Gr{\"a}fener}  \&
  {Liermann}}{{Hamann} et~al.}{2006}]{hamann06}
{Hamann} W.-R.,  {Gr{\"a}fener} G.,   {Liermann} A.,  2006, \aap, 457, 1015

\bibitem[\protect\citeauthoryear{{Lundstrom} \& {Stenholm}}{{Lundstrom} \&
  {Stenholm}}{1989}]{lundstrom89}
{Lundstrom} I.,  {Stenholm} B.,  1989, \aap, \href
  {http://adsabs.harvard.edu/abs/1989A%26A...218..199L} {218, 199}

\bibitem[\protect\citeauthoryear{{Moffat} \& {Seggewiss}}{{Moffat} \&
  {Seggewiss}}{1986}]{moffat86}
{Moffat} A.~F.~J.,  {Seggewiss} W.,  1986, \mn@doi [\apj] {10.1086/164640},
  \href {http://adsabs.harvard.edu/abs/1986ApJ...309..714M} {309, 714}

\bibitem[\protect\citeauthoryear{{Neugent}, {Massey}  \& {Morrell}}{{Neugent}
  et~al.}{2012}]{neugent12}
{Neugent} K.~F.,  {Massey} P.,   {Morrell} N.,  2012, \mn@doi [\aj]
  {10.1088/0004-6256/144/6/162}, \href
  {http://adsabs.harvard.edu/abs/2012AJ....144..162N} {144, 162}

\bibitem[\protect\citeauthoryear{{Schnurr}, {Moffat}, {St-Louis}, {Morrell}  \&
  {Guerrero}}{{Schnurr} et~al.}{2008}]{schnurr08}
{Schnurr} O.,  {Moffat} A.~F.~J.,  {St-Louis} N.,  {Morrell} N.~I.,
  {Guerrero} M.~A.,  2008, \mn@doi [\mnras] {10.1111/j.1365-2966.2008.13584.x},
  \href {http://adsabs.harvard.edu/abs/2008MNRAS.389..806S} {389, 806}

\bibitem[\protect\citeauthoryear{{Smith}, {Shara}  \& {Moffat}}{{Smith}
  et~al.}{1990}]{smith90}
{Smith} L.~F.,  {Shara} M.~M.,   {Moffat} A.~F.~J.,  1990, \mn@doi [\apj]
  {10.1086/168978}, \href {http://adsabs.harvard.edu/abs/1990ApJ...358..229S}
  {358, 229}

\bibitem[\protect\citeauthoryear{{Smith}, {Shara}  \& {Moffat}}{{Smith}
  et~al.}{1996}]{smith96}
{Smith} L.~F.,  {Shara} M.~M.,   {Moffat} A.~F.~J.,  1996, \mnras, 281, 163

\bibitem[\protect\citeauthoryear{{Tramper} et~al.,}{{Tramper}
  et~al.}{2015}]{tramper15}
{Tramper} F.,  et~al., 2015, \mn@doi [\aap] {10.1051/0004-6361/201425390},
  \href {http://adsabs.harvard.edu/abs/2015A%26A...581A.110T} {581, A110}

\bibitem[\protect\citeauthoryear{{Walborn}, {Drissen}, {Parker}, {Saha},
  {MacKenty}  \& {White}}{{Walborn} et~al.}{1999}]{walborn99}
{Walborn} N.~R.,  {Drissen} L.,  {Parker} J.~W.,  {Saha} A.,  {MacKenty} J.~W.,
    {White} R.~L.,  1999, \mn@doi [\aj] {10.1086/301038}, \href
  {http://adsabs.harvard.edu/abs/1999AJ....118.1684W} {118, 1684}

\makeatother
\end{thebibliography}

\section[]{Spectroscopic observations of IR selected candidates}

\begin{table*}
  \begin{center}
  \caption{Catalogue of candidates spectroscopically observed with NTT/SOFI, indicating emission (em) or absorption (abs) line features.}
    \begin{tabular}{l@{\hspace{2mm}}r@{\hspace{1mm}}r@{\hspace{2mm}}c@{\hspace{1mm}}c@{\hspace{2mm}}
r@{\hspace{2mm}}r@{\hspace{2mm}}r@{\hspace{2mm}}r@{\hspace{2mm}}r@{\hspace{2mm}}r@{\hspace{2mm}}r@{\hspace{2mm}}l}
      \hline
      ID & l   &  b  &  RA & Dec & J & H & K & [3.6] & [4.5] & [5.8] & [8.0] & Notes \\
         &     &     & \multicolumn{2}{c}{J2000} & mag & mag & mag & mag & mag & mag & mag & \\
      \hline
      E\#3 & 298.0981 & --0.3769 & 12:08:52.47 & --62:50:54.9 & 13.34 & 11.67 & 10.47 & 9.67 & 9.21 & 8.92 & 8.49 & WR46-18 (WC6--7) 
\\
      E\#2 & 298.1901 & --0.4807 & 12:09:31.38 & --62:57:57.7 & 14.98 & 13.40 & 12.21 & 11.13 & 10.76 & 10.48 & 10.05 & --\\ 
      E\#4 & 298.5476 & --0.4541 & 12:12:40.16& --62:59:43.7 & 13.56 & 12.41 & 11.56 & 10.76 & 10.45 & 10.12 & 9.85 & --\\ 
      E\#1 & 298.9823 &  +0.3429 & 12:17:23.44& --62:14:16.3 & 14.47 & 13.07 & 12.08 & 11.12 & 10.74 & 10.43 & 10.15 & Br$\gamma$ em\\ 
      B\#3 & 300.4069 &  +0.0016 & 12:29:20.63& --62:45:42.1 & 13.27 & 10.66 & 9.09  & 8.07  & 7.55 & 7.05 & 6.99 & Br$\gamma$ abs\\
      B\#6 & 300.7404 & --0.2304 & 12:32:05.88& --63:01:11.4 & 14.40 & 13.29 & 12.57 & 11.86 & 11.46 & 11.18 & 11.05 & Weak Br$\gamma$ em\\
      D\#1 & 302.1241 & --0.2318 & 12:44:17.82& --62:05:32.8 & 12.78 & 11.45 & 10.64 & 9.75 & 9.44 & 9.06 & 8.85 & Br$\gamma$, 10, 11, He\,{\sc i} 2.058 em\\ 
      B\#9 & 302.1126 &  +0.4871 & 12:44:22.22& --62:22:24.0 & 15.62 & 13.80 & 12.75 & 11.81 & 11.40 & 11.11 & 10.94 & weak Br$\gamma$ em\\
      B\#13& 302.8599 &  +0.4606 & 12:50:48.98& --62:24:39.8 & 13.32 & 12.92 & 11.09 & 10.39 & 9.95 & 9.70 & 9.40 & WR47-5 (WN6(h))\\
      B\#15& 303.8459 &  +0.0727 & 12:59:26.03& --62:47:05.4 & 14.64 & 12.90 & 11.65 & 10.32 & 9.89 & 9.55 & 9.15 & --\\
      B\#18& 304.2777 & --0.3314 & 13:03:22.21& --63:10:19.4 & 15.18 & 13.51 & 12.59 & 12.08 & 11.60 & 11.18 & 10.97 & Weak Br$\gamma$ em\\
      B\#21& 304.7418 & --0.4276 & 13:07:31.72& --63:14:34.3 & 13.33 & 10.85 &  9.35 & 7.61 & 7.02 & 6.78 & 6.36 & CO 2.3 abs\\
      A\#1 & 305.2360 &  +0.0224 & 13:11:35.63& --62:45:32.6 & 15.98 & 14.01 & 12.64 & 11.04 & 10.58 & 10.11 & 9.72 & --\\
      A\#2 & 305.2674 &  +0.2222 & 13:11:43.79& --62:33:26.9 & 16.07 & 13.24 & 11.33 & 9.25 & 8.62 & 8.07 & 7.70 & --\\
      B\#22& 305.2689 & --0.1301 & 13:11:59.17& --62:54:30.9 & 13.66 & 12.43 & 11.60 & 11.08 & 10.68 & 10.18 & 9.76 & Br$\gamma$, 10, 11 em  \\
      B\#26& 305.6698 & --0.0117 & 13:15:23.76& --62:45:21.7 & 12.99 & 11.69 & 10.83 & 9.96 & 9.56 & 9.23 & 9.03 & Br$\gamma$, 10, 11 em  \\
      B\#27& 305.7340 & --0.3132 & 13:16:12.44& --63:03:00.6 & 15.14 & 13.87 & 12.93 & 12.16 & 11.67 & 11.37 & 11.13 & Weak Br$\gamma$ em\\
      B\#28& 305.7956 & --0.2272 & 13:16:40.46& --62:57:30.8 & 16.19 & 13.99 & 12.50 & 10.54 & 10.03 & 9.67 & 9.49 &  -- \\
      B\#30& 305.9956 &  +0.0635 & 13:18:09.56& --62:38:56.7 & 15.17 & 13.34 & 12.16 & 11.11 & 10.63 & 10.25 & 9.86 & --\\
      A\#4 & 306.0554 & --0.0271 & 13:18:45.64& --62:43:59.1 & 13.62 & 11.65 & 10.17 & 8.67 & 8.15 & 7.71 & 7.36 & --\\
      B\#31& 306.4620 & --0.2216 & 13:22:29.34& --62:52:49.3 & 15.01 & 13.17 & 12.15 & 11.10 & 10.68 & 10.44 & 10.29 & -- \\
      B\#33& 306.7893 &  +0.2640 & 13:24:47.24& --62:21:27.7 & 15.58 & 13.27 & 11.74 & 9.85 & 9.34 & 9.00 & 8.47 &  --  \\
      D\#2 & 307.2723 & --0.1744 & 13:29:27.72& --62:43:31.3 & 12.87 & 11.60 & 10.82 & 10.00 & 9.67 & 9.32 & 9.02 & Br$\gamma$, 10, 11 em\\ 
      B\#36& 308.2364 & --0.1775 & 13:37:44.90& --62:34:16.9 & 14.58 & 13.21 & 12.38 & 11.48 & 11.01 & 10.76 & 10.64 & --\\
      B\#38& 308.7720 &  +0.3908 & 13:41:21.77& --61:54:48.5 & 14.40 & 12.84 & 11.87 & 10.91 & 10.50 & 10.32 & 10.08 & Br$\gamma$, 10, 11 em\\
      B\#37& 308.7434 & --0.0387 & 13:41:50.01& --62:20:25.8 & 14.94 & 13.21 & 12.18 & 11.31 & 10.83 & 10.66 & 10.49 & WR56-1 (WN5o)\\
      B\#40& 309.0157 &  +0.0950 & 13:43:53.90& --62:09:19.5 & 16.18 & 13.84 & 12.53 & 11.33 & 10.92 & 10.54 & 10.25 & --\\
      B\#41& 309.0367 & --0.4150 & 13:44:58.44& --62:39:01.8 & 15.42 & 13.66 & 12.46 & 10.96 & 10.48 & 10.13 & 9.69 & --\\
      B\#42& 309.2191 &  +0.0369 & 13:45:42.37& --62:10:13.7 & 14.66 & 13.37 & 12.53 & 11.72 & 11.29 & 10.99 & 10.63 & --\\
      A\#5 & 309.8974 &  +0.3827 & 13:50:41.49& --61:41:06.3 & 14.79 & 12.97 & 11.61 & 10.22 & 9.73 & 9.26 & 8.71 & --\\
      B\#45& 310.2014 & --0.3327 & 13:54:38.45& --62:18:32.7 & 15.29 & 13.03 & 11.62 & 10.48 & 10.03 & 9.74 & 9.46 & Br$\gamma$, 10, 11 em\\
      B\#46& 310.2641 & --0.1602 & 13:54:48.21& --62:07:35.6 & 14.29 & 12.89 & 12.07 & 11.34 & 10.93 & 10.62 & 10.33 & --\\
      B\#47& 310.3786 & --0.0643 & 13:55:33.11& --62:00:20.0 & 14.00 & 12.71 & 11.85 & 11.15 & 10.71 & 10.30 & 10.11 & Weak Br$\gamma$ em\\
      B\#48& 310.6160 & --0.3828 & 13:58:12.06& --62:15:15.7 & 13.78 & 12.27 & 11.34 & 10.53 & 10.03 & 9.79 & 9.53 & Br$\gamma$ em\\
      B\#50& 311.1224 &  +0.2097 & 14:01:02.41&	--61:33:00.6 & 13.91 & 12.45 & 11.53 & 10.70 & 10.28 & 9.88 & 9.59 & Br$\gamma$, 10, 11 em\\
      B\#51& 311.3533 &  +0.3626 & 14:02:33.44&	--61:20:27.2 & 13.28 & 11.57 & 10.28 & 9.19 & 8.76 & 8.46 & 8.18 & WR60-7 (WC7--8)\\
      B\#52 &311.3750 &  +0.1595 & 14:03:11.73 &--61:31:49.2 & 15.82 & 13.61 & 12.34 &11.20 &10.76 &10.48 &10.25 & Br$\gamma$, 10, 11 em\\
      A\#6  &311.4747 & --0.4529 & 14:05:26.94 &--62:05:27.2 & 15.34 & 13.69 & 12.57 &11.25 &10.85 &10.62 &10.33 & --\\
      B\#54 &312.1913 & --0.1538 & 14:10:31.18 &--61:35:44.4 & 16.19 & 13.08 & 11.25 & 9.93 & 9.50 & 9.19 & 8.94 & Br$\gamma$ abs\\
      B\#57 &312.4077 &  +0.3376 & 14:11:00.14 &--61:03:42.5 & 15.09 & 12.71 & 11.37 &10.12 & 9.64 & 9.27 & 9.04 & Br$\gamma$, He\,{\sc i} 2.058 em\\
      B\#55 &312.2737 & --0.2463 & 14:11:24.98 &--61:39:31.9 & 15.51 & 13.54 & 12.34 &11.33 &10.80 &10.58 &10.44 & Br$\gamma$ em\\
      B\#56 &312.3518 & --0.3278 & 14:12:15.19 &--61:42:45.2 & 14.90 & 12.98 & 11.76 &10.61 &10.10 & 9.72 & 9.43 & WR60-8 (WN6o)\\
      B\#58 &312.8095 &  +0.4991 & 14:13:44.39 &--60:47:04.4 & 16.43 & 14.09 & 12.65 &10.88 &10.23 & 9.85 & 9.52 & --\\
      B\#59 &313.0292 &  +0.1771 & 14:16:17.64 &--61:01:12.4 & 11.57 & 10.49 &  9.81 & 9.37 & 8.63 & 8.62 & 8.38 & Br$\gamma$, 10, 11, He\,{\sc i} 2.058 em\\
      B\#60 &313.2705 & --0.0194 & 14:18:42.65 &--61:07:37.6 & 15.10 & 13.60 & 12.70 &11.74 &11.30 &11.01 &10.87 & Br$\gamma$ em\\
      C\#6  &313.8298 &  +0.0804 & 14:22:46.4 0&--60:50:38.7 & 11.86 & 10.57 &  9.80 & 8.93 & 8.61 & 8.32 & 8.10 & Br$\gamma$, 10, 11, He\,{\sc i} 2.058 em\\
      A\#7  &314.2797 &  +0.4370 & 14:25:11.50 &--60:21:09.6 & 14.85 & 11.98 & 10.10 & 8.18 & 7.73 & 7.26 & 6.99 & --\\
      C\#3  &314.0615 & --0.3361 & 14:25:46.13 &--61:09:10.4 & 12.93 & 11.79 & 11.04 &10.31 & 9.95 & 9.68 & 9.41 & Br$\gamma$, 10, 11 em\\
      B\#66 &314.9795 &  +0.0376 & 14:31:39.06 &--60:28:07.2 & 15.99 & 14.17 & 13.03 &12.81 &12.28 &11.92 &11.38 & He\,{\sc i} 2.058 em\\
      B\#65 &314.9262 & --0.3702 & 14:32:30.51 &--60:51:58.6 & 13.93 & 12.63 & 11.85 &11.17 &10.63 &10.52 &10.24 & Br$\gamma$, 10, 11 em\\
      B\#69 &316.1646 &  +0.1717 & 14:39:59.99 &--59:52:52.5 & 14.95 & 13.57 & 12.72 &11.81 &11.31 &10.92 &10.78 & --\\
      C\#13 &316.6186 & --0.4678 & 14:45:26.51 &--60:16:28.3 & 13.72 & 12.51 & 11.74 &11.07 &10.73 &10.45 &10.12 & Br$\gamma$, 10, 11 em\\
      B\#72 &317.1176 &  +0.2706 & 14:46:31.62 &--59:23:37.5 & 14.91 & 13.18 & 12.02 &10.61 &10.06 & 9.74 & 9.49 & --\\
      B\#75 &317.4918 & --0.3007 & 14:51:09.26 &--59:44:40.5 & 15.29 & 13.19 & 11.96 &10.59 & 9.95 & 9.51 & 9.33 & --\\
      B\#76 &317.6899 &  +0.0868 & 14:51:11.84 &--59:18:33.8 & 14.56 & 12.66 & 11.37 &10.21 & 9.70 & 9.41 & 8.76 & --\\
      B\#78 &318.0224 &  +0.0688 & 14:53:35.11 &--59:10:35.3 & 15.31 & 13.24 & 11.80 &10.41 & 9.92 & 9.62 & 9.46 & --\\
      B\#80 &318.3162 & --0.0265 & 14:55:57.84 &--59:07:38.1 & 15.57 & 13.12 & 11.49 & 9.74 & 9.08 & 8.65 & 8.18 & --\\
      B\#82 &318.4866 &  +0.0168 & 14:56:59.05 &--59:00:36.7 & 15.77 & 13.67 & 12.35 &10.81 &10.26 & 9.94 & 9.53 & --\\
      B\#81 &318.4855 &  +0.0116 & 14:56:59.74 &--59:00:54.8 & 16.09 & 13.62 & 12.14 &10.93 &10.39 &10.09 & 9.90 & --\\
      B\#83 &318.5683 &  +0.0024 & 14:57:35.91 &--58:59:05.8 & 15.18 & 13.34 & 12.15 &11.25 &10.75 &10.47 &10.28 & Br$\gamma$, 10, 11 em\\
      B\#84 &318.7320 &  +0.2431 & 14:57:51.03 &--58:41:44.6 & 16.37 & 14.29 & 13.05 &12.12 &11.59 &11.11 &10.55 & --\\
      B\#86 &319.1429 &  +0.2102 & 15:00:45.20 &--58:31:52.1 & 13.21 & 12.01 & 11.32 &10.80 &10.35 &10.10 & 9.86 &Br$\gamma$, 10, 11 em \\
      B\#85 &319.0607 & --0.0724 & 15:01:14.05 &--58:49:07.4 & 14.46 & 12.88 & 11.89 &10.82 &10.31 &10.13 & 9.76 & WR64-2 (WN6o)\\
      \hline
    \end{tabular}
    \label{tab:candidates}
    \end{center}
\end{table*}      

\addtocounter{table}{-1}

\begin{table*}
  \begin{center}
  \caption{(continued)}
    \begin{tabular}{l@{\hspace{2mm}}r@{\hspace{1mm}}r@{\hspace{2mm}}c@{\hspace{1mm}}c@{\hspace{2mm}}
r@{\hspace{2mm}}r@{\hspace{2mm}}r@{\hspace{2mm}}r@{\hspace{2mm}}r@{\hspace{2mm}}r@{\hspace{2mm}}r@{\hspace{2mm}}l}
      \hline
      ID & l   &  b  &  RA & Dec & J & H & K & [3.6] & [4.5] & [5.8] & [8.0] & Notes \\
         &     &     & \multicolumn{2}{c}{J2000} & mag & mag & mag & mag & mag & mag & mag & \\
      \hline
      B\#87 &319.4120 &  +0.1536 & 15:02:46.14 &--58:27:06.5 & 12.39 & 11.04 & 10.18 & 9.30 & 8.83 & 8.68 & 8.34 & WR64-3 (WN6o) \\
      B\#88 &319.5721 &  +0.0601 & 15:04:11.15 &--58:27:21.5 & 11.35 & 10.01 &  9.10 & 8.34 & 7.88 & 7.68 & 7.41 & WR64-4 (WN6o+OB)\\
      B\#91 &320.0540 &  +0.0209 & 15:07:31.84 &--58:15:09.6 & 13.11 & 11.66 & 10.80 & 9.91 & 9.47 & 9.25 & 8.97 & WR64-5 (WN6o)\\
      B\#93 &320.5939 &  +0.0474 & 15:10:57.65 &--57:57:28.5 & 13.84 & 12.19 & 11.11 &10.12 & 9.58 & 9.31 & 8.97 & WR64-6 (WN6b)\\
      B\#95 &320.7930 & --0.4404 & 15:14:09.21 &--58:16:26.9 & 15.93 & 13.87 & 12.66 &11.58 &11.12 &10.68 &10.52 & Br$\gamma$ em\\
      C\#34 &321.1822 & --0.0202 & 15:15:00.68 &--57;42:46.0 & 14.10 & 12.67 & 11.76 &10.85 &10.48 &10.18 & 9.89 & Br$\gamma$, 10, 11 em\\
      B\#96 &322.1468 & --0.3860 & 15:22:33.62 &--57:30:26.4 & 12.67 & 10.52 &  9.07 & 7.48 & 6.91 & 6.42 & 6.09 & Br$\gamma$ em\\
      B\#100&323.5860 & --0.2326 & 15:30:42.92 &--56:34:38.5 & 12.52 & 11.21 & 10.35 & 9.46 & 9.00 & 8.80 & 8.51 & CO 2.3 abs\\
      B\#103&324.3283 &  +0.1597 & 15:33:28.03 &--55:49:48.0 & 13.16 & 11.93 & 11.13 &10.46 &10.03 & 9.87 & 9.63 & broad Br$\gamma$, 10, 11 em\\
      B\#106&324.6980 &  +0.2691 & 15:35:09.10 &--55:31:34.7 & 14.67 & 12.98 & 11.97 &10.87 &10.42 &10.17 & 9.91 & --\\
      B\#105&324.6326 & --0.4487 & 15:37:46.51 &--56:08:45.2 & 13.12 & 11.38 &  9.96 & 8.44 & 7.94 & 7.59 & 7.38 & WR70-13 (WC8d)\\
      B\#107&324.9946 & --0.3129 & 15:39:17.02 &--55:49:18.9 & 14.12 & 12.58 & 11.50 &10.35 & 9.76 & 9.58 & 9.22 & WR70-14 (WN4b)\\
      C\#44 &326.2077 &  +0.4099 & 15:43:02.30 &--54:30:55.8 & 13.01 & 11.74 & 10.94 &10.11 & 9.76 & 9.38 & 9.12 & Br$\gamma$ em?\\
      B\#110&326.6728 &  +0.4761 & 15:45:17.55 &--54:10:46.9 & 15.51 & 13.57 & 12.37 &10.93 &10.39 &10.03 & 9.56 & --\\
      C\#48 &326.0985 & --0.3665 & 15:45:43.54 &--55:11:51.5 & 12.84 & 11.42 & 10.49 & 9.55 & 9.17 & 8.84 & 8.57 & Br$\gamma$, 10, 11 em\\
      B\#116&328.3669 &  +0.2962 & 15:54:59.19 &--53:15:35.4 & 15.58 & 13.51 & 12.20 &11.29 &10.76 &10.48 &10.29 & Br$\gamma$ em\\
      B\#118&328.5562 & --0.0472 & 15:57:26.19 &--53:24:05.3 & 14.09 & 12.28 & 11.12 & 9.93 & 9.42 & 9.11 & 8.95 & broad Br$\gamma$, 10, 11 em\\
      B\#119&328.6291 & --0.0897 & 15:57:59.60 &--53:23:12.5 & 16.18 & 14.25 & 13.08 &12.02 &11.56 &11.20 &11.08 & --\\
      B\#123&329.1414 &  +0.2865 & 15:58:57.97 &--52:46:05.4 & 15.61 & 13.59 & 12.40 &11.20 &10.62 &10.34 &10.12 & WR70-15 (WN5o)\\
      B\#125&329.3266 &  +0.2204 & 16:00:10.73 &--52:41:51.6 & 12.97 & 11.48 & 10.57 & 9.75 & 9.30 & 9.08 & 8.89 & Br$\gamma$, 10, 11 em\\
      B\#122&328.9764 & --0.4398 & 16:01:17.50 &--53:25:34.2 & 14.70 & 12.69 & 10.74 & 9.37 & 8.87 & 8.40 & 8.16 & CO 2.3 abs\\
      B\#127&329.4091 &  +0.0227 & 16:01:26.78 &--52:47:35.7 & 15.56 & 13.41 & 12.11 &11.01 &10.50 &10.23 & 9.86 & broad Br$\gamma$ em\\
      B\#124&329.1877 & --0.4186 & 16:02:15.92 &--53:16:16.7 & 15.22 & 13.43 & 12.26 &10.61 &10.16 & 9.88 & 9.62 & --\\
      B\#130&330.1017 &  +0.3914 & 16:03:16.34 &--52:03:34.6 & 13.74 & 12.49 & 11.69 &10.96 &10.54 &10.27 &10.01 & Br$\gamma$, 10, 11 em\\
      C\#29 &330.8606 &  +0.4416 & 16:06:43.15 &--51:31:01.9 & 15.03 & 13.17 & 11.78 &10.33 & 9.86 & 9.53 & 9.08 & --\\
      B\#133&330.6085 &  +0.1173 & 16:06:54.91 &--51:55:36.9 & 14.33 & 12.85 & 11.95 &10.91 &10.52 &10.28 &10.11 & Br$\gamma$, 10, 11 em\\
      B\#132&330.5909 &  +0.0726 & 16:07:01.45 &--51:58:18.3 & 12.56 & 11.14 & 10.27 & 9.50 & 8.98 & 8.74 & 8.46 & WR72-5 (WN6o)\\
      B\#134&330.6751 &  +0.1604 & 16:07:02.83 &--51:51:00.7 & 15.53 & 13.58 & 12.40 &11.28 &10.88 &10.66 &10.42 & Br$\gamma$ em\\
      B\#131&330.3175 & --0.2612 & 16:07:09.70 &--52:24:09.7 & 14.75 & 12.76 & 11.45 &10.23 & 9.72 & 9.29 & 9.11 & --\\
      B\#135&330.7968 &  +0.2625 & 16:07:11.27 &--51:41:34.2 & 13.56 & 12.11 & 11.24 &10.31 & 9.77 & 9.60 & 9.22 & --\\
      C\#31 &331.0646 &  +0.3349 & 16:08:09.05 &--51:27:33.2 & 14.20 & 12.76 & 11.90 &11.03 &10.64 &10.33 &10.07 & Br$\gamma$, 10, 11 em\\
      C\#46 &331.2770 &  +0.3320 & 16:09:10.09 &--51:19:04.9 & 11.97 & 10.66 &  9.86 & 9.00 & 8.61 & 8.33 & 8.06 & Br$\gamma$, 10, 11, He\,{\sc i} 2.058 em\\
      B\#136&331.1094 &  +0.0024 & 16:09:48.50 &--51:40:26.7 & 13.53 & 12.03 & 11.16 &10.49 &10.07 & 9.84 & 9.65 & Br$\gamma$, 10, 11 em\\
      B\#137&331.2573 &  +0.1529 & 16:09:51.13 &--51:27:47.3 & 15.48 & 13.60 & 12.46 &11.69 &11.23 &10.81 &10.81 & Br$\gamma$ em\\
      A\#9  &330.7901 & --0.4539 & 16:10:17.68 &--52:13:33.6 & 12.63 & 11.10 &  9.95 & 8.60 & 8.16 & 7.78 & 7.56 & Br$\gamma$ em\\
      B\#138&331.4221 & --0.0219 & 16:11:23.45 &--51:28:45.6 & 15.71 & 13.70 & 12.43 &11.31 &10.82 &10.34 &10.31 & Br$\gamma$, 10 em\\
      C\#36 &331.2588 & --0.2259 & 16:11:31.09 &--51:44:23.8 & 12.68 & 11.48 & 10.80 &10.35 &10.04 & 9.69 & 9.12 & --\\
      B\#139&332.1603 & --0.3519 & 16:16:16.61 &--51:20:40.5 & 13.76 & 12.38 & 11.49 &10.65 &10.26 &10.00 & 9.83 & Br$\gamma$, 10, He\,{\sc i} 2.058 em\\
      A\#10 &332.7911 &  +0.0695 & 16:17:17.67 &--50:28:11.5 & 14.34 & 12.30 & 10.96 & 9.41 & 8.93 & 8.48 & 8.27 & Br$\gamma$ em\\
      B\#140&333.6774 &  +0.3817 & 16:19:52.89 &--49:37:37.1 & 15.71 & 13.71 & 12.51 &11.15 &10.51 &10.37 &10.09 & --\\
      B\#141&333.6899 & --0.1793 & 16:22:23.37 &--50:00:56.8 & 15.90 & 13.85 & 12.60 &11.27 &10.63 &10.34 & 9.86 & --\\ 
      C\#38 &334.4312 &  +0.2406 & 16:23:46.34 &--49:11:37.7 & 13.09 & 12.02 & 11.37 &10.78 &10.37 &10.38 & 9.90 & Br$\gamma$ em\\
      A\#11 &334.7557 &  +0.2255 & 16:25:13.60 &--48:58:22.3 & 15.28 & 13.34 & 12.21 &11.12 &10.66 &10.38 &10.10 & WR75-31 (WN7o) \\
      B\#143&334.8115 &  +0.1094 & 16:25:58.25 &--49:00:50.7 & 15.05 & 13.15 & 12.03 &10.91 &10.50 &10.22 & 9.97 & Br$\gamma$, 10, 11 em\\
      B\#145&335.4017 &  +0.3561 & 16:27:23.18 &--48:25:06.4 & 14.78 & 13.20 & 12.26 &11.29 &10.84 &10.56 &10.25 & --\\
      B\#144&335.0446 & --0.4287 & 16:29:19.69 &--49:13:09.3 & 15.87 & 13.91 & 12.56 &11.16 &10.60 &10.21 & 9.77 & --\\
      B\#147&335.4470 & --0.1581 & 16:29:49.32 &--48:44:28.2 & 14.73 & 13.18 & 12.23 &11.27 &10.80 &10.60 &10.36 & broad Br$\gamma$, 10, 11 em\\
      C\#41 &336.2729 &  +0.2343 & 16:31:30.87 &--47:52:16.2 & 13.68 & 12.32 & 11.48 &10.66 &10.32 & 9.98 & 9.70 & Br$\gamma$, 10, 11 em\\
      A\#12 &336.2983 &  +0.1534 & 16:31:58.23 &--47:54:28.5 & 15.66 & 13.74 & 12.50 &11.28 &10.81 &10.38 &10.40 & Br$\gamma$, 10, 11 em\\
      A\#13 &336.3959 &  +0.1395 & 16:32:25.70 &--47:50:45.8 & 15.15 & 12.93 & 11.57 &10.35 & 9.75 & 9.54 & 9.35 & WR75-30 (WN7o)\\
      C\#39 &336.4792 & --0.4420 & 16:35:19.23 &--48:10:46.5 & 11.53 & 10.42 &  9.77 & 9.16 & 8.64 & 8.49 & 8.20 & Br$\gamma$, He\,{\sc i} 2.058 em\\
      B\#148&337.2407 &  +0.1689 & 16:35:41.65 &--47:12:19.4 & 14.98 & 13.15 & 11.89 &10.38 & 9.67 & 9.31 & 9.06 & --\\
      B\#149&337.4217 &  +0.1337 & 16:36:33.84 &--47:05:42.9 & 15.12 & 13.10 & 11.89 &10.91 &10.49 &10.17 & 9.98 & broad Br$\gamma$, 10, 11 em\\
      C\#33 &337.5172 &  +0.1107 & 16:37:02.50 &--47:02:24.0 & 13.90 & 12.40 & 11.52 &10.43 &10.09 & 9.80 & 9.56 & Br$\gamma$, 10, 11, He\,{\sc i} 2.058 em\\
      B\#150&337.7587 & --0.0231 & 16:38:34.35 &--46:57:00.2 & 14.61 & 12.56 & 11.12 & 9.61 & 9.13 & 8.74 & 8.61 & Br$\gamma$, 10, 11, He\,{\sc i} 2.058 em\\
      B\#151&337.9344 & --0.1758 & 16:39:55.49 &--46:55:14.6 & 14.95 & 12.85 & 11.56 &10.48 & 9.98 & 9.57 & 9.44 & Br$\gamma$, 10, 11, He\,{\sc i} 2.058 em\\
      B\#152&337.9619 & --0.1518 & 16:39:55.61 &--46:53:03.2 & 14.31 & 12.31 & 10.85 & 9.32 & 8.77 & 8.28 & 7.89 & --\\
      B\#154&338.5451 &  +0.2997 & 16:40:12.92 &--46:08:54.0 & 15.05 & 13.15 & 11.98 &10.88 &10.43 &10.10 & 9.97 & WR76-11 (WN7o)\\
      B\#153&338.2202 & --0.1241 & 16:40:48.38 &--46:40:21.0 & 13.61 & 11.68 & 10.48 & 8.85 & 8.29 & 7.75 & 7.76 & CO 2.3 abs\\
      B\#155&338.9494 & --0.0691 & 16:43:21.24 &--46:05:16.5 & 15.14 & 13.20 & 12.03 &10.90 &10.48 &10.13 & 9.97 & Br$\gamma$, 10, 11, He\,{\sc i} 2.058 em\\
      B\#156&339.1638 &  +0.0831 & 16:43:30.11 &--45:49:34.6 & 14.04 & 12.46 & 10.85 & 9.13 & 8.66 & 8.24 & 8.31 & --\\
      C\#43 &338.9622 & --0.4951 & 16:45:15.96 &--46:21:24.7 & 12.93 & 11.81 & 11.03 &10.21 & 9.88 & 9.61 & 9.37 & Br$\gamma$, 10, 11, He\,{\sc i} 2.058 em\\
      B\#157&339.8284 & --0.0272 & 16:46:27.58 &--45:23:40.3 & 14.96 & 13.29 & 12.10 &10.92 &10.47 &10.04 & 9.94 & Br$\gamma$, 10, 11, He\,{\sc i} 2.058 em\\
      C\#42 &339.3198 & --0.4615 & 16:46:27.72 &--46:03:48.1 & 12.91 & 11.73 & 10.94 &10.16 & 9.84 & 9.51 & 9.25 & Br$\gamma$, 10, 11 em\\
      \hline
    \end{tabular}
    \end{center}
\end{table*}      


\bsp

\label{lastpage}

\end{document}